\documentclass[aps,prb,twocolumn,amsmath,amssymb,nofootinbib,superscriptaddress,floatfix,eqsecnum]{revtex4-1}

\usepackage{amsmath}
\usepackage{amssymb}
\usepackage{amsthm}
\usepackage[dvipdf]{color}
\usepackage{graphicx}
\usepackage{dcolumn}
\usepackage{bm}

\begin{document}
\title{A Microscopic Field Theory for the Universal Shift of Sound Velocity and Dielectric Constant in Low-Temperature Glasses}

  \author{Di Zhou}
\email{dizhou@umich.edu}  

 \affiliation{
 Department of Physics,
  University of Michigan, Ann Arobor, 
 MI 48109-1040, USA
 }

 \affiliation{
 Department of Physics,
  University of Illinois at Urbana-Champaign,
  Urbana, Illinois 61801, USA
 }

\date{\today}

\begin{abstract}
In low-temperature glasses, the sound velocity changes as the logarithmic function of temperature below $10$K: $[ c(T)-c(T_0)]/c(T_0)=\mathcal{C}\ln (T/T_0)$. With increasing temperature starting from $T=0$K, the sound velocity does not increase monotonically, but reaches a maximum at a few Kelvin and decreases at higher temperatures. Tunneling-two-level-system\cite{Phillips1987} (TTLS) model successfully explained the $\ln T$ dependence of sound velocity shift. According to TTLS model, the slope ratio of $\ln T$ dependence of sound velocity shift between lower temperature increasing regime (known as resonance regime) and higher temperature decreasing regime (known as relaxation regime) is $\mathcal{C}^{\rm res }:\mathcal{C}^{\rm rel }=1:-\frac{1}{2}$. In this paper we develop a new glass model, namely the generic coupled block model, to prove that the slope ratio of sound velocity shift between two regimes is $\mathcal{C}^{\rm res }:\mathcal{C}^{\rm rel }=1:-1$ rather than $1:-\frac{1}{2}$, which agrees with the majority of the sound velocity measurements in low-temperature glasses. On the other hand, the dielectric constant shift in low-temperature glasses, $[\epsilon_r(T)-\epsilon_r(T_0)]/\epsilon_r(T_0)$, has a similar logarithmic temperature dependence below $10$K: $[ \epsilon(T)-\epsilon(T_0)]/\epsilon(T_0)=\mathcal{C}\ln (T/T_0)$. Starting from $T=0$K, the dielectric constant does not decrease monotonically, but reaches a minimum at a few Kelvin and increases at higher temperatures. According to TTLS model, the slope ratio between lower temperature resonance regime and higher temperature relaxation regime is $\mathcal{C}^{\rm res}:\mathcal{C}^{\rm rel}=-1:\frac{1}{2}$. In this paper we apply the electric dipole-dipole interaction, to prove that the slope ratio between two regimes is $\mathcal{C}^{\rm res}:\mathcal{C}^{\rm rel}=-1:1$ rather than $-1:\frac{1}{2}$. Our result agrees with the dielectric constant measurements in low-temperature glasses. By developing a real space renormalization technique for glass non-elastic and dielectric susceptibilities, we show that these universal properties of low-temperature glasses essentially come from the $1/r^3$ long range interactions, independent of the materials' microscopic properties. 
\end{abstract}

\maketitle

\section{Introduction}
One of the unambiguous experiments presented by Zeller and Pohl\cite{Zeller1971} is that the low temperature heat capacity of glasses differs significantly from that of crystalline solids. In pure and defect-free insulating crystals the heat capacity is proportional to $T^3$ below $1$K, which comes from the long wavelength phonon vibrational modes. In glasses, however, the heat capacity is the summation of two parts: long wavelength phonon contibution from Debye's theory, and an excess specific heat known as the glass excitations approximated by $C_{\rm excess}=c_1T^{1+\delta}+c_2T^3$, where $\delta$ is of order $0.1$, and $c_1, c_2$ vary for different materials\cite{Lasjaunias1975}. Anderson, Halperin and Varma's\cite{Anderson1972} group and Phillips\cite{Phillips1987} independently developed a microscopic phenomenological model which was later known as tunneling-two-level-system (TTLS) model. It successfully explained the excess heat capacity of glasses, together with several other universal properties such as saturation, echoes etc. 

To further verify the existence of two-level-systems, L. Pich${\rm\acute{e}}$, R. Maynard, S. Hunklinger and J. J$\rm \ddot{a}$ckle\cite{{Hunklinger1974}} studied the influence of two-level-systems on the sound velocity in vitreous silica Suprasil I at temperatures around $0.28{\rm K}<T<4.2 {\rm K}$ and frequencies around $30{\rm MHz}<f<150{\rm MHz}$. Below $10$K, the shift of sound velocity was observed to be the logarithmic function of temperature, in both of the lower temperature resonance regime and higher temperature relaxation regime: $[c(T)-c(T_0)]/c(T_0)|_{\rm res}=\mathcal{C}^{\rm res}\ln(T/T_0)$, $[c(T)-c(T_0)]/c(T_0)|_{\rm rel}=\mathcal{C}^{\rm rel}\ln(T/T_0)$. In the high frequency lower temperature resonance regime, with $\omega\tau\gg1$ ($\tau$ is the effective thermal relaxation time, please refer to section 2(A) for detailed discussions) the sound velocity increases with the increase of temperature: $\mathcal{C}^{\rm res}>0$. The slope $\mathcal{C}^{\rm res}$ is independent of the input phonon frequency. In the low frequency higher temperature relaxation regime with $\omega\tau\ll1$, the sound velocity decreases with increasing temperature: $\mathcal{C}^{\rm rel}<0$. Such increase-decrease transition of sound velocity occurs at the transition point $\omega\tau(T_c)\approx 1$, which means the transition temperature $T_c$ is the function of phonon frequency. However, as long as the sound velocity shift $[c(T)-c(T_0)]/c(T_0)$ enters into relaxation regime, the slope $\mathcal{C}^{\rm rel}$ turns out to be independent of frequency as well. The main purpose of this paper is to discuss the slope of $\ln T$ dependence of sound velocity shift in relaxation and resonance regimes separately, so we assume that the slopes $\mathcal{C}^{\rm rel}$ and $\mathcal{C}^{\rm res}$ are independent of frequency in both of relaxation and resonance regimes. Such universal property has been observed in amorphous materials such as vitreous silica, lithium-doped KCl\cite{Weiss1999} and silica based microscopic cover glasses\cite{Hunklinger1982}, etc.. 

In TTLS model, the phonon strain fields are coupled to two-level-systems. If we calculate phonon-phonon correlation function (phonon Green's function), the pole of Green's function tells us that the dispersion relation of long wavelength phonon is no longer proportional to the wave number $k$, but to receive a slight shift due to the coupling between phonon strain field and two-level-systems. The shift is  proportional to the real part of two-level-system susceptibility. If we use Kramers-Kronig relation to convert the real part of two-level-system susceptibility into the imaginary part and average over the configurations of two-level-systems, we will find that the sound velocity changes as the logarithmic of temperature\cite{Phillips1987, Hunklinger1986}. TTLS model also proves that in resonance regime the sound velocity increases as the increase of temperature, while in relaxation regime it decreases as the increase of temperature. The slope ratio between resonance regime and relaxation regime is $\mathcal{C}^{\rm res }:\mathcal{C}^{\rm rel }=1:-{1}/{2}$, which agrees quite well with the measurements of silica based microscopic cover glass\cite{Hunklinger1982}. However, at least to the author's knowledge, this is the only amorphous material with the slope ratio $\mathcal{C}^{\rm res }:\mathcal{C}^{\rm rel }=1:-{1}/{2}$. Other materials, present the absolute value of $\mathcal{C}^{\rm rel}$ equal or slightly greater than $\mathcal{C}^{\rm res}$: vitreous silica Suprasil I\cite{{Hunklinger1974}}, PdSiCu\cite{Hunklinger1984}, Zr-Nb\cite{Weiss1981}, lithium-doped KCl\cite{Weiss1999}, vitreous silica\cite{Hunklinger2000}, metallic glasses\cite{Bellessa1977} Ni$_{81}$P$_{19}$, etc. (in metallic glass Ni$_{81}$P$_{19}$, the electron-TTLS coupling is relatively weak compared to phonon-TTLS coupling, so conducting electrons are not strong enough to affect the behaviors of sound velocity\cite{Hunklinger1986}). According to S. Hunklinger and C. Enss\cite{Hunklinger2005}, most of the slope ratios of sound velocity shifts in low-temperature glasses are rather $1$ to $-1$, probably due to the mutual interactions between tunneling systems. In this paper our main goal is to set up a generic coupled block model to prove the universal slope ratio of sound velocity shift $\mathcal{C}^{\rm res}:\mathcal{C}^{\rm rel}=1:-1$ in low-temperature glasses. 
\begin{figure}[h]
\includegraphics[scale=0.3]{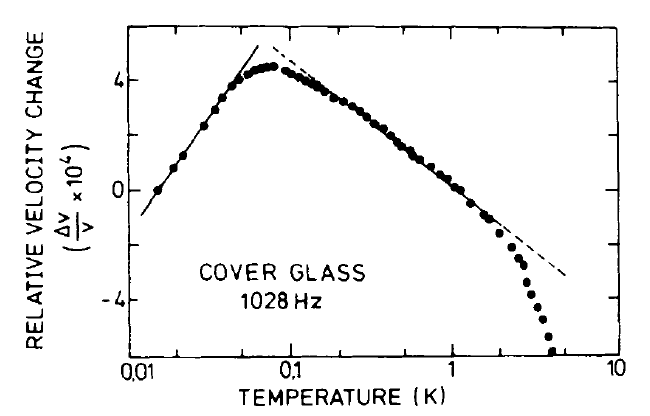}
\caption{The sound velocity shift of silica based microscope cover glass at 1028Hz by S. Hunklinger and A. K. Raychaudhuri\cite{Hunklinger1986}. So far, among the 12 sound velocity shift measurements in low-temperature glasses\cite{Weiss1981, Weiss1999, Hunklinger1976, Hunklinger2000, Hunklinger1976, Cahill1996, Bellessa1977, Hunklinger1974, Hunklinger1984, Hunklinger1982}, this is the only material with the abosulte value of $\mathcal{C}^{\rm rel}$ (the negative slope on the r.h.s.) smaller than $\mathcal{C}^{\rm res}$ (the positive slope on the l.h.s.).}  
\end{figure}

\begin{figure}[h]
\includegraphics[scale=0.3]{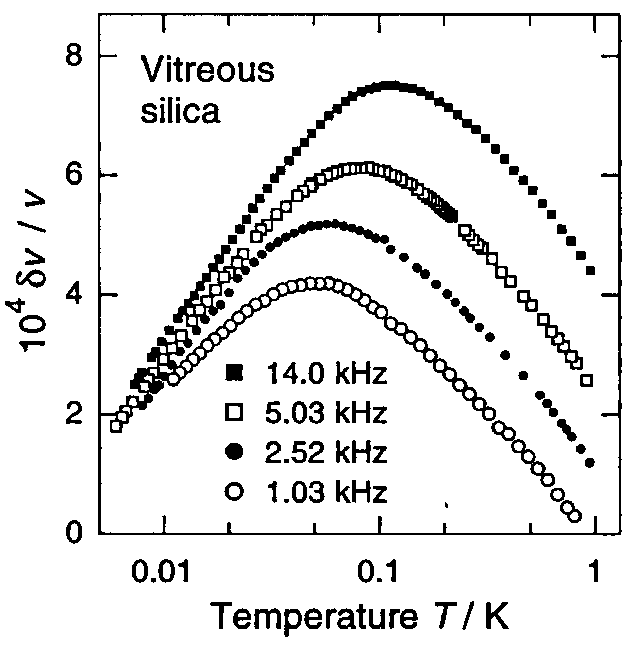}
\caption{The sound velocity shift of Vitreous silica\cite{Hunklinger2005}. The slope ratio of sound velocity shift is $\mathcal{C}^{\rm res}:\mathcal{C}^{\rm rel}=1:-1$ rather than $1:-\frac{1}{2}$ predicted by TTLS model. }  
\end{figure}

In low-temperature glasses below $10$K, the dielectric constant changes as the logarithmic function of temperature as well, in both of the lower temperature resonance regime and higher temperature relaxation regime: $[\epsilon_r(T)-\epsilon_r(T_0)]/\epsilon_r(T_0)|_{\rm res}=\mathcal{C}^{\rm res}\ln(T/T_0)$, $[\epsilon_r(T)-\epsilon_r(T_0)]/\epsilon_r(T_0)|_{\rm rel}=\mathcal{C}^{\rm rel}\ln(T/T_0)$.  In the high frequency lower temperature resonance regime, the dielectric constant decreases with the increase of temperature: $\mathcal{C}^{\rm res}<0$. In the low frequency higher temperature relaxation regime, the dielectric constant increases with increasing temperature: $\mathcal{C}^{\rm rel}>0$. By assuming that the electric field couples to two-level-systems\cite{Hunklinger1981}, TTLS model successfully proved the logarithmic temperature dependence of dielectric constant shift. According to TTLS model, the slope ratio between resonance regime and relaxation regime is $\mathcal{C}^{\rm res}:\mathcal{C}^{\rm rel}=-1:+\frac{1}{2}$. Up to now, we only find 3 dielectric constant measurements in low-temperature glasses: vitreous silica Suprasil W and vitreous As$_2$S$_3$\cite{Schickfus1977}, vitreous silica Suprasil I\cite{Hunklinger1977} and borosilicate glasses (BK7)\cite{Schickfus1990}. All of the measurements show that the slope ratio is $\mathcal{C}^{\rm res }:\mathcal{C}^{\rm rel }=-1:+1$ rather than $-1:+\frac{1}{2}$. In chapter 5 we apply electric dipole-dipole interaction to prove the universal shift of dielectric constant in low-temperature glasses.

\begin{figure}[h]
\includegraphics[scale=0.3]{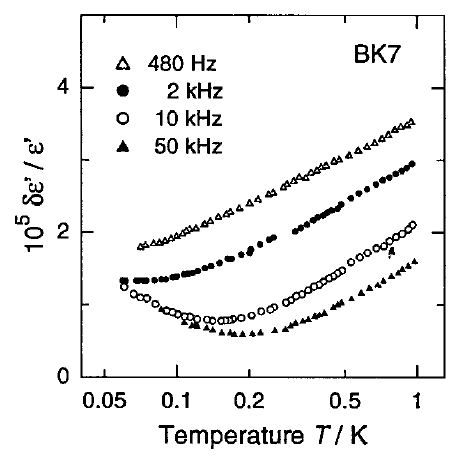}
\caption{The dielectric constant shift of BK7\cite{Hunklinger2005}. The slope ratio of dielectric constant shift is $\mathcal{C}^{\rm res}:\mathcal{C}^{\rm rel}=-1:+1$ rather than $-1:+\frac{1}{2}$ predicted by TTLS model. }  
\end{figure}

Besides the above discrepancies, there are several other problems in TTLS model. First, TTLS model cannot explain other universal properties between $1{\rm K}<T<50{\rm K}$\cite{Leggett2011}, e.g., the universal internal friction $Q^{-1}$ between $10{\rm K}<T<50{\rm K}$\cite{Pohl2002}. Second, there are too many adjustable parameters in TTLS model. For example, the random distribution function $f(\epsilon, \Delta)$ of parameters for diagonal and off-diagonal matrix elements $\epsilon, \Delta$ in two-level-system Hamiltonian, the phonon strain field-TTLS coupling constants $\gamma_{l,t}$, etc.. Third, according to the suggestions by S. Hunklinger and C. Enss, two-level-systems must generate a mutual RKKY-type interaction\cite{Joffrin1976} due to the couplings between two-level-systems and intrinsic phonon strain fields. Taking this virtual-phonon exchange interaction into account may not only change the results in theory, but also question the validity of TTLS model. 

In this paper we focus on the universal shift of sound velocity and dielectric constant in low-temperature glasses. We start by expanding the non-elastic part of glass Hamiltonian in orders of intrinsic phonon strain field $e_{ij}(\vec x)$ to introduce the concept of non-elastic stress tensors. We derive the non-elastic stress-stress susceptibility by using linear response theory. By putting inthe mutual RKKY-type interaction, we derive the renormalization relation between large and small length scale non-elastic stress-stress susceptibilities. The renormalization equation presents a  non-trivial fixed point, which leads to the universal slope ratio $\mathcal{C}^{\rm res}:\mathcal{C}^{\rm rel}=1:-1$ of sound velocity shift in low-temperature glasses. By considering the electric dipole-dipole interaction, we use the same renormalization technique to prove the universal shift of dielectric constant as well.

The organization of this paper is as follows: in section 2 we generalize the glass two-level-system model to multiple-level-system model. Then we derive our generic coupled block model by introducing the mutual RKKY-type interaction between multiple-level-systems. We introduce the most important concept of this paper, namely non-elastic stress-stress susceptibility at the end of section 2(A). In section 3 we treat RKKY-type interaction as a perturbation, to set up the renormalization relation between small and large length scale non-elastic stress-stress susceptibilities. By repeating such recursion relation, we eventually carry out non-elastic stress-stress susceptibility at macroscopic length scales. In section 4 we derive the sound velocity shift in terms of non-elastic relaxation/resonance susceptibilities to prove the universal sound velocity shift in low-temperature glasses. In section 5 we derive the dielectric constant shift in terms of glass dielectric susceptibility. By applying the same real space renormalization technique we prove the universal shift of dielectric constant in low-temperature glasses.

\section{The Model}
\subsection{The Definitions of Glass Non-elastic Hamiltonian, Stress Tensor and Susceptibility}
Let us consider a block of amorphous material (glass). Our purpose is to discuss the universal properties in amorphous materials at low-temperatures below 10K. To explore these properties, we begin our discussion from the famous tunneling-two-level-system model (TTLS model)\cite{Phillips1987}. In this model we assume that there are a group of TTLSs randomly embedded in the glass material, with the location $\vec x_i$ for the $i$-th TTLS. The effective glass Hamiltonian $\hat{H}^{\rm tot}$ in TTLS theory is the summation of long wavelength phonon Hamiltonian $\hat{H}^{\rm el}$, tunneling-two-level-system Hamiltonian, and the coupling between TTLS and strain field (phonon field):
\begin{eqnarray}\label{1}
\hat{H}^{\rm tot} & = & \hat{H}^{\rm el}+\sum_i\frac{1}{2}\left(
\begin{array}{cc}
E_i & 0\\
0 & -E_i\\
\end{array}
\right)\nonumber \\
 & {} & +\frac{1}{2}\sum_i \left(
\begin{array}{cc}
D_i & M_i\\
M_i & -D_i\\
\end{array}
\right)\cdot\bm{e}(\vec x_i)
\end{eqnarray}
where the first, second and third terms stand for long wavelength phonon Hamiltonian (we will also call it ``purely elastic Hamiltonian $\hat{H}^{\rm el}$"), the Hamiltonian of a group of two-level-systems, and the coupling between every two-level-system and intrinsic phonon strain field at corresponding position $\vec x_i$, respectively. The two-level-system Hamiltonian is written in the representation of energy eigenvalue basis, with $E_i=\sqrt{\Delta_i^2+\Delta_{0i}^2}$ the energy splitting; $D_i=\Delta_i/E_i$ and $M_i=\Delta_{0i}/E_i$ are diagonal and off-diagonal matrix elements of the coupling between two-level-system and strain field, and by definition they are no greater than 1; $\bm{e}(\vec x_i)$ is the local intrinsic strain field at the position of the $i$-th two-level-system. 

The purpose of this subsection is to set up our multiple-level-system model from the generalization of 2-level-system model. At this moment, we have not applied any external strain field yet. We will consider external strain field in subsection 2(D). We begin our model by considering a single block of glass with the length scale $L$ much greater than the atomic distance $a\sim 10\AA$. In the subsection 2(C), we will combine a group of such single blocks to form a ``super block". We will consider the RKKY-type interaction between these single blocks, which is generated by virtual phonon exchange process. For now, we do not consider RKKY-type interaction and focus on the Hamiltonian of single block glass only. 

We define intrinsic strain field $e_{ij}(\vec x)$ at position $\vec x$: if $\vec u(\vec x)$ denotes the displacement relative
to some arbitrary reference frame of the matter at point $\vec x$, then strain field is defined as follows
\begin{eqnarray}\label{2}
e_{ij}(\vec x)=\frac{1}{2}\left(\frac{\partial u_i(\vec x)}{\partial x_j}+\frac{\partial u_j(\vec x)}{\partial x_i}\right)
\end{eqnarray}
We write down our general glass Hamiltonian as $\hat{H}^{\rm tot}$. Let us separate out from the glass general Hamiltonian $\hat{H}^{\rm tot}$, the purely elastic contribution $\hat{H}^{\rm el}$. It can be represented by phonon creation-annihilation operators as follows:
\begin{eqnarray}\label{3}
\hat{H}^{\rm el}
 = \sum_{k\alpha}\hbar\omega_{k\alpha}\left(\hat{a}_{k\alpha}^{\dag}\hat{a}_{k\alpha}+\frac{1}{2}\right)
\end{eqnarray}
where $\alpha=l,t$ represents longitudinal and transverse phonon polarizations. We will discuss the purely elastic Hamiltonian $\hat{H}^{\rm el}$ in details in subsection 2(B).

Subtracting the purely elastic part of Hamiltonian $\hat{H}^{\rm el}$, we name the left-over glass Hamiltonian $(\hat{H}^{\rm tot}-\hat{H}^{\rm el})$ as ``the non-elastic part of glass Hamiltonian, $\hat{H}^{\rm non}$". We expand the left-over Hamiltonian $\hat{H}^{\rm non}$ up to the first order expansion of long wavelength intrinsic phonon strain field. We name the coefficient of the first order expansion to be ``non-elastic stress tensor $\hat{T}_{ij}^{\rm non}(\vec x)$", defined as follows:
\begin{eqnarray}\label{4}
 & {} & \hat{H}^{\rm non} = \hat{H}^{\rm non}_0+\int d^3x\sum_{ij}e_{ij}(\vec x)\hat{T}^{\rm non}_{ij}(\vec x)+\mathcal{O}(e_{ij}^2)\nonumber \\
 & {} & \hat{T}_{ij}^{\rm non}(\vec x) = {\delta \hat{H}^{\rm non}}/{\delta e_{ij}(\vec x)}
\end{eqnarray}
Now let us stop for a moment and compare Eq.(\ref{4}) with Eq.(\ref{1}): $\hat{H}^{\rm tot}$ and $\hat{H}^{\rm el}$ in Eq.(\ref{4}) corresponds to the glass total Hamiltonian and purely elastic Hamiltonian in Eq.(\ref{1}) of TTLS model, respectively; the zeroth order expansion of non-elastic Hamiltonian $\hat{H}^{\rm non}$ with respect to strain field $e_{ij}$, $\hat{H}^{\rm non}_0$, is the generalization from two-level-system Hamiltonian to multiple-level-system Hamiltonian; non-elastic stress tensor $\hat{T}_{ij}^{\rm non}$ is the multiple-level generalization of the $2\times 2$ matrix which couples to strain field in TTLS model. In the rest of this paper, we denote $\hat{H}^{\rm non}_0$ to be the non-elastic Hamiltonian excluding the coupling between intrinsic phonon strain $e_{ij}(\vec x)$ and non-elastic stress tensor $\hat{T}_{ij}^{\rm non}(\vec x)$. We denote $\hat{H}^{\rm non}$ to be the non-elastic Hamiltonian including stress-strain coupling (see the first equation of Eqs.(\ref{4})).

Let us denote $|m\rangle$ and $E_m$ to be the $m$-th eigenstate and eigenvalue of $\hat{H}_0^{\rm non}$. Such a set of eigenbasis $|m\rangle$ is a generic multiple-level-system. Now we can define the most important quantity of this paper, namely the non-elastic stress-stress susceptibility (i.e., linear response function). Let us apply an external infinitesimal testing strain, $e_{ij}(\vec x, t)=e_{ij}(\vec k)e^{i(\vec k\cdot \vec x-\omega_k t)}$. The non-elastic Hamiltonian $\hat{H}^{\rm non}$ will provide a stress response $\langle \hat{T}_{ij}^{\rm non}\rangle(\vec x, t)=\langle \hat{T}_{ij}^{\rm non}\rangle e^{i(\vec k\cdot \vec x-\omega_kt)}$. Then we are ready to define the non-elastic stress-stress susceptibility (complex response function\cite{Anderson1986}) $\chi_{ijkl}^{\rm non}(\vec x, \vec x'; t, t')$
\begin{eqnarray}\label{5}
\chi_{ijkl}^{\rm non}(\vec x, \vec x'; \omega, \omega') & = & \int dtdt' \,e^{i\omega t+i\omega't'}\chi_{ijkl}^{\rm non}(\vec x, \vec x'; t, t') \nonumber \\
\chi_{ijkl}^{\rm non}(\vec x, \vec x'; t, t') & = & \frac{\delta  \langle\hat{T}^{\rm non}_{ij}\rangle(\vec x, t)}{\delta e_{kl}(\vec x',t')}
\end{eqnarray}
In the rest of this paper we will always use $\hat{H}$, $\hat{H}_0$, $\chi_{ijkl}$ and $\hat{T}_{ij}$ to represent non-elastic Hamiltonians $\hat{H}^{\rm non}$, $\hat{H}_0^{\rm non}$, susceptibility $\chi_{ijkl}^{\rm non}$ and stress tensor $\hat{T}^{\rm non}_{ij}$ respectively, while we use $\hat{H}^{\rm el}$, $\chi_{ijkl}^{\rm el}$ and $\hat{T}_{ij}^{\rm el}$ to represent the elastic Hamiltonian, susceptibility and stress tensor, respectively.

In Eq.(\ref{5}) the stress response of non-elastic Hamiltonian, $\langle \hat{T}_{ij}\rangle(\vec x, t)$, is defined as follows: (please note from now on we use $\hat{T}_{ij}$ to stand for $\hat{T}_{ij}^{\rm non}$)
\begin{eqnarray}\label{6}
\langle \hat{T}_{ij}\rangle(\vec x, t) & = & \frac{\delta F(t)}{\delta e_{ij}(\vec x, t)}=-\frac{1}{\beta}\frac{\delta }{\delta e_{ij}(\vec x, t)}\ln \mathcal{Z}(t)\nonumber \\
 & = & \sum_m\frac{e^{-\beta E_m(t)}}{\mathcal{Z}(t)}\langle m_I,t|\hat{T}_{ij,(I)}(\vec x, t)|m_I,t\rangle\quad
\end{eqnarray}
where $\mathcal{Z}(t)=\sum_ne^{-\beta E_n(t)}$ is the time-dependent partition function of non-elastic Hamiltonian. With the presence of external testing strain field $e_{ij}(\vec x, t)$, the amorphous material receives a time-dependent perturbation $\int d^3x\,\sum_{ij}e_{ij}(\vec x, t)\hat{T}_{ij}(\vec x)$. In the representation in which $\hat{H}_0$ is diagonal, the perturbation has both of diagonal and off-diagonal matrix elements. The diagonal matrix elements of external perturbation shift the energy eigenvalues: $E_n(t)= E_n+\int d^3x\,e_{ij}(\vec x, t)\langle n|\hat{T}_{ij}(\vec x) |n\rangle$, resulting in the probability function and partition function shift: $e^{-\beta E_n(t)}/\mathcal{Z}(t)$, $\mathcal{Z}(t)=\sum_ne^{-\beta E_n(t)}$. The off-diagonal matrix elements change the eigenstate wavefunctions: $|m_I, t\rangle=\mathcal{T}e^{\frac{1}{i\hbar}\int^t d^3x \, e_{ij}(\vec x, t')\hat{T}_{ij,(I)}(\vec x, t')dt'}|m\rangle$. $\hat{T}_{ij,(I)}(\vec x, t)$ is the stress tensor operator in the interaction picture: $\hat{T}_{ij,(I)}(\vec x, t)=e^{i\hat{H}_0t/\hbar}\hat{T}_{ij}(\vec x)e^{-i\hat{H}_0t/\hbar}$.

In the rest of this paper, we use ``non-elastic susceptibility" to short for ``non-elastic stress-stress susceptibility". According to the definitions in Eq.(\ref{5}) and Eq.(\ref{6}), the non-elastic susceptibility is the function of temperature. However, for notational simplicity we write $\chi_{ijkl}(\vec x, \vec x'; t, t'; T)$ as $\chi_{ijkl}(\vec x,\vec x'; t, t')$ in the rest of this paper. By using linear response theory, we expand $\langle \hat{T}_{ij}\rangle(\vec x, t)$ up to the first order of perturbation $\int d^3x\,e_{ij}(\vec x, t)\hat{T}_{ij}(\vec x)$, to calculate non-elastic susceptibility Eq.(\ref{5}) with the following process:

We use the same language as TTLS model, that the non-elastic susceptibility can be expressed in the relxation and resonance parts. We use $\chi_{ijkl}^{\rm rel}$ and $\chi_{ijkl}^{\rm res}$ to stand for the relaxation and resonance susceptibilities respectively. Let us denote $\tau$ to be the effective thermal relaxation time for the glass single block Hamiltonian. The non-elastic susceptibility is therefore expressed as follows,
\widetext
\begin{eqnarray}\label{7}
\chi_{ijkl}(\vec x, \vec x'; \omega, \omega') & = & (2\pi)\delta(\omega+\omega')\chi_{ijkl}(\vec x, \vec x'; \omega)\nonumber \\
\chi_{ijkl}(\vec x, \vec x'; \omega) & = &  \frac{1}{1-i\omega\tau}{\chi_{ijkl}^{\rm rel}(\vec x, \vec x')}+ \chi_{ijkl}^{\rm res}(\vec x, \vec x'; \omega)\nonumber \\
\chi_{ijkl}^{\rm rel}(\vec x, \vec x') & = & \beta\bigg(\sum_{nm}P_nP_m\langle n|\hat{T}_{ij}(\vec x)|n\rangle\langle m|\hat{T}_{kl}(\vec x')|m\rangle-\sum_n P_n\langle n|\hat{T}_{ij}(\vec x)|n\rangle \langle n|\hat{T}_{kl}(\vec x')|n\rangle\bigg)\nonumber \\
\chi_{ijkl}^{\rm res}(\vec x, \vec x'; \omega) & = & -\frac{1}{\hbar}\sum_{n}\sum_{m\neq n}P_m\frac{\langle n|\hat{T}_{ij}(\vec x)|m\rangle\langle m|\hat{T}_{kl}(\vec x')|n\rangle}{\omega+(E_n-E_m)/\hbar+i\eta}
+
\frac{1}{\hbar}\sum_{m}\sum_{n\neq m}P_n\frac{\langle n|\hat{T}_{ij}(\vec x)|m\rangle\langle m|\hat{T}_{kl}(\vec x')|n\rangle}{\omega+(E_n-E_m)/\hbar+i\eta}
\end{eqnarray}
\endwidetext
where $P_n=e^{-\beta E_n}/\mathcal{Z}$ is the $n$-th eigenstate probability function, $\mathcal{Z}$ is the partition function, $\beta$ is the inverse of temperature $\beta=(k_BT)^{-1}$, and $\eta$ is a phenomenological parameter to represent higher order corrections of non-elastic susceptibility due to the coupling between intrinsic strain field and non-elastic stress tensor: $\sum_{ij}e_{ij}(\vec x)\hat{T}_{ij}(\vec x)$. 

Please note it is an approximation that we use the parameter $\tau$ to represent the effective thermal relaxation time of glass non-elastic part of Hamiltonian $\hat{H}_0$: in principle, the relaxation process of the $n$-th state is the summation of all relaxation processes between $m$-th state and $n$-th state, $\forall m\neq n$. The effective thermal relaxation time $\tau_n$ is therefore different for different quantum number $n$. Generally speaking, one cannot use a simple parameter $\tau$ to stand for the thermal relaxation process for an arbitrary multiple-level-system. However, in this paper we discuss the renormalization equation for non-elastic susceptibility in relaxation and resonance regimes separately. In resonance regime with $\omega\tau_n\gg 1$, the factor $(1-i\omega\tau_n)^{-1}\ll 1$ makes the relaxation susceptibility negligible compared to the resonance susceptibility, while in relaxation regime with $\omega\tau_n\ll 1$, the factor $(1-i\omega\tau_n)^{-1}\approx 1$, which means the relaxation susceptibility is no longer negligible compared to the resonance susceptibility. The only regime which is sensitive to $\tau_n$ is the relaxation-resonance cross-over regime with $\omega\tau_n\approx 1$, and, we are not interested in it within this paper. Therefore in the relaxation susceptibility of Eq.(\ref{7}), we use a simple parameter $\tau$ to tell the difference between resonancec regime ($\omega\tau\gg 1$) and relaxation regime ($\omega\tau\ll 1$). It does not harm our theory to replace $\tau_n$, $\forall n=0,1,2,...$ with a simple parameter $\tau$ to represent the effective thermal relaxation process of our multiple-level-system.

According to the definition of non-elastic susceptibility in Eq.(\ref{7}), the imaginary part of non-elastic susceptibility ${\rm Im\,}\chi_{l,t}(\omega)$ is always negative for arbitrary $\omega>0$.
The negativity of ${\rm Im\,}\chi_{l,t}(\omega>0)$ will be very useful in the discussions of the renormalization behavior of non-elastic susceptibility in section 3.

We further define the space-averaged non-elastic susceptibility for a single block of glass with the volume $V=L^3$, which will be very useful in later discussions,
\begin{eqnarray}\label{7.1}
& {} & \chi_{ijkl}(\omega) = \frac{1}{V}\int_{V} d^3xd^3x'\,\chi_{ijkl}(\vec x, \vec x'; \omega)
\end{eqnarray}
It is very useful to apply the assumption in the rest of this paper, that our amorphous material is rotationally invariant under real space SO(3) group. Therefore, the non-elastic susceptibility obeys the generic form of an arbitrary isotropic 4-indice quantity: $\chi_{ijkl}=(\chi_l-2\chi_t)\delta_{ij}\delta_{kl}+\chi_t(\delta_{ik}\delta_{jl}+\delta_{il}\delta_{jk})$, where $\chi_l$ is the compression modulus and $\chi_t$ is the shear modulus.

In this subsection we define the most important quantities of this paper, namely the non-elastic Hamiltonian, stress tensor and susceptibility. In the following subsection we will set up the problem of universal sound velocity shift in amorphous materials.

\subsection{Phonon-Phonon Correlation Function: the Sound Velocity Shift in Glass}
Now, let us focus on the purely elastic Hamiltonian, Eq.(\ref{3}). Let us consider the phonon-phonon correlation function (Green's function), $\chi^{\rm ph}_{ijkl}(\vec x-\vec x'; t-t')=-\frac{i}{\hbar}\Theta(t-t')\sum_m\frac{e^{-\beta E_m}}{\mathcal{Z}}\langle m|\left[e_{ij}(\vec x, t),e_{kl}(\vec x', t')\right]|m\rangle
$, where $\Theta(t-t')$ is Heaviside step function, and $\mathcal{Z}=\sum_m e^{-\beta E_m}$ is partition function of phonon energy levels. The full phonon-phonon correlation function containing higher order corrections of stress-strain coupling is obtained by Dyson equation:
\begin{eqnarray}\label{10}
\left(\chi^{\rm ph}_{ijkl}\right)^{-1}(\omega)=\left(\chi^{\rm ph}_{ijkl}\right)_0^{-1}(\omega)-\chi_{ijkl}(\omega)
\end{eqnarray}
where $\chi_{ijkl}(\omega)$ is the space-averaged non-elastic susceptibility. The bare longitudinal/transverse phonon-phonon correlation functions are given by $\chi^{\rm ph}_{l,t}(\omega)=({\rho c_{l,t}^2})^{-1}{\omega_{k; l,t}^2}/({\omega^2-\omega_{k; l,t}^2})$. According to Eq.(\ref{10}), the phonon dispersion relation is shifted away from the linear dispersion relation $\omega_k=c_{l,t}k$ due to the correction of non-elastic susceptibility:
\begin{eqnarray}\label{11}
\frac{\Delta \omega_{k; l,t}}{\omega_{k; l,t}}=\frac{{\rm Re}\,\chi_{l,t}(\omega)+i{\,\rm Im\,}\chi_{l,t}(\omega)}{2\rho c_{l,t}^2}
\end{eqnarray}
In the above result, the real part of frequency shift $\Delta \omega_{k; l,t}$ corresponds to the sound velocity shift, while the imaginary part corresponds to the phonon mean free path $l$.

Since the glass non-elastic susceptibility can be separated into the relaxation and resonance parts, we can calculate the contributions to sound velocity shift from the relaxation and resonance susceptibilities, separately. In Eq.(\ref{7}) the relaxation susceptibility is multiplied by a factor of $(1-i\omega\tau)^{-1}$. In low temperature high frequency resonance regime, $\omega \tau\gg 1$ so the prefactor of relaxation susceptibility $(1-i\omega\tau)^{-1}$ makes it negligible compared to the resonance one. The sound velocity shift $\Delta c/c$ is dominated by non-elastic resonance susceptibility. In high temperature low frequency relaxation regime $\omega\tau\ll 1$, so $(1-i\omega\tau)^{-1}\approx 1$, relaxation susceptibility is no longer much smaller than the resonance one. Both of the relaxation and resonance susceptibilities contribute to the sound velocity shift $\Delta c/c$ in resonance regime:
\begin{eqnarray}\label{13}
\frac{\Delta c_{l,t}}{c_{l,t}} & = & \frac{{\rm Re}\,\left(\chi^{\rm res}_{l,t}(k,\omega)+\chi^{\rm rel}_{l,t}(k,\omega)\right)}{2\rho c_{l,t}^2}\quad {\rm relaxation \,\, regime}\nonumber \\
\frac{\Delta c_{l,t}}{c_{l,t}} & = & \frac{{\rm Re}\,\chi^{\rm res}_{l,t}(k,\omega)}{2\rho c_{l,t}^2}\qquad\qquad\quad\qquad\,\,\,\, {\rm resonance \,\, regime}\nonumber \\
\end{eqnarray}
In order to explain the universal property of sound velocity shift in low-temperature glasses, we need to investigate the properties of non-elastic relaxation and resonance susceptibilities.

\subsection{Virtual Phonon Exchange Interactions}
Within single-block considerations, non-elastic stress tensor $\hat{T}_{ij}(\vec x)$ and non-elastic part of glass Hamiltonian $\hat{H}_0$ are simply generalizations from 2-level-system to multiple-level-system, so nothing non-trivial will be obtained within single block considerations. However, if we combine a set of $N_0^3$ single blocks together to form a ``super block", the interactions between single blocks must be taken into account. Since the non-elastic stress tensors are coupled to intrinsic strain field, if we allow virtual phonons to exchange with each other, it will generate a RKKY-type many-body interaction between single blocks. This RKKY-type interaction is the product of stress tensors of single blocks at different locations:
\begin{eqnarray}\label{14}
\hat{V}=\int d^3xd^3x'\sum_{ijkl}\Lambda_{ijkl}(\vec x-\vec x')\hat{T}_{ij}(\vec x)\hat{T}_{kl}(\vec x')
\end{eqnarray}
where the coefficient $\Lambda_{ijkl}(\vec x-\vec x')$ was first derived by J. Joffrin and A. Levelut\cite{Joffrin1976}. We further give a detailed correction to this coefficient, please see Appendix (A):
\begin{eqnarray}\label{15}
 & {} & {\Lambda}_{ijkl}(\vec x-\vec x') = \sum_{\vec k}\Lambda_{ijkl}(\vec k)e^{i\vec k\cdot (\vec x-\vec x')}\nonumber \\
 & {} & \Lambda_{ijkl}(\vec k) = \frac{\alpha}{2\rho c_t^2}\left(\frac{k_ik_jk_kk_l}{k^4}\right)\nonumber \\
 & {} & -\frac{1}{8\rho c_t^2}\left(\frac{k_jk_l\delta_{ik}+k_jk_k\delta_{il}
+k_ik_l\delta_{jk}+k_ik_k\delta_{jl}
}{k^2}\right)\nonumber \\
\end{eqnarray}
where $\alpha=1-{c_t^2}/{c_l^2}$. $\Lambda_{ijkl}(\vec k)$ is the coefficient of RKKY-type interaction in momentum space. $\rho$ is the mass density of amorphous material. $i,j,k,l$ are cartesian coordinate indices which run over $1, 2, 3$. In the rest of this paper we name Eq.(\ref{14}) the non-elastic stress-stress interaction.

By combining $N_0^3$ identical $L\times L\times L$ single blocks, we get a $N_0L\times N_0L\times N_0L$ super block. The non-elastic part of super block Hamiltonian (without the presence of external strain field) is given by the summation of single block non-elastic Hamiltonian and non-elastic stress-stress interaction:
\begin{eqnarray}\label{16}
\hat{H}^{\rm sup} = \sum_{s=1}^{N_0^3}\hat{H}^{(s)}
+\int d^3xd^3x'\sum_{ijkl}\Lambda_{ijkl}(\vec x-\vec x')\hat{T}_{ij}(\vec x)\hat{T}_{kl}(\vec x')\nonumber \\
\end{eqnarray}
where $\hat{H}^{(s)}$ represents the non-elastic part of Hamiltonian for the $s$-th glass single block, including the coupling between non-elastic stress tensor and intrinsic strain field: $\hat{H}^{(s)}=\hat{H}_0^{(s)}+\int_{V^{(s)}} d^3x\sum_{ij}e_{ij}(\vec x)\hat{T}_{ij}(\vec x)$.

\subsection{Glass Super Block Hamiltonian and Susceptibility with the Presence of External Strain}
In previous subsections, we used the notation $\bm{e}(\vec x)$ to stand for intrinsic strain field. In this subsection we turn on external weak strain field as a perturbation, and use the notation $\bm{e}(\vec x, t)$ to stand for it. We consider glass single block and super block non-elastic Hamiltonians $\hat{H}^{(s)}(\bm{e}(\vec x, t))$ and $\hat{H}^{\rm sup}(\bm{e}(\vec x, t))$ with the presence of external weak strain $\bm{e}(\vec x, t)$ in this subsection. It seems the Hamiltonian Eq.(\ref{16}) simply adds a stress-strain coupling $\int d^3x \,\sum_{ij}e_{ij}(\vec x, t)\hat{T}_{ij}(\vec x)$. However, more questions arise with the presence of $\bm{e}(\vec x, t)$.   

First of all, stress tensor operator $\hat{T}_{ij}(\vec x)$ might be modified by $\bm{e}(\vec x,t)$. A familiar example is that external strain field can modify electric dipole moments by changing positive-negative charge pairs' relative positions: $\Delta p_i(t)=\sum_{j}(\partial u_i(t)/\partial x_j)p_j$. Let's denote the change of $\hat{T}_{ij}(\vec x)$ is $\Delta \hat{T}_{ij}(\vec x)$. We further define a new stress tensor operator $\hat{T}_{ij}(\vec x, \bm{e}(\vec x, t))$ as follows,
\begin{eqnarray}\label{17}
\hat{T}_{ij}(\vec x, \bm{e}(\vec x, t)) & = & \hat{T}_{ij}(\vec x)+\Delta \hat{T}_{ij}(\vec x)={\delta \hat{H}^{(s)}(\bm{e}(\vec x, t)})/{\delta e_{ij}(\vec x)}\nonumber \\
\end{eqnarray}
The above result means the new tensor operator $\hat{T}_{ij}(\vec x, \bm{e}(\vec x, t))$ is actually the non-elastic stress tensor with the presence of external strain field ${\bm e}(\vec x, t)$. The stress-intrinsic strain field coupling is then given by $ e_{ij}(\vec x)\hat{T}_{ij}(\vec x, \bm{e}(\vec x, t))$. The non-elastic susceptibility in Eq.(\ref{7}) is given by replacing $\hat{T}_{ij}(\vec x)$ with $\hat{T}_{ij}(\vec x, \bm{e}(\vec x, t))$. Virtual phonon exchange process gives non-elastic stress-stress interaction $\hat{V}=\int d^3xd^3x'\sum_{ijkl}\Lambda_{ijkl}(\vec x-\vec x')\hat{T}_{ij}(\vec x, \bm{e}(\vec x, t))\hat{T}_{kl}(\vec x', \bm{e}(\vec x, t))$. In the rest of this paper, we still use $\chi_{ijkl}(\vec x, \vec x'; \omega)$, $\hat{T}_{ij}(\vec x)$, $\hat{H}^{(s)}$ and $\hat{H}_0^{(s)}$ to stand for $\chi_{ijkl}(\vec x, \vec x'; \omega; \bm{e}(\vec x, t))$, $\hat{T}_{ij}(\vec x, \bm{e}(\vec x, t))$, $\hat{H}^{(s)}(\bm{e}(\vec x, t))$ and $\hat{H}_0^{(s)}(\bm{e}(\vec x, t))$ for simplicity. Such simplicity does not harm the result of our theory. 

Second, the relative positions between different single blocks $\vec x-\vec x'$ could be changed by external strain field, so the coefficient of non-elastic stress-stress interaction $\Lambda_{ijkl}(\vec x-\vec x')$ is changed to be $\Lambda_{ijkl}(\vec x-\vec x'; \bm{e}(\vec x, t))$. Thus the glass total non-elastic Hamiltonian with the presence of external strain is given by 
\begin{eqnarray}\label{18} 
\hat{H}^{\rm sup} & = & \sum_{s}\hat{H}^{(s)}+\int d^3x\sum_{ij}e_{ij}(\vec x, t)\hat{T}_{ij}(\vec x)\nonumber \\
 & + & \int d^3x d^3x'\sum_{ijkl}\Lambda_{ijkl}(\vec x-\vec x'; \bm{e}(\vec x, t))\hat{T}_{ij}(\vec x)\hat{T}_{kl}(\vec x')\nonumber \\
\end{eqnarray}
According to the definition of stress tensor operator, the super block non-elastic stress tensor is given by the derivative of super block non-elastic Hamiltonian with respect to intrinsic strain field: $\hat{T}_{ij}^{\rm sup}(\vec x)={\delta  \hat{H}^{\rm sup}}/{\delta e_{ij}(\vec x)}$. Because $\Lambda_{ijkl}(\vec x-\vec x'; \bm{e}(\vec x))$ is also the function of intrinsic strain field, an extra term appears in the super block non-elastic stress tensor:
\begin{eqnarray}\label{19}
\hat{T}_{ij}^{\rm sup}(\vec x) & = & \hat{T}_{ij}(\vec x)\nonumber \\
 & + & \int d^3x_sd^3x_s'\sum_{abcd}\frac{ \delta\Lambda_{abcd}(\vec x_s-\vec x_s')}{\delta e_{ij}(\vec x)}\hat{T}_{ab}(\vec x_s)\hat{T}_{cd}(\vec x_s')\nonumber \\
\end{eqnarray}
In Eq.(\ref{23}) we will prove that the contribution of the second term in Eq.(\ref{19}) to the renormalization equation of non-elastic susceptibility is negligible. Therefore, it is reasonable to drop the second term in Eq.(\ref{19}). For now, we still want to keep it in super block stress tensor $\hat{T}_{ij}^{\rm sup}(\vec x)$. We will prove that it is negligible in Eq.(\ref{23}). Finally, we are able to rewrite super block Hamiltonian as the summation of single block Hamiltonian $\sum_{s=1}^{N_0^3}\hat{H}^{(s)}$, non-elastic stress-stress interaction $\hat{V}=\int d^3xd^3x'\sum_{ijkl}\Lambda_{ijkl}(\vec x-\vec x')\hat{T}_{ij}(\vec x)\hat{T}_{kl}(\vec x')$ and externally applied time-dependent perturbation $\hat{H}'(t)=\int d^3x\sum_{ij}e_{ij}(\vec x, t)\hat{T}_{ij}^{\rm sup}(\vec x)$ as follows: 

\begin{eqnarray}\label{20}
 & {} & \hat{H}^{\rm sup}(t) = \sum_{s=1}^{N_0^3}\hat{H}^{(s)}+\hat{V}+\hat{H}'(t)\nonumber \\
& {} & \hat{V} =  \int d^3x_sd^3x_s'\sum_{ijkl}\Lambda_{ijkl}(\vec x-\vec x')\hat{T}_{ij}(\vec x)\hat{T}_{kl}(\vec x')\nonumber \\
& {} & \hat{H}'(t) = \int d^3x\sum_{ij}e_{ij}(\vec x, t)\hat{T}_{ij}^{\rm sup}(\vec x)
\end{eqnarray}
With the input of external weak strain field $\bm{e}(\vec x, t)$, the super block glass non-elastic Hamiltonian $\hat{H}^{\rm sup}(t)$ provides a corresponding stress response $\langle \hat{T}_{ij}^{\rm sup}\rangle(\vec x, t)$. We can define the super block glass non-elastic susceptibility is as follows
\begin{eqnarray}\label{21}
\chi_{ijkl}^{\rm sup}(\vec x, \vec x'; t, t')=\frac{\delta \langle \hat{T}_{ij}^{\rm sup}\rangle (\vec x, t)}{\delta e_{kl}(\vec x', t')}
\end{eqnarray}

To calculate super block non-elastic susceptibility, let us denote $|n^{\rm sup}\rangle$ and $E_n^{\rm sup}$ to be the $n^{\rm sup}$-th eigenstate and eigenvalue of super block non-elastic Hamiltonian $\sum_s\hat{H}_0^{(s)}+\hat{V}$, and use linear response theory for the perturbation $\int d^3x\sum_{ij}e_{ij}(\vec x, t)\hat{T}_{ij}^{\rm sup}(\vec x)$. Please note that the in the following super block non-elastic relaxation susceptibility, the ``effective thermal relaxation time", $\tau^{\rm sup}$ should be different from single block relaxation time $\tau$. However, since we will discuss the renormalization behaviors of non-elastic susceptibility in relaxation regime (with $\omega\tau$ and $\omega\tau^{\rm sup}\ll 1$) resonance regime (with $\omega\tau$ and $\omega\tau^{\rm sup}\gg 1$) separately, the exact relation between $\tau$ and $\tau^{\rm sup}$ is not important. We still use $\tau$ to stand for the super block relaxation time for convenience. In the following super block non-elastic resonance susceptibility, $\eta$ is a phenomenological parameter which stands for higher order corrections due to the coupling between super block stress tensor $\hat{T}_{ij}^{\rm sup}(\vec x)$ and intrinsic phonon strain field $e_{ij}(\vec x)$. The super block non-elastic susceptibility is therefore given by 
\widetext
\begin{eqnarray}\label{22}
\chi_{ijkl}^{\rm sup}(\omega) & = & \frac{1}{(N_0L)^3}\frac{\beta}{1-i\omega\tau}\int d^3xd^3x'\bigg(\sum_{n^{\rm sup}m^{\rm sup}}\frac{e^{-\beta(E_n^{\rm sup}+E_m^{\rm sup})}}{\mathcal{Z}^{\rm sup 2}}\langle n^{\rm sup}|\hat{T}_{ij}^{\rm sup}(\vec x)|n^{\rm sup}\rangle\langle m^{\rm sup}|\hat{T}_{kl}^{\rm sup}(\vec x')|m^{\rm sup}\rangle\nonumber \\
 & {} & \qquad\qquad\qquad\qquad\qquad\qquad\,\, -\sum_{n^{\rm sup}}\frac{e^{-\beta E_n^{\rm sup}}}{\mathcal{Z}^{\rm sup}}\langle n^{\rm sup}|\hat{T}_{ij}^{\rm sup}(\vec x)|n^{\rm sup}\rangle\langle n^{\rm sup}|\hat{T}_{kl}^{\rm sup}(\vec x')|n^{\rm sup}\rangle\bigg)\nonumber \\
 & + & \frac{1}{(N_0L)^3}\frac{2}{\hbar}\int d^3xd^3x'\sum_{n^{\rm sup}l^{\rm sup}}\frac{e^{-\beta E_n^{\rm sup}}}{\mathcal{Z}^{\rm sup}}\frac{(E_l^{\rm sup}-E_n^{\rm sup})/\hbar}{(\omega+i\eta)^2-(E_l^{\rm sup}-E_n^{\rm sup})^2/\hbar^2}\langle l^{\rm sup}|\hat{T}_{ij}^{\rm sup}(\vec x)|n^{\rm sup}\rangle\langle n^{\rm sup}|\hat{T}_{kl}^{\rm sup}(\vec x')|l^{\rm sup}\rangle\nonumber \\
\end{eqnarray}
\endwidetext
where in the above result, the first and second lines stand for the relaxation part of super block susceptibility $\frac{1}{1-i\omega\tau}\chi^{\rm sup\, rel}_{ijkl}(\omega)$, and the third line stands for the resonance part of super block susceptibility $\chi^{\rm sup\, res}_{ijkl}(\omega)$. $\mathcal{Z}^{\rm sup}=\sum_{n^{\rm sup}}e^{-\beta E_n^{\rm sup}}$ is the partition function of the super block non-elastic Hamiltonian $\sum_s\hat{H}_0^{(s)}+\hat{V}$.

\section{Renormalization Procedure of Glass Non-Elastic Susceptibility}
We have defined the single and super block non-elastic Hamiltonians, stress tensor operators and non-elastic susceptibilities. In this section our main purpose is to set up the recursion relation (i.e., real space renormalization equation) between single block and super block non-elastic susceptibilities. Let us combine $N_0^3$ glass single blocks with the dimension $L\times L\times L$ to form a super block with the dimension $N_0L\times N_0L\times N_0L$. Because the super block is $N_0^3$ times the volume of single block, repeating such process from microscopic length scale $L_1$ will eventually carry out the glass Hamiltonian, non-elastic stress tensor and susceptibility at experimental length scale $R$. According to the argument by D. C. Vural and A. J. Leggett\cite{Leggett2011}, the suggested starting microscopic length scale of renormalization procedure is, for example, $L_1\sim 50\AA$. Since the final result only logarithmically depends on this choice, it will not be sensitive. In the $n$-th step of renormalization, we combine $N_0^3$ identical blocks with the side $L_n$ to form a $n$-th step super block with the side $N_0L_n$. The single block length scale in the next step is $L_{n+1}=N_0L_n$. On the other hand, the experimental length scale is the phonon wave length $R=2\pi/q$, which means throughout the entire renormalization process, the super block length scale $L_{n+1}$ is always smaller than the phonon wave length. We always have the important relations $\vec q\cdot (\vec x_s-\vec x_s')\ll 1$ and $e^{i\vec q\cdot (\vec x_s-\vec x_s')}\approx 1$, where $\vec x_s-\vec x_s'$ is the relative position between arbitrary single blocks in the arbitrary renormalization step.

We begin with such a group of non-interacting single blocks with the non-elastic part of Hamiltonian $\hat{H}_0=\sum_{s=1}^{N_0^3}\hat{H}_0^{(s)}$, eigenstates $|n\rangle = \prod_{s=1}^{N_0^3}|n^{(s)}\rangle$ and eigenvalues $E_n=\sum_{s=1}^{N_0^3}E_n^{(s)}$, where $|n^{(s)}\rangle$ and $E_n^{(s)}$ stand for the $n^{(s)}$-th eigenstate and eigenvalue for the $s$-th single block Hamiltonian $\hat{H}_0^{(s)}$. The partition function of these non-interacting single blocks is $\mathcal{Z}=\prod_s\mathcal{Z}^{(s)}$. We combine them to form a super block, and turn on non-elastic stress-stress interaction $\hat{V}$. We assume $\hat{V}$ is relatively weak compared to $\sum_{s=1}^{N_0^3}\hat{H}_0^{(s)}$, so it can be treated as a perturbation. Because the length scale dependence of $\hat{V}$ is $\hat{V}\propto 1/|\vec x-\vec x'|^3$, at small length scales it always dominates glass Hamiltonian. However, if the non-elastic susceptibility decreases logarithmically as the increase of length scale (which will be proved in Eq.(\ref{24.3})), then that means $\hat{V}$ can be treated as a perturbation at late stages. The assumption that $\hat{V}$ can be treated as a perturbation is qualitatively correct. In the last section we denote the $n^{\rm sup}$-th eigenstate and eigenvalue of super block Hamiltonian $\sum_s\hat{H}_0^{(s)}+\hat{V}$ to be $|n^{\rm sup}\rangle$ and $E_n^{\rm sup}$. Their relations with $|n\rangle$ and $E_n$ are given as follows, 

\begin{eqnarray}\label{23.1}
|n^{\rm sup}\rangle & = & |n\rangle+\sum_{p\neq n}\frac{\langle p|\hat{V}|n\rangle}{E_n-E_p}|p\rangle+\mathcal{O}(V^2)\nonumber \\
E_n^{\rm sup} & = & E_n+\langle n|\hat{V}|n\rangle +\mathcal{O}(V^2)
\end{eqnarray}
With the help of Eq.(\ref{23.1}) one can rewrite the super block non-elastic susceptibility in terms of single block susceptibilities: we expand super block relaxation and resonance susceptibilities (see Eq.(\ref{22})) up to the first order of interaction $\hat{V}$. Up
to the first order of $\hat{V}$ we can write the expansions of super block non-elastic susceptibility $\chi_{ijkl}^{\rm sup}(\omega)$ in terms of single block non-elastic susceptibilities $\chi_{ijkl}(\omega)$. The recursion relation for single block and super block non-elastic
susceptibilities are given as follows:

\widetext
\begin{eqnarray}\label{23}
\chi_{ijkl}^{\rm sup}(\omega)
 & = & 
\bigg(\frac{\chi_{ijkl}^{\rm rel}}{1-i\omega\tau}+\mathcal{K}_1\bigg)+\bigg(\chi_{ijkl}^{\rm res}(\omega) +\mathcal{K}_2\bigg)+\bigg(\mathcal{K}_3+\mathcal{K}_4\bigg)\nonumber \\
\nonumber \\
\mathcal{K}_1 & = & 
 -\frac{1}{(N_0L_n)^3(1-i\omega\tau)}\int d^3x_sd^3x_s'd^3x_ud^3x_u'\left[-\sum_{abcd}\Lambda_{abcd}(\vec x_u-\vec x_u')\right]\nonumber \\
 & {} & \bigg[\chi_{ijab}^{\rm rel}(\vec x_s, \vec x_u)\chi_{cdkl}^{\rm rel}(\vec x_u', \vec x_s')+\chi_{ijab}^{\rm rel}(\vec x_s, \vec x_u)\chi_{cdkl}^{\rm res}(\vec x_u', \vec x_s'; 0)+\chi_{ijab}^{\rm res}(\vec x_s, \vec x_u; 0)\chi_{cdkl}^{\rm rel}(\vec x_u', \vec x_s')\bigg]\nonumber \\
\mathcal{K}_2 & = & -\frac{1}{(N_0L_n)^3}\int d^3x_sd^3x_s'd^3x_ud^3x_u'\left[-\sum_{abcd}\Lambda_{abcd}(\vec x_u-\vec x_u')\right] \chi_{ijab}^{\rm res}(\vec x_s, \vec x_u; \omega)\chi_{cdkl}^{\rm res}(\vec x_u', \vec x_s'; \omega)\nonumber \\
\mathcal{K}_3 & = & \frac{2\beta^{-1}}{(N_0L_n)^3(1-i\omega\tau)}\sum_{abcdefgh}\int d^3xd^3x'd^3x_sd^3x_s'd^3x_ud^3x_u'\frac{\delta\Lambda_{abcd}(\vec x_s-\vec x_s')}{\delta e_{ij}(\vec x)}\frac{\delta\Lambda_{efgh}(\vec x_u-\vec x_u')}{\delta e_{kl}(\vec x')}\nonumber \\
 & {} & \bigg[\chi_{abef}^{\rm rel(1)}(\vec x_s, \vec x_u)\chi_{cdgh}^{\rm rel(1)}(\vec x_u', \vec x_s')-\chi_{abef}^{\rm rel(2)}(\vec x_s, \vec x_u)\chi_{cdgh}^{\rm rel(2)}(\vec x_u', \vec x_s')\bigg]\nonumber \\
\mathcal{K}_4 & = &  -  \frac{2}{(N_0L_n)^3}\sum_{abcdefgh}\int d^3xd^3x'd^3x_sd^3x_s'd^3x_ud^3x_u'\frac{1}{\pi^2}\frac{\delta \Lambda_{abcd}(\vec x_s-\vec x_s')}{\delta e_{ij}(\vec x)}\frac{\delta \Lambda_{efgh}(\vec x_u-\vec x_u')}{\delta e_{kl}(\vec x')}\nonumber \\
 & {} & \bigg\{ \int\frac{(1-e^{-\beta\hbar(\omega_s+\omega_s')})}{(1-e^{-\beta\hbar\omega_s})(1-e^{-\beta \hbar\omega_s'})}\frac{{\rm Im}\,\chi_{abef}^{\rm res}(\vec x_s, \vec x_u; \omega_s)
{\rm Im}\,\chi_{cdgh}^{\rm res}(\vec x_u', \vec x_s'; \omega_s') }{\hbar\omega_{s}+\hbar\omega_{s'}-\hbar\omega}d(\hbar\omega_s)d(\hbar\omega_s')\nonumber \\
 & {} & + i(1-e^{-\beta\hbar\omega})\pi\int\,\frac{{\rm Im}\,\chi_{abef}^{\rm res}(\vec x_s, \vec x_u; \omega_s){\rm Im}\,\chi_{cdgh}^{\rm res}(\vec x_u', \vec x_s'; \omega-\omega_s)}{(1-e^{-\beta\hbar\omega_s})(1-e^{-\beta\hbar(\omega-\omega_s)})}d(\hbar\omega_s)\bigg\}
\end{eqnarray}
\endwidetext

For details of the calculations of obtaining Eq.(\ref{23}), please see Appendix (B). In the terms $\mathcal{K}_3$ and $\mathcal{K}_4$ of Eq.(\ref{23}), $\chi_{ijkl}^{\rm rel(1)}, \chi_{ijkl}^{\rm rel(2)}$ are defined as follows:

\begin{eqnarray}\label{X1}
\chi_{ijkl}^{\rm rel(1)}(\vec x, \vec x')
 & = & 
\beta\sum_{nm}P_nP_m\langle n|\hat{T}_{ij}(\vec x)|n\rangle \langle m|\hat{T}_{kl}(\vec x')|m\rangle\nonumber \\
\chi_{ijkl}^{\rm rel(2)}(\vec x, \vec x')
 & = & 
\beta\sum_nP_n\langle n|\hat{T}_{ij}(\vec x)|n\rangle \langle n|\hat{T}_{kl}(\vec x')|n\rangle  \nonumber \\
\frac{\chi_{ijkl}^{\rm rel}(\vec x, \vec x')}{1-i\omega\tau}
 & = & 
\frac{1}{1-i\omega\tau}\left(\chi_{ijkl}^{\rm rel(1)}(\vec x, \vec x')-\chi_{ijkl}^{\rm rel(2)}(\vec x, \vec x')\right)\nonumber \\
\end{eqnarray}


Compared to the other terms in Eq.(\ref{23}), $\mathcal{K}_3$ and $\mathcal{K}_4$ decrease ($\propto L^{-3}$) much faster with the increase of length scale $L$, due to the $1/L^3$ decreasing behavior of the coefficient of non-elastic stress-stress interaction $\Lambda_{ijkl}(\vec x_s-\vec x_s')$. We first qualitatively investigate the length scale dependence of $\mathcal{K}_3$. With the symmetry assumption that ${\chi}^{\rm rel(1,2)}_{ijkl}=({\chi}^{\rm rel(1,2)}_l-2{\chi}^{\rm rel(1,2)}_t)\delta_{ij}\delta_{kl}+{\chi}^{\rm rel(1,2)}_t(\delta_{ik}\delta_{jl}+\delta_{il}\delta_{jk})$, $\mathcal{K}_3$ can be simplified as $\beta^{-1}{\rm C}_{l,t}^{(1,2)}(\chi_{l,t}^{\rm rel(1,2)})^2/(1-i\omega\tau)\rho^2 c_{l,t}^4L_n^3$, where ${\rm C}_{l,t}^{(1,2)}$ are dimensionless constants of order 1 with the upper indices $(1,2)$ stand for the first and second parts of non-elastic relaxaton susceptibilities (see Eq.(\ref{X1})). If we require that there is a critical length scale $L_c^{\rm rel}$, beyond which $\mathcal{K}_3$ is smaller than $\mathcal{K}_1$ and $\mathcal{K}_2$, then the upper limit of $L_c^{\rm rel}$ is given as follows
\begin{eqnarray}\label{23.5}
L_c^{\rm rel}< \left({k_BT}/{\rho c_{l,t}^2}\right)^{\frac{1}{3}}
\end{eqnarray}
We further let the temperature $T$ to take an extremely high value, $T=10^4$K (much greater than the melting temperature of glass), then the upper limit of $L_c^{\rm rel}$ is $4.6\AA$, which is still much smaller than $50\AA$, the microscopic starting length scale of our generic coupled block model. Therefore throughout the entire renormalization process, $\mathcal{K}_3$ is always negligible compared to $\mathcal{K}_1$ and $\mathcal{K}_2$.

Next, we want to investigate the length scale dependence of $\mathcal{K}_4$. We use the assumption that the reduced imaginary resonance suceptibility ${\rm Im\,}\tilde{\chi}^{\rm res}_{ijkl}(\omega)={\rm Im\,}\chi^{\rm res}_{ijkl}(\omega)/(1-e^{-\beta\hbar\omega})$  is approximately a constant up to the frequency of $\omega_c$ (which will be discussed later) and temperatures of the order $10$K. Integrating over the frequency variables $\omega_s, \omega_s'$ in $\mathcal{K}_4$ gives us $\mathcal{K}_4\approx -{\rm C}_{l,t}\,[\hbar\omega_c\ln(\frac{\omega_c}{\omega})-i\pi\hbar\omega]{({\rm Im}\,\tilde{\chi}_{l,t}^{\rm res})^2}/{\rho^2 c_{l,t}^4 L_n^3}$, where we obtain this result by using the symmetry property of resonance susceptibility ${\chi}^{\rm res}_{ijkl}(\omega)=\left({\chi}^{\rm res}_l(\omega)-2{\chi}^{\rm res}_t(\omega)\right)\delta_{ij}\delta_{kl}+{\chi}^{\rm res}_t(\omega)(\delta_{ik}\delta_{jl}+\delta_{il}\delta_{jk})$. C$_{l,t}$ is a positive constant of order 1, and $\omega$ is the experimentally input frequency\cite{Hunklinger1974, Pohl2002} of order $\omega\sim 10^7$rad/s. If we require that there is a critical length scale $L_c^{\rm res}$, beyond which $\mathcal{K}_4$ is smaller than $\mathcal{K}_1$ and $\mathcal{K}_2$, then we need to calculate the order of magnitude of $L_c^{\rm res}$. The upper limit of $L_c^{\rm res}$ can be obtained by letting $\omega_c$ to be an extremely high value, $\omega_c\sim 10^{15}$Hz which corresponds to $T_c=\hbar\omega_c/k_B\sim 10^4$K:
\begin{eqnarray}\label{23.5}
L_c^{\rm res}< \left(\frac{1}{\rho c_{l,t}^2}\frac{\hbar\omega_c}{\ln\left({\omega_c}/{\omega}\right)}\right)^{\frac{1}{3}}\sim 1.7\AA< L_1=50\AA
\end{eqnarray}
The above result implies that even if we choose an unreasonably high cut-off $\omega_c$, the upper limit of $L_c^{\rm res}$ is still much smaller than the microscopic starting length scale of our generic coupled block model. Throughout the entire renormalization procedure $\mathcal{K}_3$ and $\mathcal{K}_4$ are always negligible compared to $\mathcal{K}_1$ and $\mathcal{K}_2$. This agrees with the conclusion by D. Zhou and A. J. Leggett\cite{Zhou2015-1}, that  the contribution of resonance phonon energy absorption from $\delta\Lambda_{ijkl}(\vec x_s-\vec x_s')/\delta e(\vec x)$ is renormalization irrelevant. Dropping $\mathcal{K}_3$ and $\mathcal{K}_4$, the renormalization equations for non-elastic relaxation and resonance susceptibilities are simplified as follows 
\begin{eqnarray}\label{24}
 & {} & \chi_{l,t}^{\rm sup \,\, rel}  =  \chi_{l,t}^{\rm rel}-\frac{\ln N_0}{\rho c_{l,t}^2}\left[\left(\chi_{l,t}^{\rm rel }\right)^2+2\chi_{l,t}^{\rm rel}\chi_{l,t}^{\rm res}(0)\right]\nonumber \\
 & {} & \chi_{l,t}^{\rm sup \,\, res}(\omega)  =  \chi_{l,t}^{\rm res}(\omega)-\frac{\ln N_0}{\rho c_{l,t}^2}\left[\chi_{l,t}^{\rm res }(\omega)\right]^2
\end{eqnarray}
Eqs.(\ref{24}) are the most important results of this paper.

\subsection{The Renormalization Behavior of Non-elastic Resonance Susceptibility $\chi_{l,t}^{\rm res}(\omega)$}
We now examine the renormalization behaviors of non-elastic relaxation and resonance susceptibilities in Eqs.(\ref{24}). First of all, we focus on the renormalization behavior of non-elastic resonance susceptibility. Since $\chi_{l,t}^{\rm res}(\omega)={\rm Re}\,\chi_{l,t}^{\rm res}(\omega)+i\,{\rm Im\,}\chi_{l,t}^{\rm res}(\omega)$, there are two renormalization equations for non-elastic resonance susceptibility:
\begin{eqnarray}\label{24.1}
\delta\,{\rm Re\,}\chi_{l,t}^{\rm res}(\omega) & = & -\frac{\ln N_0}{\rho c_{l,t}^2}\left[\left({\rm Re\,}\chi_{l,t}^{\rm res}(\omega)\right)^2-\left({\rm Im\,}\chi_{l,t}^{\rm res}(\omega)\right)^2\right]\nonumber \\
\delta\,{\rm Im\,}\chi_{l,t}^{\rm res}(\omega) & = & -\frac{2\ln N_0}{\rho c_{l,t}^2}\,{\rm Re\,}\chi_{l,t}^{\rm res}(\omega )\,{\rm Im}\,\chi_{l,t}^{\rm res}(\omega)
\end{eqnarray}
where we define the ``change of non-elastic resonance susceptibility" $\delta\,{\rm Re\,}\chi_{l,t}^{\rm res}(\omega)=\,{\rm Re\,}\chi_{l,t}^{\rm sup\, res}(\omega)-\,{\rm Re\,}\chi_{l,t}^{\rm res}(\omega)$ and $\delta\,{\rm Im\,}\chi_{l,t}^{\rm res}(\omega)=\,{\rm Im\,}\chi_{l,t}^{\rm sup\, res}(\omega)-\,{\rm Im\,}\chi_{l,t}^{\rm res}(\omega)$. Solve the differential equation between $\,{\rm Re\,}\chi_{l,t}^{\rm res}(\omega)$ and $\,{\rm Im\,}\chi_{l,t}^{\rm res}(\omega)$, the renormalization flow curve is given by 
\begin{eqnarray}\label{24.2}
\left(\frac{\,{\rm Re\,}\chi_{l,t}^{\rm res}(\omega)}{\rho c_{l,t}^2}\right)^2+\left(\frac{\,{\rm Im\,}\chi_{l,t}^{\rm res}(\omega)}{\rho c_{l,t}^2}+C\right)^2=C^2
\end{eqnarray}
where $C$ is a constant determined by microscopic starting length scale non-elastic resonance susceptibility $\chi_{l,t}^{\rm res}(\omega, L_1)$. Let us plot the RG flow for ${\rm Re\,}\chi_{l,t}^{\rm res}(\omega)$ v.s. ${\rm Im\,}\chi_{l,t}^{\rm res}(\omega)$ in the following Fig.4, where the flow direction indicates the increase of length scales.
\begin{figure}[h]
\includegraphics[scale=0.4]{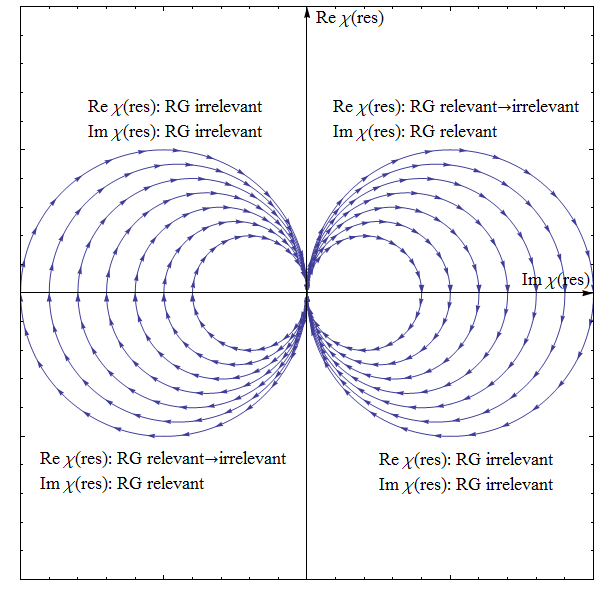}
\caption{The renormalization flow plot for ${\rm Re\,}\chi_{l,t}^{\rm res}(\omega)$ v.s. ${\rm Im\,}\chi_{l,t}^{\rm res}(\omega)$. In the first and third quadrants, the absolute value of ${\rm Re\,}\chi_{l,t}^{\rm res}(\omega)$ increases first, then decreases, while the absolute value of ${\rm Im\,}\chi_{l,t}^{\rm res}(\omega)$ keeps increasing; in the second and fourth quadrants, the absolute values of ${\rm Re\,}\chi_{l,t}^{\rm res}(\omega)$ and ${\rm Im\,}\chi_{l,t}^{\rm res}(\omega)$ keeps decreasing.}  
\end{figure}

According to Eq.(\ref{7}), the imaginary part of non-elastic resonance susceptibility is always negative for arbitrary positive $\omega$. Therefore we are interested in the second and third quadrants of Fig.4. The renormalization behavior of ${\rm Re\,}\chi_{l,t}^{\rm res}(\omega)$ and ${\rm Im\,}\chi_{l,t}^{\rm res}(\omega)$ depends on the sign of ${\rm Re\,}\chi_{l,t}^{\rm res}(\omega)$. If ${\rm Re\,}\chi_{l,t}^{\rm res}(\omega)<0$, then both of ${\rm Re\,}\chi_{l,t}^{\rm res}(\omega)$ and ${\rm Im\,}\chi_{l,t}^{\rm res}(\omega)$ increase with the increase of length scale. The qualitative increasing behavior of non-elastic resonance susceptibility is approximately given by $|\chi_{l,t}^{\rm res}(\omega, R)|=|\chi_{l,t}^{\rm res}(\omega, L_1)|+\frac{|\chi_{l,t}^{\rm res}(\omega, L_1)|^2}{\rho c_{l,t}^2}\ln\left(\frac{R}{L_1}\right)$.

If ${\rm Re\,}\chi_{l,t}^{\rm res}(\omega)>0$, then both of ${\rm Re\,}\chi_{l,t}^{\rm res}(\omega)$ and ${\rm Im\,}\chi_{l,t}^{\rm res}(\omega)$ decrease with the increase of length scale. By solving Eq.(\ref{24.1}), the real and imaginary parts of non-elastic resonance susceptibility are given as follows:
\begin{eqnarray}\label{24.3}
\,{\rm Re\,}\chi_{l,t}^{\rm res}(\omega, R) & = & \frac{{\rho c_{l,t}^2}}{\ln\left({R}/{L_1}\right)}\nonumber \\
\,{\rm Im\,}\chi_{l,t}^{\rm res}(\omega, R) & = & -\frac{1}{2C} \frac{{\rho c_{l,t}^2}}{\ln^2\left({R}/{L_1}\right)}
\end{eqnarray}
where we use the approximation that ${\rm Im}\,\chi_{l,t}^{\rm res}(\omega, R)\ll {\rm Re\,}\chi_{l,t}^{\rm res}(\omega, R)$, and we assume $({\rm Re\,}\chi_{l,t}^{\rm res}(\omega, L_1))^{-1}\ll ( {\rho c_{l,t}^2})^{-1}{\ln\left({R}/{L_1}\right)}$. 

According to the previous discussions, the renormalization behaviors of ${\rm Re\,}\chi_{l,t}^{\rm res}(\omega)$ and ${\rm Im}\,\chi_{l,t}^{\rm res}(\omega)$ are significantly different when ${\rm Re\,}\chi_{l,t}^{\rm res}(\omega)$ changes sign. It is obvious that for large enough $\omega$, ${\rm Re\,}\chi_{l,t}^{\rm res}(\omega)>0$, so both of ${\rm Re\,}\chi_{l,t}^{\rm res}(\omega)$ and ${\rm Im\,}\chi_{l,t}^{\rm res}(\omega)$ decrease with the increase of length scale, i.e., they are renormalization irrelevant. However, at the first glance, $\lim_{\omega \to 0^+}{\rm Re\,}\chi_{l,t}^{\rm res}(\omega)<0$, which means in zero-frequency limit both of ${\rm Re\,}\chi_{l,t}^{\rm res}(\omega)$ and ${\rm Im\,}\chi_{l,t}^{\rm res}(\omega)$ are renormalization relevant, i.e., they increase with the increase of length scale. This is an unphysical conclusion, because when the decreasing $\omega$ passes through some critical value $\omega_{\rm crit}$, the real space renormalization irrelevant quantities ${\rm Re\,}\chi_{l,t}^{\rm res}(\omega)$ and ${\rm Im\,}\chi_{l,t}^{\rm res}(\omega)$ suddenly become renormalization relevant. At experimental length scale $R$, $\chi_{l,t}^{\rm res}(\omega>\omega_c)$ is logarithmically small, while $\chi_{l,t}^{\rm res}(\omega<\omega_c)$ is logarithmically large. At least to the author's knowledge, such unusual phenomena of glass mechanical response function was never reported\cite{Lasjaunias1975, Hunklinger1981, Hunklinger1982, Schickfus1990, Pohl2002, Cahill1996}. A more severe problem is that when the non-elastic susceptibility is in static limit, the logarithmically huge, negative $\chi_{ijkl}^{\rm res}(\omega=0)$ implies that any amorphous material is mechanically unstable against arbtrary infinitesimal static perturbation (see the non-elastic part of free energy $F^{\rm non}(e)=F^{\rm non}_0+\frac{1}{2}\int dxdx'\chi(x-x') e(x, t)e(x', t)$). Because of the above two unreasonable implications, we would like to argue that for arbitrary $\omega$, ${\rm Re\,}\chi_{l,t}^{\rm res}(\omega)$ is always positive (including ${\rm Re\,}\chi_{l,t}^{\rm res}(\omega=0)>0$). This is because the phonon-phonon correlation function $\chi_{l,t}^{\rm ph}(\omega)$ will provide a positive real part of self-energy correction in the Dyson equation of $\chi_{l,t}^{\rm res}(\omega)$, so that $\lim_{\omega\to 0}{\rm Re\,}\chi_{l,t}^{\rm res}(\omega)$ is prevented from being negative. For details of discussions, please see Appendix (C). Finally, we would like to argue, that $\chi_{l,t}^{\rm res}(\omega)$ is always renormalization irrelevant for arbitrary $\omega$. The length scale dependences of ${\rm Re\,}\chi_{l,t}^{\rm res}(\omega)$ and ${\rm Im\,}\chi_{l,t}^{\rm res}(\omega)$ are given by Eqs.(\ref{24.3}).

\subsection{The Renormalization Behavior of Non-elastic Relaxation Susceptibility $\chi_{l,t}^{\rm rel}(\omega)$}
Based on the renormalization irrelevant behavior of non-elastic resonance susceptibility, we are able to discuss the non-elastic relaxation susceptibility now. According to Eq.(\ref{24}), the non-elastic relaxation susceptibility has a non-trivial stable fixed point and a trivial stable fixed point:
\begin{eqnarray}\label{26}
\chi_{l,t}^{\rm rel }(R)=-2\chi_{l,t}^{\rm res}(\omega=0, R)
\end{eqnarray}
\begin{eqnarray}\label{26.1}
\chi_{l,t}^{\rm rel }(R)=0
\end{eqnarray}
Eq.(\ref{26}) is the main result to explain the universal shift of sound velocity in glasses. The non-trivial stable fixed point, Eq.(\ref{26}) indicates that even if the non-elastic relaxation and resonance susceptibilities are entirely different at microscopic level, at experimental large length scales the relaxation susceptibility always flows to $-2$ of resonance susceptibility in zero-frequency limit. On the other hand, however, the trivial stable fixed point Eq.(\ref{26.1}) suggests that the relaxation susceptibility will approach zero as the length scale increases, with the speed ($\chi^{\rm rel}\propto L_1/R$) much faster than that of resonance susceptibility ($\chi^{\rm res}\propto 1/\ln(R/L_1)$). These two stable fixed points seem to contradict with each other. Let us discuss this problem as follows.

First, in the zero temperature limit, the non-elastic relaxation susceptibility $\lim_{T\to 0}\chi^{\rm rel}(L)$ is always 0 (see the non-elastic relaxation susceptibility in Eq.(\ref{7})). In the trivial fixed point, Eq.(\ref{26.1}), $\chi^{\rm rel}(L)$ approaches zero extremely fast with the increase of length scale, $\chi^{\rm rel}(R)\propto L_1/R \sim 10^{-10}$, which intuitively agrees with the behavior of $\chi^{\rm rel}(L)$ in the $T=0$ limit. In the non-trivial fixed point, the relaxation susceptibility $\chi^{\rm rel}(L)$ approaches zero relatively slow: $\chi^{\rm rel }(R)=-2\chi^{\rm res}(\omega=0, R)\propto 1/\ln(R/L_1)\sim 1/20$. The non-trivial fixed point Eq.(\ref{26}) does not agree with the behavior of $\lim_{T\to 0}\chi^{\rm rel}(L)$ intuitively. It seems that the non-trivial fixed point, Eq.(\ref{26}) is unphysical. To explain this dilemma, we plot the renormalization flow of $\chi^{\rm rel}$ v.s. $\chi^{\rm res}$ in Fig.5, where the flow direction represents the increase of length scale.
\begin{figure}[h]
\includegraphics[scale=0.35]{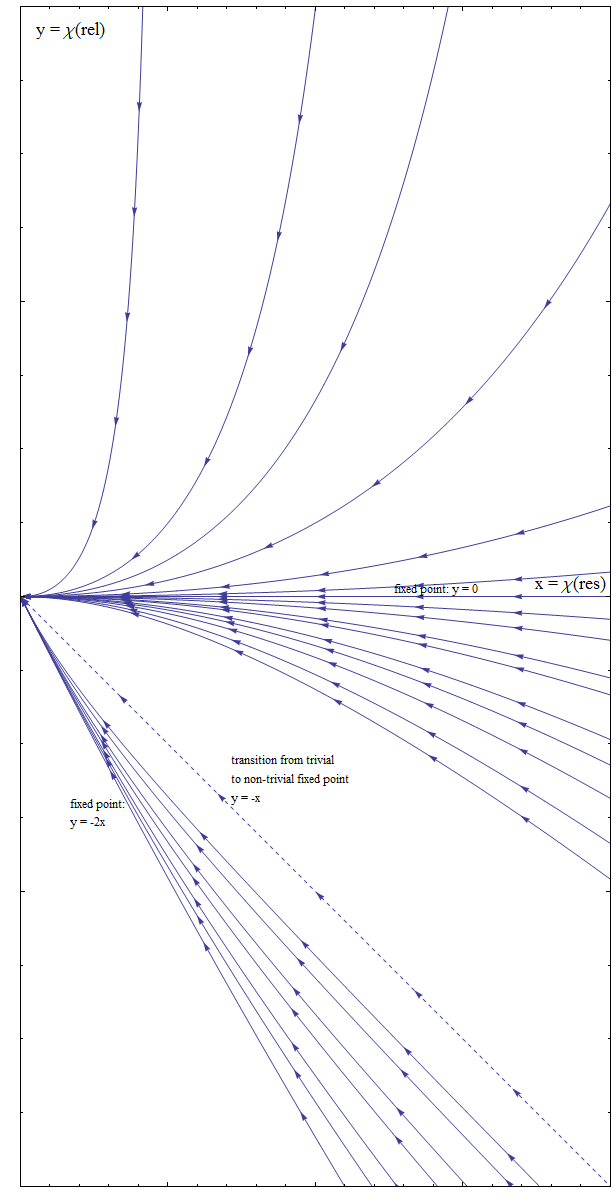}
\caption{RG flow plot for $y=\chi^{\rm rel}$ v.s. $x=\chi^{\rm res}$. We will mainly discuss three important lines: $y=0$, $y=-x$ and $y=-2x$ in the following discussions. }  
\end{figure}

In Fig.5 we see two fixed points (fixed lines): (1) trivial fixed point $y=0$ ($\chi^{\rm rel}=0$), which means $\chi^{\rm rel}(L)$ will always flow to 0 much faster than the logarithmic decreasing behavior of $\chi^{\rm res}(L)$ as the function of length scale $L$. It agrees with the zero-temperature behavior of $\chi^{\rm rel}(L)$, that the non-elastic relaxation susceptibility $\lim_{T\to 0}\chi^{\rm rel}(L)\to 0$; (2) the non-trivial fixed point $y=-2x$ ($\chi^{\rm rel}=-2\chi^{\rm res}$) implies that the non-elastic relaxation susceptibility will flow to $-2$ of resonance susceptibility; and (3) the dashed line, $y=-x$ ($\chi^{\rm rel}=-\chi^{\rm res}$), it is the transition line between the trivial fixed point and the non-trivial fixed point. At microscopic starting length scale $L_1$, if the non-elastic relaxation susceptibility is greater than $-1$ of resonance susceptibility: $\chi^{\rm rel}(L_1)>-\chi^{\rm res}(L_1)$ (for example, in zero-temperature limit, $\lim_{T\to 0}\chi^{\rm rel}(L_1)=0>-\chi^{\rm res}(L_1)$), then the renormalization flow chooses trivial stable fixed point, that $\chi^{\rm rel}$ flows to 0 much faster than $\chi^{\rm res}$. On the other hand, if the starting non-elastic relaxation susceptibility is smaller than $-1$ of resonance susceptibility, $\chi^{\rm rel}(L_1)<-\chi^{\rm res}(L_1)$, the renormalization flow will choose non-trivial fixed point $\chi^{\rm rel}(R)=-2\chi^{\rm res}(R)$. From Fig.5 we see the slope of renormalization flow slowly approaches to $-2$ when $\chi^{\rm res}(R)$ logarthmically approaches 0 with the increase of length scale.

To further verify the stability of Eq.(\ref{26}) and Eq.(\ref{26.1}) we consider the quantity $|\chi_{l,t}^{\rm rel }+\chi_{l,t}^{\rm res}(\omega=0)|$, which has the same length scale behavior as $|\chi_{l,t}^{\rm res}(\omega=0)|$. At the experimental length scale $R$ they are given as follows
\begin{eqnarray}\label{26.5}
|\chi^{\rm rel}_{l,t}(R)+\chi^{\rm res}_{l,t}(\omega=0, R)|=|\chi_{l,t}^{\rm res}(\omega=0, R)|=\frac{\rho c_{l,t}^2}{\ln\left({R}/{L_1}\right)}\nonumber \\
\end{eqnarray}
Eq.(\ref{26.5}) has two solutions, $\chi_{l,t}^{\rm rel }(R)=0$ which stands for the trivial fixed point, and $\chi_{l,t}^{\rm rel }(R)=-2\chi_{l,t}^{\rm res}(\omega=0, R)$ to represent the non-trivial fixed point. To illustrate the stability of the trivial and non-trivial fixed points, we further plot the renormalization flow of $|\chi_{l,t}^{\rm rel }+\chi_{l,t}^{\rm res}(\omega=0)|^{-1}$ v.s. $|\chi_{l,t}^{\rm res}(\omega=0)|^{-1}$ in Fig.6.
\begin{figure}[h]
\includegraphics[scale=0.45]{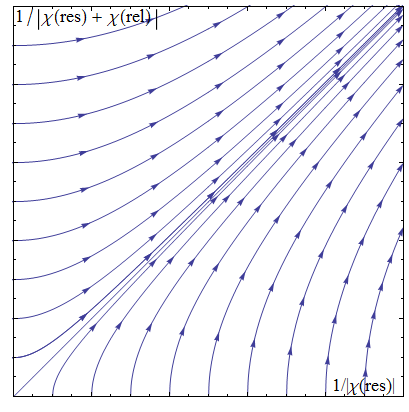}
\caption{At starting microscopic length scale $L_1$, for arbitrary relation between $y=|\chi_{l,t}^{\rm res}(0)+\chi_{l,t}^{\rm rel}|^{-1}$ and $x=|\chi_{l,t}^{\rm res}(\omega=0)|^{-1}$,  they always flow to the asymptote $y=x$ when length scale becomes large.}  
\end{figure}

Next we give two qualitative arguments on the existence of the non-trivial stable fixed point: $\chi^{\rm rel }(R)=-2\chi^{\rm res}(\omega=0, R)$. At microscopic length scale $L_1=50\AA$, there is no specific reason that the non-elastic relaxation and resonance susceptibilities must have such a simple relation. In fact, relaxation susceptibility is the product of diagonal matrix elements $\langle n|\hat{T}_{ij}(\vec x)|n\rangle\langle m|\hat{T}_{kl}(\vec x')|m\rangle$, while the resonance one is the product of off-diagonal matrix elements $\langle n|\hat{T}_{ij}(\vec x)|m\rangle\langle m|\hat{T}_{kl}(\vec x')|n\rangle$. These diagonal and off-diagonal elements $\langle n|\hat{T}_{ij}(\vec x)|m\rangle$ are entirely random for different quantum numbers $n, m$ and spacial coordinates $\vec x, \vec x'$.

First, with the increase of length scale, more and more single blocks join in the glass total Hamiltonian. Before turning on non-elastic stress-stress interaction $\hat{V}$, super block non-elastic stress tensor equals to the single block stress tensor $\hat{T}_{ij}^{\rm sup}(\vec x)=\hat{T}_{ij}(\vec x)$, super block eigenstates are the direct product of single block eigenstates $|n\rangle=\prod_{s=1}^{N_0^3}|n^{(s)}\rangle$. Hence the diagonal and off-diagonal matrix elements of super block stress tensors are the same as the those of single block stress tensors: $\langle n|\hat{T}_{ij}^{\rm sup}(\vec x)|n\rangle=\langle n|\hat{T}_{ij}(\vec x)|n\rangle$, $\langle n|\hat{T}_{ij}^{\rm sup}(\vec x)|m\rangle=\langle n|\hat{T}_{ij}(\vec x)|m\rangle$. Without non-elastic stress-stress interaction, non-elastic relaxation and resonance susceptibilities still have no specific relation at experimental length scales. The presence of non-elastic stress-stress interaction allows super block eigenstates to be the mixture of single block eigenstates. Besides, super block stress tensor receives an extra term proportional to $\delta\Lambda_{ijkl}(\vec x_s-\vec x_s')/\delta e_{ab}(\vec x)$ (see Eq.(\ref{19})). It further mixes the diagonal and off-diagonal matrix elements of stress tensors. It is at this point that the non-elastic relaxation and resonance susceptibilities could possibly have a relation at macroscopic length scales.

Second, the criteria to distinguish the relaxation and resonance processes for a pair of eigenstates $|n\rangle$ and $|m\rangle$ is whether $E_n-E_m\ll \hbar/\tau$ or not. At small length scale when glass Hamiltonian has a distinct set of eigenvalues, energy spacing is so large compared to $\hbar/\tau$ that the off-diagonal matrix elements of stress tensors must contribute to the transition process between different eigenstates. As the glass size grows, the increasing number of single blocks and non-elastic stress-stress interactions greatly increase the super block Hamiltonian density of states. At experimental length scales, the ``diagonal matrix element of stress tensor $\langle n|\hat{T}_{ij}|n\rangle$" is no longer a clear definition since the criteria whether $E_n-E_m$ is much smaller than $\hbar/\tau$ depends on the values of thermal relaxation time $\tau$ and level spacing $E_n-E_m$.  Furthermore, according to the TTLS calculations by J. J$\ddot{\rm a}$ckle\cite{Jackle1972}, $\hbar/\tau_{mn} \propto T(E_n-E_m)^2$ roughly increases with temperature, which means the diagonal matrix elements in low frequency higher temperature relaxation regime could be off-diagonal when the experimental condition enters in high frequency lower temperature resonance regime. Since a huge amount of eigenstate pairs $|m\rangle$, $|n\rangle$ take part in both of the resonance and relaxation processes, it is possible to have a simple relation between relaxation and resonance susceptibilities.

\section{Universal Sound Velocity Shift of Low-Temperature Glasses}
In this section we want to discuss the universal temperature dependence of longitudinal and transverse ultrasound velocity $c_{l,t}(T)$, in relaxation and resonance regimes separately. It is convenient to set up a reference frequency shift $\Delta \omega_{k; l,t}(T_0)$ at some reference temperature $T_0$, then consider the relative phonon frequency shift $\Delta \omega_{k; l,t}(T)$ at arbitrary temperature $T$. Since one can always write phonon frequency shift as $\Delta \omega_{k; l,t}(T)=k\Delta c_{l,t}(T)$, we get the relative sound velocity shift as follows:
\begin{eqnarray}\label{27}
\frac{\Delta c_{l,t}(T)-\Delta c_{l,t}(T_0)}{c_{l,t}}
 & = & \frac{{\rm Re}\,\chi_{l,t}(\omega, T)-{\rm Re}\,\chi_{l,t}(\omega, T_0)}{2\rho c_{l,t}^2}\nonumber \\
\end{eqnarray}
The behavior of sound velocity shift is different in relaxation and resonance regimes. In resonance regime the only contribution to the sound velocity shift is the real part of non-elastic resonance susceptibility, while in relaxation regime both of the real part of resonance and relaxation susceptibilities contribute to the sound velocity shift. The real part of non-elastic resonance susceptibility can be transformed into the imaginary part by Kramers-Kronig relation. Using the assumption that the reduced imaginary resonance susceptibility ${\rm Im}\,\tilde{\chi}_{l,t}^{\rm res}(\omega, T)=(1-e^{-\beta\hbar\omega})^{-1}{\rm Im}\,{\chi}_{l,t}^{\rm res}(\omega, T)$ is approximately a constant of frequency and temperature, up to the frequency of order $\omega_c\sim 10^{15}$Hz and around the temperature of order 10K\cite{Pohl2002}, we obtain the logarithmic temperature dependence of relative sound velocity shift as follows:
\begin{eqnarray}\label{28}
 & {} & \frac{\Delta c_{l,t}(T)-\Delta c_{l,t}(T_0)}{c_{l,t}( T_0)}\bigg|_{\rm res}\nonumber \\
 & = & \frac{2}{2\pi\rho c_{l,t}^2}\mathcal{P}\int_0^{\infty}\frac{\Omega\left({\rm Im}\,\chi_{l,t}^{\rm res}(\Omega, T)-{\rm Im}\,\chi_{l,t}^{\rm res}(\Omega, T_0)\right)}{\Omega^2-\omega^2}d\Omega\nonumber \\
 & = & \mathcal{C}_{l,t}\ln\left(\frac{T}{T_0}\right)
\end{eqnarray}
where $\mathcal{C}_{l,t}=-{\,{\rm Im\,}\tilde{\chi}_{l,t}^{\rm res}}/{2\pi\rho c_{l,t}^2}$ is a positive constant proportional to reduced imaginary resonance susceptibility. For details of calculations please refer to Appendix (D). Eq.(\ref{28}) is a multiple-level-system generalization of TTLS model derivation on logarithmic temperature dependence of sound velocity shift\cite{Hunklinger1974}. The constant $\mathcal{C}_{l,t}$ is not the functional of phonon frequency. We analytically continue this result from high frequency $\omega\tau\gg1$ regime to $\omega\to 0^+$, and carry out it's contribution in relaxation regime.

Next we discuss the sound velocity shift in relaxation regime. Both of the real part resonance and relaxation susceptibilities contribute to the sound velocity shift in this regime. By analytical continuation, the contribution of real part resonance susceptibility  in relaxation regime is still $\mathcal{C}_{l,t}\ln\left({T}/{T_0}\right)$. On the other hand, from Eq.(\ref{26}) the relaxation susceptibility equals to $-2$ of zero-frequency resonance susceptibility: $\Delta{\rm Re}\,\chi^{\rm rel}_{l,t}(\omega, T)/(-2\rho c_{l,t}^2)=-2\mathcal{C}_{l,t}\ln(T/T_0)$. Finally, the sound velocity shift in relaxation regime is given by the summation of both contributions
\begin{eqnarray}\label{29}
\frac{\Delta c_{l,t}(T)-\Delta c_{l,t}(T_0)}{c_{l,t}}\bigg|_{\rm rel}
 & = & \frac{\Delta{\rm Re}\,\left(\chi^{\rm rel}_{l,t}(\omega, T)+\chi^{\rm res}_{l,t}(\omega, T)\right)}{2\rho c_{l,t}^2}\nonumber \\
 & = & -\mathcal{C}_{l,t}\ln\left(\frac{T}{T_0}\right)
\end{eqnarray}
where we denote $\chi_{l,t}^{\rm rel}(\omega)=\frac{\chi_{l,t}^{\rm rel}}{1-i\omega\tau}$. Combining Eqs.(\ref{28}, \ref{29}), the theoretical temperature dependence of sound velocity shift in different regimes is summarized in Fig.8. The slopes in different regimes $\mathcal{C}_{l,t}^{\rm rel, res}$ are frequency independent, with the ratio $\mathcal{C}_{l,t}^{\rm res}:\mathcal{C}_{l,t}^{\rm rel}=1:-1$.

Up to now, we only find 12 sound velocity shift measurements\cite{Weiss1981, Weiss1999, Hunklinger1976, Hunklinger2000, Hunklinger1976, Cahill1996, Bellessa1977, Hunklinger1974, Hunklinger1984, Hunklinger1982}. Only one of them supports TTLS model expectation (Fig.7): the slope ratio is $\mathcal{C}_{l,t}^{\rm res}:\mathcal{C}_{l,t}^{\rm rel}=1:-\frac{1}{2}$. Nine of them support our generic coupled block model (Fig.8): the slope ratio is $\mathcal{C}_{l,t}^{\rm res}:\mathcal{C}_{l,t}^{\rm rel}=1:-1$. Two of them even have slightly greater ratio (Fig.9). We do now know how to explain such great ratio. 
\begin{figure}[h]
\includegraphics[scale=0.3]{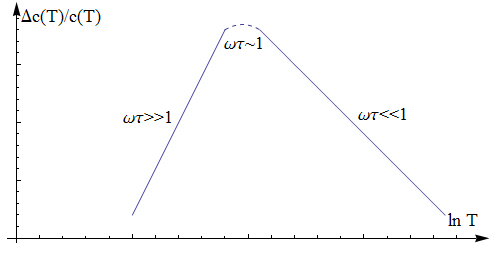}
\caption{The theoretical prediction of sound velocity shift by TTLS model, $\mathcal{C}_{l,t}^{\rm res}:\mathcal{C}_{l,t}^{\rm rel}=1:-\frac{1}{2}$. It agrees with the data of silica based microscopic cover glass measurements\cite{Hunklinger1982} with $f=1028$Hz.}  
\end{figure}

\begin{figure}[h]
\includegraphics[scale=0.3]{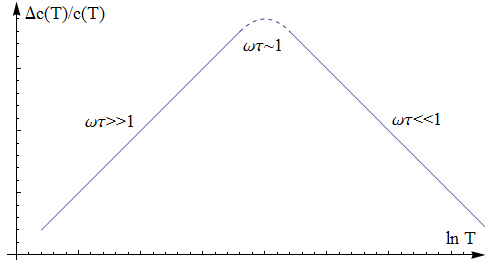}
\caption{The theoretical result of sound velocity shift by our generic coupled block model, the slope ratio $\mathcal{C}_{l,t}^{\rm res}:\mathcal{C}_{l,t}^{\rm rel}=1:-1$. It agrees with the data of Zr-Nb\cite{Weiss1981}, lithium-doped KCl\cite{Weiss1999}, vitreous silica\cite{Hunklinger2000} with $f\sim 10$kHz, BK7\cite{Hunklinger1976} with $f\sim 100{\rm MHz}$, (ZrO$_2$)$_{0.89}$(CaO)$_{0.11}$ with $f=83$kHz, (CaF$_2$)$_{0.74}$(LaF$_3$)$_{0.26}$, PS with $f=87$kHz, epoxy with $f=84$kHz\cite{Cahill1996} and metallic glass\cite{Bellessa1977} Ni$_{81}$P$_{19}$ with $f=150$MHz.}  
\end{figure}

\begin{figure}[h]
\includegraphics[scale=0.25]{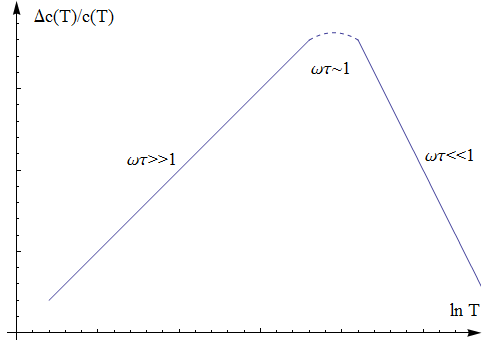}
\caption{Several amorphous materials' sound velocity shifts even have slightly greater slope ratios: vitreous silica Suprasil I\cite{{Hunklinger1974}} with $f=90$MHz and PdSiCu\cite{Hunklinger1984} with $f=960$MHz, 1030Hz.}  
\end{figure}

\section{Universal Dielectric Constant Shift $\Delta \epsilon_r/\epsilon_r$ of Low-Temperature Glasses}
In this section we want to discuss the universal property of dielectric constant shift $\Delta \epsilon_r/\epsilon_r$ in low-temperature glasses (see Fig.3). We still consider a single block of glass with the size much greater than atomic distance $L\gg a$. We write down the general single block glass Hamiltonian as $\hat{H}^{\rm tot}$. Let us separate out from the glass general Hamiltonian $\hat{H}^{\rm tot}$, the purely electro-magnetic field Hamiltonian $\hat{H}_{EM}$. Subtracting the purely electro-magnetic field Hamiltonian, we name the left-over Hamiltonian $(\hat{H}^{\rm tot}-\hat{H}_{EM})$ as the ``dielectric part of glass Hamiltonian $\hat{H}_{EM}^{\rm non}$".


We expand the dielectric Hamiltonian $\hat{H}^{\rm non}_{EM}$ up to the first order expansion of long wavelength intrinsic electric field. We name the coefficient of the first order expansion to be ``electric dipole moment $\hat{p}_i(\vec x)$", defined as follows, 
\begin{eqnarray}\label{34}
 & {} & \hat{H}^{\rm non}_{EM}=\hat{H}^{\rm tot}-\hat{H}_{EM}\nonumber \\
 & {} & \hat{H}^{\rm non}_{EM} = \hat{H}^{\rm non}_{EM; 0}-\int d^3x \sum_iE_i(\vec x)\hat{p}_{i}(\vec x)+\mathcal{O}(E^2)\nonumber \\
 & {} & \hat{p}_{i}(\vec x) =- \frac{\delta\hat{H}_{EM}^{\rm non}}{\delta E_i(\vec x)}
\end{eqnarray} 
Let us compare Eq.(\ref{34}) with TTLS model in Eq.(\ref{1}): the zero-th order expansion of dielectric Hamiltonian $\hat{H}^{\rm non}_{EM}$ with respect to intrinsic electric field $E_i$, $\hat{H}_{EM; 0}^{\rm non}$, is the generalization from two-level-system Hamiltonian to multiple-level-system Hamiltonian; dipole moment $\hat{p}_i(\vec x)$ is the multiple-level generalization of $2\times 2$ matrix which couples to electric field in TTLS model. 

Let us denote $|m\rangle$ and $E_m$ to be the $m$-th eigenstate and eigenvalue of the glass dielectric Hamiltonian $\hat{H}_{EM; 0}^{\rm non}$. Such a set of eigenbasis $|m\rangle$ is a generic multiple-level-system. Let us apply an external infinitesimal electric field, $E_i(\vec x, t)=E_i(\vec k)e^{i\vec k\cdot \vec x-i\omega_k t}$. The glass dielectric Hamiltonian $\hat{H}_{EM}^{\rm non}$ receives a time-dependent perturbation $\hat{H}'(t)=-\int d^3x\,\sum_iE_i(\vec x, t)\hat{p}_i(\vec x)$, and provides a dipole response, $\langle \hat{p}_i\rangle(\vec x, t)=\langle \hat{p}_i\rangle e^{i\vec k\cdot \vec x-i\omega_kt}$. Then we are able to define the complex response function (dielectric susceptibility) $\chi_{ij}^{\rm non}(\vec x, \vec x'; t, t')$ as follows, 
\begin{eqnarray}\label{34.1}
& {} & \chi_{ij}^{\rm non}(\vec x, \vec x'; \omega, \omega')=\int dtdt' \, e^{i\omega t+i\omega't'}\chi_{ij}^{\rm non}(\vec x, \vec x'; t, t') \nonumber \\
& {} & \chi_{ij}^{\rm non}(\vec x, \vec x'; t, t') = \frac{\delta \left<\hat{p}_i\right>(\vec x, t)}{\delta E_j(\vec x', t')}
\end{eqnarray} 
In the rest of this paper we use $\chi_{ij}(\vec x, \vec x'; \omega, \omega')$ to stand for $\chi_{ij}^{\rm non}(\vec x, \vec x'; \omega, \omega')$. In Eq.(\ref{34.1}) the dipole response of dielectric part of glass Hamiltonian, $\langle \hat{p}_i\rangle(\vec x, t)$, is defined as follows:
\begin{eqnarray}\label{34.2}
\left<\hat{p}_i\right>(\vec x, t)=\sum_m\frac{e^{-\beta E_m(t)}}{\mathcal{Z}(t)}\langle m_I, t|\hat{p}_{i, (I)}(\vec x, t)|m_I,t\rangle
\end{eqnarray} 
where $|m_I, t\rangle =\mathcal{T}e^{\frac{1}{i\hbar}\int^t_{-\infty}\hat{H}_I'(t)dt'}|m\rangle$ is the interaction picture wave function, and $\hat{p}_{i, (I)}(\vec x, t)=e^{i\hat{H}_{EM, 0}^{\rm non}t/\hbar}\hat{p}_{i}(\vec x)e^{-i\hat{H}_{EM, 0}^{\rm non}t/\hbar}$ is the interaction picture dipole operator. We expand $\langle p_i\rangle(\vec x, t)$ up to the first order of perturbation $\hat{H}'(t)$ to calculate the dielectric susceptibility $\chi_{ij}(\vec x, \vec x'; \omega, \omega')$ in Eq.(\ref{34.1}). We use the same language as TTLS mode, that the non-elastic susceptibility can be expressed in the relaxation and resonance parts. We use $\chi^{\rm res}_{ij}$ and $\chi^{\rm rel}_{ij}$ to stand for them. In the representation in which $\hat{H}_{EM, 0}^{\rm non}$ is diagonal, the perturbation $\hat{H}'(t)$ has both of diagonal and off-diagonal matrix elements. The diagonal matrix elements corresponds to the relaxation susceptibility, while the off-diagonal matrix elements corresponds to the resonance one. Let us denote $\tau$ to be the effective thermal relaxation time for the glass single block dielectric Hamiltonian at temperature $T$. We use the same argument in non-elastic susceptibility, that we are only interested in the relaxation and resonance regimes separately. Using a simple relaxation time $\tau$ to represent the effective thermal relaxation process does not harm our theory. The dielectric susceptibility is therefore given by 
\widetext
\begin{eqnarray}\label{36}
 & {} & \chi_{ij}(\vec x, \vec x'; \omega, \omega')=2\pi\delta(\omega+\omega')\chi_{ij}(\vec x, \vec x'; \omega)\nonumber \\
 & {} & \chi_{ij}(\vec x, \vec x'; \omega)=\frac{1}{1-i\omega\tau}\chi_{ij}^{\rm rel}(\vec x, \vec x')+\chi_{ij}^{\rm res}(\vec x, \vec x'; \omega)\nonumber \\
 & {} & \chi_{ij}^{\rm rel}(\vec x, \vec x') =
\beta\sum_n\bigg(\sum_{m}P_nP_m\langle n|\hat{p}_{i}(\vec x)|n\rangle \langle m|\hat{p}_{j}(\vec x')|m\rangle -P_n\langle n|\hat{p}_{i}(\vec x)|n\rangle \langle n|\hat{p}_{j}(\vec x')|n\rangle  \bigg)\nonumber \\
& {} & \chi_{ij}^{\rm res}(\vec x, \vec x'; \omega) =
\sum_{nl}(P_n-P_l)\frac{\langle n|\hat{p}_{i}(\vec x)|l\rangle \langle l|\hat{p}_{j}(\vec x')|n\rangle }{\omega+(E_n-E_l)/\hbar+i\eta}
\end{eqnarray}
\endwidetext
Where $P_n=e^{-\beta E_n}/\mathcal{Z}$ stands for the partition function of the $n$-th eigenstate of glass dielectric Hamiltonian.

We further define the space-averaged dielectric susceptibility for a single block of glass with the volume $V=L^3$: $\chi_{ij}(\omega)=V^{-1}\int d^3xd^3x'\,\chi_{ij}(\vec x, \vec x'; \omega)$. Since the dielectric susceptibility must be invariant under SO(3) rotational group transformations, it has the generic isotropic form of an arbitrary 2-indice quantity, $\chi_{ij}(\omega)=\chi(\omega)\delta_{ij}$. Similar with the phonon frequency shift, the photon frequency can be shifted by dielectric susceptibility $\chi(\omega)$: 
\begin{eqnarray}\label{39}
\frac{\Delta \omega_k}{\omega_k}=\frac{\chi(\omega)}{2\epsilon}
\end{eqnarray}
where the real part of frequency shift corresponds to the dielectric constant shift, and the imaginary part of frequency shift corresponds to the dielectric loss $\alpha$. 
\begin{eqnarray}\label{40}
\frac{\Delta \epsilon_r}{\epsilon_r} & = & -\frac{{\rm Re}\left(\chi^{\rm res}(\omega)+\chi^{\rm rel}(\omega)\right)}{\epsilon}\quad\quad {\rm relaxation\,\, regime}\nonumber \\
\frac{\Delta \epsilon_r}{\epsilon_r} & = & -\frac{{\rm Re}\,\chi^{\rm res}(\omega)}{\epsilon}\qquad\qquad\qquad\quad\, {\rm resonance\,\, regime}\nonumber \\
\end{eqnarray}
To explore the universal dielectric shift, we want to calculate the temperature dependence of the real part of dielectric susceptibility in relaxation and resonance regimes.

One may realize that the dielectric Hamiltonian Eq.(\ref{34}) is incomplete, because electric dipole moments $\hat{p}_i(\vec x)$ can interact with each other via $1/r^3$ dipole-dipole interaction. In fact, if we combine a set of single blocks together and allow virtual photons to exchange with each other, we will be able to obtain the electric dipole-dipole interaction:

\begin{eqnarray}\label{41}
 & {} & \hat{U}= \sum_{i,j=1}^3\int d^3xd^3x'\,\mu_{ij}(\vec x-\vec x')\hat{p}_i(\vec x)\hat{p}_j(\vec x')\nonumber \\
 & {} & \mu_{ij}(\vec x-\vec x') = \frac{\delta_{ij}-3n_in_j}{8\pi\epsilon |\vec x-\vec x'|^3}
\end{eqnarray}
where $n_i$ is the $i$-th component of unit vector of $\vec x-\vec x'$. Since we combine $N_0^3$ copies of $L\times L\times L$ glass single blocks to form a $N_0L\times N_0L\times N_0L$ super block, dipole-dipole interaction between single blocks will affect glass super block dielectric Hamiltonian. With the presence of external electric field, the glass super block dielectric Hamiltonian is given by
\begin{eqnarray}\label{42}
& {} &\hat{H}_{EM}^{\rm sup\, non}(t) = \sum_{s=1}^{N_0^3}\hat{H}_{EM}^{\rm non\,(s)}+\hat{U}+\hat{H}'(t)\nonumber \\
& {} &\hat{U} = \sum_{ij}\int d^3xd^3x'\mu_{ij}(\vec x-\vec x')\hat{p}_i(\vec x)\hat{p}_j(\vec x')\nonumber \\
& {} &\hat{H}'(t) = -\int d^3x\sum_{i=1}^3E_i(\vec x, t)\hat{p}_i(\vec x)
\end{eqnarray}
Please note that different from the previous phonon strain fields, this time the electric field is not a collection of real particle oscillations. Therefore the presence of external electric field $\vec E(\vec x, t)$ cannot modify the relative positions of particles in amorphous materials, which means the coefficient of electric dipole-dipole interaction, $\mu_{ij}(\vec x-\vec x')$ will not be modified by external electric field. Hence the super-block electric dipole moment $\hat{p}_i^{\rm sup}(\vec x)=-\delta \hat{H}_{EM}^{\rm sup\,non}/\delta E_i(\vec x)$ is the same as single block electric dipole moments: $\hat{p}_i^{\rm sup}(\vec x)=\hat{p}_i(\vec x)$.

With the input of external weak electric field $\vec E(\vec x, t)$, the super block glass dielectric Hamiltonian receives a time-dependent perturbation $\hat{H}'(t)$. It provides a corresponding dipole response $\langle \hat{p}_i^{\rm sup}\rangle(\vec x, t)$. The super block glass dielectric susceptibility is therefore defined as follows, 
\begin{eqnarray}\label{43}
\chi_{ij}^{\rm sup}(\vec x, \vec x'; t,t')=\frac{\delta\langle \hat{p}_{i}^{\rm sup}\rangle(\vec x, t)}{\delta E_j(\vec x', t')}
\end{eqnarray}
To calculate super block super block dielectric susceptibility, let us denote $|n^{\rm sup}\rangle$ and $E_n^{\rm sup}$ to be the $n^{\rm sup}$-th eigenstate and eigenvalue for super block unperturbed dielectric Hamiltonian: $\sum_{s=1}^{N_0^3}\hat{H}_{EM; 0}^{\rm non\, (s)}+\hat{U}$, and use linear response theory for the perturbation $-\int d^3x\,\sum_i E_i(\vec x, t)\hat{p}_i(\vec x)$. Please note when calculating the super block dielectric relaxation susceptibility, the ``effective thermal relaxation time" $\tau^{\rm sup}$ should be different from the single block relaxation time $\tau$. However, we use the same argument in the phonon strain field case, that we will be interested in the relaxation regime $\omega\tau, \omega\tau^{\rm sup}\ll 1$ and resonance regime $\omega\tau, \omega\tau^{\rm sup}\gg 1$ separately, the exact relation between $\tau$ and $\tau^{\rm sup}$ is not important. We use $\tau$ to stand for the supre block relaxation time for simplicity. The super block space-averaged dielectric susceptibility is given by 
\widetext
\begin{eqnarray}\label{44}
\chi_{ij}^{\rm sup}(\omega) 
 & = & 
\frac{1}{(N_0L)^3}\frac{\beta}{1-i\omega\tau}\int d^3xd^3x'\bigg(\sum_{n^{\rm sup}m^{\rm sup}}\frac{e^{-\beta \left(E_n^{\rm sup}+E_m^{\rm sup}\right)}}{\mathcal{Z}^{\rm sup\,2}}\langle n^{\rm sup}|\hat{p}_{i}(\vec x)|n^{\rm sup}\rangle \langle m^{\rm sup}|\hat{p}_{j}(\vec x')|m^{\rm sup}\rangle\nonumber \\
 & {} & \qquad\qquad\qquad\qquad \qquad\qquad\,\,\,-\sum_{n^{\rm sup}}\frac{e^{-\beta E_n^{\rm sup}}}{\mathcal{Z}^{\rm sup}}\langle n^{\rm sup}|\hat{p}_{i}(\vec x)|n^{\rm sup}\rangle \langle n^{\rm sup}|\hat{p}_{j}(\vec x')|n^{\rm sup}\rangle  \bigg)\nonumber \\
 & + & \frac{1}{(N_0L)^3}\frac{2}{\hbar}\int d^3xd^3x'\sum_{n^{\rm sup}l^{\rm sup}}\frac{e^{-\beta E_n^{\rm sup}}}{\mathcal{Z}^{\rm sup}}
\frac{(E_l^{\rm sup}-E_n^{\rm sup})/\hbar}{(\omega+i\eta)^2-(E_l^{\rm sup}-E_n^{\rm sup})^2/\hbar^2}
 \langle l^{\rm sup}|\hat{p}_{i}(\vec x)|n^{\rm sup}\rangle \langle n^{\rm sup}|\hat{p}_{j}(\vec x')|l^{\rm sup}\rangle \nonumber \\
\end{eqnarray}
\endwidetext
The first and second terms in Eq.(\ref{44}) are super block dielectric relaxation and resonance susceptibilities, respectively. We want to sep up the recursion relation between single block and super block dielectric susceptibities. Since the length scales of single and super blocks differ by a factor of $N_0$, repeating this recursion relation from microscopic length scale will eventually carry out the experimental length scale dielectric susceptibility. Let us still choose the starting microscopic length scale to be $L_1\sim 50\AA$. In the $n$-th step renormalization, the single and super block length scales are $L_n$ and $N_0L_n$.

We begin with a group of non-interacting single blocks with the dielectric Hamiltonian $\sum_{s=1}^{N_0^3}\hat{H}_{EM; 0}^{\rm non\,(s)}$, eigenstates $|n\rangle = \prod_{s=1}^{N_0^3}|n^{(s)}\rangle$ and eigenvalues $E_n=\sum_{s=1}^{N_0^3}E_n^{(s)}$, where $|n^{(s)}\rangle$ and $E_n^{(s)}$ represent the $n^{(s)}$-th eigenstate and eigenvalue for the $s$-th single block dielectric Hamiltonian $\hat{H}_{EM}^{\rm non\, (s)}$. We combine them to form a super block, and turn on electric dipole-dipole interaction. We assume that $\hat{U}$ is relatively weak compared to $\sum_{s=1}^{N_0^3}\hat{H}_{EM; 0}^{\rm non\,(s)}$, so it can be treated as a perturbation. The relations between $|n^{\rm sup}\rangle$, $E_n^{\rm sup}$ and $|n\rangle$, $E_n$ are therefore given by 
\begin{eqnarray}\label{45}
|n^{\rm sup}\rangle & = & |n\rangle +\sum_{m\neq n}\frac{\langle m|\hat{U}|n\rangle}{E_n-E_m}|m\rangle+\mathcal{O}(U^2)\nonumber \\
E_n^{\rm sup} & = & E_n+\langle n|\hat{U}|n\rangle +\mathcal{O}(U^2)
\end{eqnarray}
With the help of Eq.(\ref{45}) one can expand the super block relaxation and resonance susceptibilities up to the first order of $\hat{U}$. The first order expansions can be exactly written in terms of single block relaxation and resonance susceptibilities:
\widetext
\begin{eqnarray}\label{46}
\chi_{ij}^{\rm sup}(\omega)
 & = & \frac{\chi_{ij}^{\rm sup\, rel}}{1-i\omega\tau}+\chi_{ij}^{\rm sup\, res}(\omega)=\bigg(\frac{\chi_{ij}^{\rm rel}}{1-i\omega\tau}+\mathcal{R}_1\bigg)+\bigg(\chi_{ij}^{\rm res}(\omega)+\mathcal{R}_2\bigg)\nonumber \\
\mathcal{R}_1 & = & -\frac{1}{(N_0L_n)^3(1-i\omega\tau)}\int d^3x_sd^3x_s'd^3x_ud^3x_u'\left[-\sum_{ab}\mu_{ab}(\vec x_u-\vec x_u')\right]\nonumber \\
 & {} & \left(\chi_{ia}^{\rm rel}(\vec x_s, \vec x_u)\chi_{bj}^{\rm rel}(\vec x_u', \vec x_s')+\chi_{ia}^{\rm rel}(\vec x_s, \vec x_u)\chi_{bj}^{\rm res}(\vec x_u', \vec x_s'; 0)+\chi_{ia}^{\rm res}(\vec x_s, \vec x_u; 0)\chi_{bj}^{\rm rel}(\vec x_u', \vec x_s')\right)\nonumber \\
\mathcal{R}_2 & = & -\frac{1}{(N_0L_n)^3}\int d^3x_sd^3x_s'd^3x_ud^3x_u'\left[-\sum_{ab}\mu_{ab}(\vec x_u-\vec x_u')\right] \chi_{ia}^{\rm res}(\vec x_s, \vec x_u; \omega)\chi_{bj}^{\rm res}(\vec x_u',\vec x_s'; \omega)
\end{eqnarray}
\endwidetext
where $\frac{1}{1-i\omega\tau}\chi_{ij}^{\rm sup\, rel}$ and $\chi_{ij}^{\rm sup\, res}(\omega)$ represent super block dielectric relaxation and resonance susceptibilities. Applying the symmetry property of dielectric susceptibility $\chi_{ij}=\chi\delta_{ij}$, the renormalization equations can be further simplified as follows: 
\begin{eqnarray}\label{47}
 & {} & \chi^{\rm sup\, rel} = \chi^{\rm rel}-\frac{\ln N_0}{3\epsilon}\chi^{\rm rel}\left[\chi^{\rm rel}+2\chi^{\rm res}(0)\right]\nonumber \\
 & {} & \chi^{\rm sup \, res}(\omega) = \chi^{\rm res}(\omega)-\frac{\ln{N_0}}{3\epsilon}\left[\chi^{\rm res}(\omega)\right]^2
\end{eqnarray}
where $\chi^{\rm rel}$ and $\chi^{\rm res}(\omega)$ are space-averaged dielectric susceptibilities. Eq.(\ref{47}) is very similar to the renormalization equations of non-elastic stress-stress susceptibility.

First of all, for arbitrary positive $\omega$, the imaginary part of dielectric resonance susceptibility ${\rm Im\,}\chi^{\rm res}(\omega)$ is always negative, so the renormalization behaviors of $\chi^{\rm res}(\omega)$ depends on the sign of ${\rm Re\,}\chi^{\rm res}(\omega)$ (see Fig.4). It is obvious that for large enough $\omega$, ${\rm Re\,}\chi^{\rm res}(\omega)>0$. However, at the first glance, $\lim_{\omega\to 0^+}{\rm Re\,}\chi^{\rm res}(\omega)<0$. More specifically, there seems to be a critical frequency $\omega_c$, below which ${\rm Re\,}\chi^{\rm res}(\omega)$ is negative. Such behaviors imply that at experimental length scale $R$, $\chi^{\rm res}(\omega>\omega_c)$ is logarithmically small, while $\chi^{\rm res}(\omega<\omega_c)$ is logarithmically large as the function of length scale. At least to the author's knowledge, such abnormal phenomena of glass dielectric constant was never reported\cite{Hunklinger2005}. We would like to borrow the same argument in non-elastic susceptibility, that for arbitrary positive $\omega$, ${\rm Re\,}\chi^{\rm res}(\omega)$ is always positive (including ${\rm Re\,}\chi^{\rm res}(\omega=0)>0$), due to the positive real part of self-energy correction from the electro-magnetic field correlation function. For details of discussions, please see Appendix (C). We would like to argue, that the dielectric resonance susceptibility always decreases logarithmically as the function of length scale for arbitrary $\omega$. The length scale dependence of dielectric resonance susceptibility is given as follows, 
\begin{eqnarray}\label{47.1}
{\rm Re\,}\chi^{\rm res}(\omega, R) & = & \frac{3\epsilon}{\ln (R/L_1)}\nonumber \\
{\rm Im\,}\chi^{\rm res}(\omega, R) & = & -\frac{1}{2C}\frac{3\epsilon}{\ln^2 (R/L_1)}
\end{eqnarray}
where the constant $C$ is determined by starting microscopic length scale dielectric susceptibility. The ``experimental length scale $R$" is the minimum of the sample length scale $L$ and input electric field wavelength $\lambda$: $R={\rm min}(L,\lambda)$. In the problem of universal sound velocity shift in glasses, we have $L>\lambda$. In the problem of universal dielectric constant shift, for input electric field frequency $480{\rm Hz}<f<50{\rm kHz}$\cite{Schickfus1990} we have $R=L< \lambda$, while for input frequency $f=10$GHz by M. v. Schickfus\cite{Schickfus1977}, we have $L>\lambda=R$.

Second, according to Eq.(\ref{47}), the dielectric relaxation susceptibility has a non-trivial fixed point and a trivial fixed point:
\begin{eqnarray}\label{47.2}
\chi^{\rm rel}(R) & = & -2\chi^{\rm res}(\omega=0, R)
\end{eqnarray}
\begin{eqnarray}\label{47.3}
\chi^{\rm rel}(R) & = & 0
\end{eqnarray}
Eq.(\ref{47.2}) is the main result to explain the universal shift of dielectric constant in glasses. The non-trivial fixed point, Eq.(\ref{47.2}) indicates that even if the dielectric relaxation susceptibility and resonance susceptibility are entirely different at starting microscopic length scale, at experimental length scale the relaxation susceptibility always flow to $-2$ of zero-frequency resonance susceptibility. On the other hand, the trivial fixed point Eq.(\ref{47.3}) seems to contradict with Eq.(\ref{47.2}), that the dielectric relaxation susceptibility approaches zero with the speed much faster than that of resonance susceptibility. To solve this problem, we use the same argument discussed in non-elastic susceptibility (see Fig.5). At microscopic length scale $L_1$, if the dielectric relaxation susceptibility is greater than $-1$ of resonance susceptibility: $\chi^{\rm rel}(L_1)>-\chi^{\rm res}(L_1)$, then the renormalization flow chooses trivial stable fixed point. $\chi^{\rm rel}$ flows to zero much faster than $\chi^{\rm res}$. On the other hand, if the dielectric relaxation susceptibility is smaller than $-1$ of resonance susceptibility, $\chi^{\rm rel}(L_1)<-\chi^{\rm res}(L_1)$, the renormalization flow will choose the non-trivial fixed point $\chi^{\rm rel}(R)=-2\chi^{\rm res}(R)$, and this result is observed in the experiment of universal dielectric constant shift.

Now we can discuss the universal properties of dielectric constant shift in low-temperature glasses. Since the dielectric susceptibility is functional of temperature, it is convenient to set up a reference dielectric shift $\Delta\epsilon_r(T_0)$ at some reference temperature $T_0$, then calculate the relative dielectric constant shift $\Delta\epsilon_r(T)$ at arbitrary temperature $T$. The relative shift of dielectric constant at temperature $T$ is 
\begin{eqnarray}\label{49}
\frac{\Delta \epsilon_r(T)-\Delta\epsilon_r(T_0)}{\epsilon_r} & = & -\frac{{\rm Re}\,\chi(\omega, T)-{\rm Re}\,\chi(\omega, T_0)}{\epsilon}\nonumber \\
\end{eqnarray}
The behavior of dielectric constant shift is different in relaxation and resonance regimes. In resonance regime the only contribution to the dielectric constant shift is the resonance susceptibility, while in relaxation regime, both of the resonance and relaxation susceptibilities contribute to the dielectric constant shift. The real part of dielectric resonance susceptibility can be transformed into the imaginary part by Kramers-Kronig relation. We use the same assumption in the sound velocity shift discussions, that the reduced imaginary resonance susceptibility ${\rm Im}\,\tilde{\chi}^{\rm res}(\omega, T)=(1-e^{-\beta\hbar\omega})^{-1}\,{\rm Im}\,\chi^{\rm res}(\omega, T)$ is approximately the constant of frequency and temperature, up to the frequency of order $\omega_c\sim 10^{15}$Hz and around the temperature of order 10K\cite{Hunklinger1982}, we obtain the logarithmic temperature dependence of relative dielectric constant shift as follows: 
\begin{eqnarray}\label{50}
 & {} & \frac{\Delta \epsilon_r(T)-\Delta\epsilon_r(T_0)}{\epsilon_r(T_0)}\bigg|_{\rm res}\nonumber \\
 & = & -\frac{2}{\pi\epsilon}\mathcal{P}\int_0^{\infty}\frac{\Omega\left({\rm Im}\,\chi^{\rm res}(\Omega, T)-{\rm Im}\,\chi^{\rm res}(\Omega, T_0)\right)}{\Omega^2-\omega^2}d\Omega\nonumber \\
 & = & -\mathcal{C}\ln\left(\frac{T}{T_0}\right)
\end{eqnarray}
where $\mathcal{C}=-{\rm \,Im\,}\tilde{\chi}^{\rm res}/\pi\epsilon$ is a positive constant proportional to the reduced imaginary resonance susceptibility. For details of calculations please refer to Appendix (D). The constant $\mathcal{C}$ is independent of frequency $\omega$, so we analytically continue it from high frequency $\omega\tau\gg 1$ regime to $\omega\to 0^+$ relaxation regime: $-\lim_{\omega\to 0^+}\Delta {\rm Re\,}\chi^{\rm res}(\omega, T)/\epsilon=-\mathcal{C}\ln(T/T_0)$. In relaxation regime, both of the real part of resonance and relaxation susceptibilities contribute to the dielectric constant shift. By analytical continuation, the contribution of resonance susceptibility in relaxation regime is still $-\mathcal{C}\ln\left({T}/{T_0}\right)$. On the other hand, the non-trivial stable fixed point, Eq.(\ref{47.2}) indicates that the relaxation susceptibility will flow to $-2$ of zero-frequency resonance susceptibility, so the contribution to dielectric constant shift is $-\Delta\,{\rm Re\,}\chi^{\rm rel}(\omega, T)/\epsilon=2\mathcal{C}\ln(T/T_0)$. The dielectric constant shift in relaxation regime is the summation of resonance and relaxation susceptibilities:
\begin{eqnarray}\label{51}
\frac{\Delta \epsilon_r(T)-\Delta\epsilon_r(T_0)}{\epsilon_r}\bigg|_{\rm rel}
 & = & -\frac{\Delta \,{\rm Re\,}\left(\chi^{\rm rel}(\omega, T)+\chi^{\rm res}(\omega, T)\right)}{\epsilon}\nonumber \\
 & = & \mathcal{C}\ln\left(\frac{T}{T_0}\right)
\end{eqnarray}
where $\chi^{\rm rel}(\omega)=\frac{\chi^{\rm rel}}{1-i\omega\tau}$. Summarize Eq.(\ref{50}, \ref{51}), we prove the logarithmic temperature dependence of dielectric constant shift, in both of relaxation and resonance regimes. The slope ratio between resonance regime and relaxation regime is $\mathcal{C}^{\rm res}:\mathcal{C}^{\rm rel}=-1:1$.

Up to now, we only find 3 dielectric constant shift measurements\cite{Schickfus1990, Schickfus1977}. All of them support our generic coupled block model (Fig.10). The slope ratio between resonance and relaxation regimes is $\mathcal{C}^{\rm res}:\mathcal{C}^{\rm rel}=-1:1$.
\begin{figure}[h]
\includegraphics[scale=0.29]{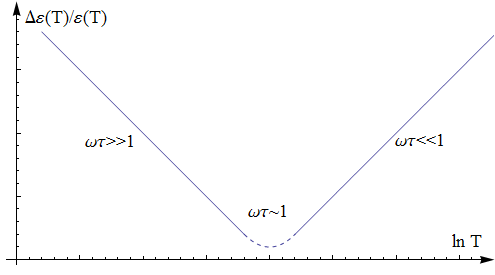}
\caption{The theoretical result of dielectric constant shift from our generic coupled block mode. The slope ratio $\mathcal{C}^{\rm res}:\mathcal{C}^{\rm rel}=-1:1$. It agrees with the data of vitreous silica Suprasil W and vitreous As$_2$S$_3$\cite{Schickfus1977} with $f=10$GHz, and BK7\cite{Schickfus1990} with $480{\rm Hz}<f<50{\rm kHz}$.}  
\end{figure}


\section{Conclusion}
In this paper we develop a generic coupled block model to explore the universal properties of low-temperature glass sound velocity shift and dielectric constant shift. We expand the glass Hamiltonian in orders of long wavelength phonon strain field and electric field. The coefficient of the first order expansion is defined as the non-elastic stress tensor $\hat{T}_{ij}(\vec x)$ and electric dipole moment $\hat{p}_i(\vec x)$. We combine a set of single blocks together and allow the virtual phonons to exchange with each other. The exchange of virtual phonon and photon processes will generate long-range interactions with $1/r^3$ behaviors. With the presence of non-elastic stress-stress interaction and electric dipole-dipole interaction, we use renormalization technique to iterate non-elastic and dielectric susceptibilities from small length scale to experimental length scale. The final result is logarithmically dependent on the starting microscopic length scale $L_1$, so the result is not sensitive to the choice of $L_1$.

We investigate 12 sound velocity shift measurements in glass materials, most of them agree with our theoretical results quite well, that the slope ratio between resonance regime and relaxation regime is $\mathcal{C}^{\rm res}:\mathcal{C}^{\rm rel}=1:-1$. We also investigate 3 dielectric constant shift measurements in glass materials, all of them agree with our theoretical prediction, that the slope ratio between resonance regime and relaxation regime is $\mathcal{C}^{\rm res}:\mathcal{C}^{\rm rel}=-1:1$. We believe that the universal shift of sound velocity and dielectric constant essentially come from virtual phonon and photon exchange interactions, independent of the materials' microscopic nature.

The resonance susceptibility logarithmically decreases with the increase of length scale, while the relaxation susceptibility has a non-trivial stable fixed point: at experimental length scales, the relaxation susceptibility must flow to $-2$ of resonance susceptibility at zero-frequency limit. At microscopic length scales, the relaxation and resonance susceptibilities have no specific relation. However, the presence of many body interaction mixes diagonal and off-diagonal matrix elements. Eventually at experimental length scales, the relaxation and resonance susceptibilities have a simple $-2$ relation. It is at this point that we able to explain the universal sound velocity and dielectric constant shifts.

\section{Acknowledgement}
D. Zhou wishes to express his deepest thanks for his advisor Anthony J. Leggett. D. Zhou also thanks Xueda Wen, Bo Han, Tianci Zhou, Yizhou Xin, Xianhao Xin and Yiruo Lin for their insightful discussions and constant encouragements. This work is supported by the National Science Foundation under Grant No. NSF-DMR 09-06921 at the University of Illinois.

\widetext
\appendix
\section{The Coefficient of Non-elastic Stress-Stress Interaction $\Lambda_{ijkl}(\vec x-\vec x')$}
It was Joffrin and Levelut\cite{Joffrin1976} who first gave the detailed derivation of non-elastic stress-stress interaction coefficient $\Lambda_{ijkl}(\vec x-\vec x')$ in amorphous solids. We give a further correction to their results. We will derive $\Lambda_{ijkl}(\vec x-\vec x')$ starting from amorphous solid Hamiltonian written in the summation of long wavelength phonon Hamiltonian, phonon strain field-stress tensor coupling and the non-elastic part of glass Hamiltonian:
\begin{eqnarray}\label{A2}
\hat{H}= \sum_{\vec k\mu}\left(\frac{| {p}_{\mu}(\vec k)|^2}{2m}+\frac{1}{2}m\omega^2_{\vec k\mu}| {u}_{\mu}(\vec k)|^2\right)+\int d^3x\sum_{ij}e_{ij}(\vec x)\hat{T}_{ij}(\vec x)+\hat{H}_{0}^{\rm non}
\end{eqnarray}
where $\mu$ is phonon polarization, i.e., longitudinal and transverse phonon modes; $\vec k$ is the phonon wave number and $m$ is the mass of elementary glass block, $ {p}_{\mu}(\vec k)$ and $ {u}_{\mu}(\vec k)$ are momentum and displacement operators, respectively. Phonon strain field $e_{ij}(\vec x)$ is defined as $e_{ij}(\vec x)=\frac{1}{2}\left(\partial u_{i}(\vec x)/{\partial x_j}+{\partial u_{j}(\vec x)}/{\partial x_i}\right)$. The relation of displacement operator $\vec u(\vec x)$ and $\vec u_{\mu}(\vec k)$ is set up by Fourier transformation:
\begin{eqnarray}\label{A3}
u_{i}(\vec x)=\frac{1}{\sqrt{N}}\sum_{\vec k\mu}u_{\mu}(\vec k){\rm e}_{\mu i}(\vec k)e^{i\vec k\cdot\vec x}
\end{eqnarray}
where $\vec {\rm e}_{\mu }(\vec k)$ is the unit vector representing the direction of vibrations, $N$ is the number of particles in an elementary cell in the sample. For longitudinal phonon mode with $\mu=l$,
$
{\rm e}_{li}(\vec k)={k_i}/{|\vec k|}
$, whereas for transverse modes with $\mu=t_1$ and $t_2$, we have, 
\begin{eqnarray}\label{A4}
 & {} & \vec {\rm e}_{t_1}(\vec k)\cdot\vec k=\vec {\rm e}_{t_2}(\vec k)\cdot \vec k=\vec {\rm e}_{t_1}(\vec k)\cdot \vec {\rm e}_{t_1}(\vec k)=0\nonumber \\
 & {} & \sum_{\mu=t_1,t_2}{\rm e}_{\mu i}(\vec k){\rm e}_{\mu j}(\vec k)=\delta_{ij}-\frac{k_ik_j}{k^2}
\end{eqnarray}
the strain field is therefore expressed as 
$
e_{ij}(\vec x)=\frac{1}{2\sqrt{N}}\sum_{\vec k\mu}iu_{\mu}(\vec k)e^{i\vec k\cdot \vec x}[k_j{\rm e}_{\mu i}(\vec k)+k_{i}{\rm e}_{\mu j}(\vec k)]
$. For an arbitrary function $f(\vec k)$ we always have the following relation, 
$
\sum_{\vec k}f(\vec k)=\sum_{\vec k}\frac{1}{2}[f(\vec k)+f(-\vec k)]
$. The displacement $u_{i}(\vec x)$ is real, i.e., $u_{i}(\vec x)=u^*_i(\vec x)$, we have $
u_{\mu i}(\vec k)=u_{\mu i}^*(-\vec k)
$.
With these properties of $u_{\mu}(\vec k)$, we can rewrite the stress-strain coupling term as follows, 
\begin{eqnarray}\label{A5}
\int d^3x\sum_{ij}e_{ij}(\vec x)\hat{T}_{ij}(\vec x)
 & = & \frac{1}{4\sqrt{N}}\sum_{ij}\int d^3x\sum_{\vec k\mu}\left[\left(iu_{\mu}(\vec k)e^{i\vec k\cdot \vec x}\right)+\left(iu_{\mu}(\vec k)e^{i\vec k\cdot \vec x}\right)^*\right](k_j{\rm e}_{\mu j}(\vec k)+k_j{\rm e}_{\mu i}(\vec k))\hat{T}_{ij}(\vec x)
\end{eqnarray}
Because the stress-strain coupling term is linear in displacement operators $u_{\mu}(\vec k)$, we can absorb it into the term which is quadratic in $u_{\mu}(\vec k)$, by ``completing the square". 
\begin{eqnarray}\label{A6}
\hat{H}=\sum_{\vec k\mu}\left(\frac{|p_{\mu}(\vec k)|^2}{2m}+\frac{m\omega^2_{\vec k\mu}}{2}|u_{\mu}(\vec k)-u_{\mu}^{(0)}(\vec k)|^2-\frac{m\omega^2_{\vec k\mu}}{2}|u_{\mu}^{(0)}(\vec k)|^2\right)+\hat{H}^{\rm non}
\end{eqnarray}
where the ``equilibrium position" $u_{\mu}^{(0)}(\vec k)$ is 
\begin{eqnarray}\label{A7}
u_{\mu}^{(0)}(\vec k)=\frac{i}{2\sqrt{N}m\omega_{\vec k\mu}^2}\sum_{ij}\int d^3x\bigg[k_j{\rm e}_{\mu i}(\vec k)+k_{i}{\rm e}_{\mu j}(\vec k)\bigg]\hat{T}_{ij}(\vec x)e^{-i\vec k\cdot \vec x}
\end{eqnarray}
The extra term left out after completing the square is the effective interaction between non-elastic stress tensors. It can be rewritten into two parts, the first part represents non-elastic stress-stress interaction within the same block, while the second part represents the interaction between different blocks:
\begin{eqnarray}\label{A8}
 & {} & -\sum_{\vec k\mu}\left(\frac{m\omega^2_{\vec k\mu}}{2}|u_{\mu}^{(0)}(\vec k)|^2\right)\nonumber \\
 & = & -\sum_{\vec k\mu}\frac{1}{8Nm\omega^2_{\vec k\mu}}\sum_{ijkl}\bigg[k_{j}{\rm e}_{\mu i}(\vec k)+k_i{\rm e}_{\mu j}(\vec k)\bigg]
\bigg[k_{k}{\rm e}_{\mu l}(\vec k)+k_l{\rm e}_{\mu k}(\vec k)\bigg]\int d^3x \,\hat{T}_{ij}(\vec x)\hat{T}_{kl}(\vec x')\nonumber \\
 & {} & -\sum_{\vec k\mu}\frac{1}{8Nm\omega^2_{\vec k\mu}}\sum_{ijkl}\bigg[k_{j}{\rm e}_{\mu i}(\vec k)+k_i{\rm e}_{\mu j}(\vec k)\bigg]
\bigg[k_{k}{\rm e}_{\mu l}(\vec k)+k_l{\rm e}_{\mu k}(\vec k)\bigg]\int d^3x d^3x'\,\hat{T}_{ij}(\vec x)\hat{T}_{kl}(\vec x') \cos(\vec k\cdot (\vec x-\vec x'))
\end{eqnarray}
We denote the second term in Eq.(\ref{A8}) as $\hat{V}$, non-elastic stress-stress interaction. Applying the properties of unit vector for longitudinal and transverse phonons, it is further simplified as 
\begin{eqnarray}\label{A9}
 & {} & \hat{V} 
= \sum_{ijkl}\int d^3xd^3x'\,\Lambda_{ijkl}(\vec x-\vec x')\hat{T}_{ij}(\vec x)\hat{T}_{kl}(\vec x')\nonumber \\
 & {} & \Lambda_{ijkl}(\vec x-\vec x') = \frac{1}{a^3}\sum_{\vec k}e^{i\vec k\cdot (\vec x-\vec x')}\Lambda_{ijkl}(\vec k)\nonumber \\
 & {} & \Lambda_{ijkl}(\vec k) = \frac{1}{2\rho}\left(\frac{1}{c_t^2}-\frac{1}{c_l^2}\right)\left(\frac{k_ik_jk_kk_l}{k^4}\right)-\frac{1}{8\rho c_t^2}\left(\frac{k_jk_l\delta_{ik}+k_jk_k\delta_{il}
+k_ik_l\delta_{jk}+k_ik_k\delta_{jl}}{k^2}\right)
\end{eqnarray}
where $\rho =Nm/a^3$, with $a$ the length scale of elementary glass block.

In this section we want to give a detailed discussion about the coefficient of non-elastic stress-stress interaction which appears in Eq.(\ref{23}). Since our purpose is to combine single blocks to form a super block from the starting microscopic length scale $L_1$ to phonon wave length scale $R=2\pi/q$, throughout the entire renormalization process, the dimention of super block is always smaller than the phonon wave length. We always have the important relations $\vec q\cdot (\vec x_s-\vec x_s')\ll 1$ and $e^{i\vec q\cdot (\vec x_s-\vec x_s')}\approx 1$, where $\vec x_s-\vec x_s'$ is the relative position between an arbitrary pair of single blocks.


The super block length in the $n$-th step renormalization is $N_0L_n$. We always have $|\vec x_s-\vec x_s'|\le N_0L_{n}$ for arbitrary blocks at positions $\vec x_s$ and $\vec x_s'$ (that is, the distance between single blocks within a super block must be no greater than $N_0L_n$). Let us write the coefficient $\Lambda_{ijkl}(\vec x-\vec x')$ into two parts:
\begin{eqnarray}\label{A1}
\Lambda_{ijkl}(\vec x-\vec x') & = & \frac{1}{a^3}\left(\sum_{|\vec k|=2\pi/R}^{2\pi/N_0L_n}+\sum_{|\vec k|=2\pi/N_0L_n}^{2\pi/L_1}\right)\Lambda_{ijkl}(\vec k)e^{i\vec k\cdot (\vec x-\vec x')}
\end{eqnarray}
where $R$ is the experimental length scale. Therefore, the first part of the above summation, can be simplified as $\sum_{|\vec k|=2\pi/R}^{2\pi/N_0L_n}\Lambda_{ijkl}(\vec k)e^{i\vec k\cdot (\vec x-\vec x')}\approx \sum_{|\vec k|=2\pi/R}^{2\pi/N_0L_n}\Lambda_{ijkl}(\vec k)$.
\begin{eqnarray}\label{A2}
\Lambda_{ijkl}(\vec x-\vec x') & = & \frac{1}{a^3}\sum_{|\vec k|=2\pi/R}^{2\pi/N_0L_n}\Lambda_{ijkl}(\vec k)+\frac{1}{a^3}\sum_{|\vec k|=2\pi/N_0L_n}^{2\pi/L_1}\Lambda_{ijkl}(\vec k)e^{i\vec k\cdot (\vec x-\vec x')}
\end{eqnarray}

After summing over different directions of momentum $\vec k$, the first part of Eq.(\ref{A2}) is given by 
\begin{eqnarray}\label{A3}
\frac{1}{a^3}\sum_{|\vec k|=2\pi/R}^{2\pi/N_0L_n}\Lambda_{ijkl}(\vec k)=\frac{4\pi}{3}\frac{1}{(N_0L_n)^3}\left[\frac{\alpha}{30\rho c_t^2}\left(\delta_{ij}\delta_{kl}+\delta_{ik}\delta_{jl}+\delta_{il}\delta_{jk}\right)-\frac{1}{4\rho c_t^2}\left(\delta_{jl}\delta_{ik}+\delta_{jk}\delta_{il}\right)\right]
\end{eqnarray}
where $\alpha=1-c_t^2/c_l^2$. The second term of Eq.(\ref{A2}) is obtained by D. Zhou and A. J. Leggett\cite{Zhou2015-1}:
\begin{eqnarray}\label{A4}
 & {} & \frac{1}{a^3}\sum_{|\vec k|=2\pi/N_0L_n}^{2\pi/L_{1}}\Lambda_{ijkl}(\vec k)e^{i\vec k\cdot (\vec x-\vec x')}= -\frac{\tilde{\Lambda}_{ijkl}}{8\pi\rho c_t^2|\vec x-\vec x'|^3}\nonumber \\
 & {} & \tilde{\Lambda}_{ijkl}=\frac{1}{4}\bigg\{(\delta_{jl}-3n_jn_l)\delta_{ik}+(\delta_{jk}-3n_jn_k)\delta_{il}+(\delta_{ik}-3n_in_k)\delta_{jl}
+(\delta_{il}-3n_in_l)\delta_{jk}\bigg\}\nonumber \\
 & {} & +\frac{1}{2}\alpha\bigg\{-(\delta_{ij}\delta_{kl}+\delta_{ik}\delta_{jl}+\delta_{jk}\delta_{il})+3(n_in_j\delta_{kl}+n_in_k\delta_{jl}+n_in_l\delta_{jk}+n_jn_k\delta_{il}+n_jn_l\delta_{ik}+n_kn_l\delta_{ij})-15n_in_jn_kn_l\bigg\}\nonumber \\
\end{eqnarray}
Finally, the coefficient $\Lambda_{ijkl}(\vec x-\vec x')$ which appears in Eq.(\ref{23}) is the summation of Eq.(\ref{A3}) and Eq.(\ref{A4}).

\section{Derivation of Renormalization Equation of Non-elastic Stress-Stress Susceptibility}
In this appendix we want to give a detailed derivation in obtaining the real space renormalization equation of non-elastic stress-stress susceptibility, Eq.(\ref{23}). That is, we want to set up the relation between super block non-elastic susceptibility Eq.(\ref{22}) and single block non-elastic susceptibility Eq.(\ref{7}). For notation simplicity, in this section we use $|n\rangle$, $E_n$ to represent the eigenstate and eigenvalue of Hamiltonian $\sum_s\hat{H}_0^{(s)}$, and use $|n^{\rm sup}\rangle$, $E_n^{\rm sup}$ to represent super block static Hamiltonian $\sum_{s}\hat{H}_0^{(s)}+\hat{V}$. We also use $\hat{T}_{ij}$, $\hat{T}_{ij}^{\rm sup}$, $\chi_{ijkl}$, $\chi_{ijkl}^{\rm sup}$ to represent single block and super block non-elastic stress tensors and susceptibilities. We treat non-elastic stress-stress interaction $\hat{V}$ as perturbation. By using perturbation theory, we obtain the following relations between $|n\rangle$, $E_n$ and $|n^{\rm sup}\rangle$, $E_n^{\rm sup}$
\begin{eqnarray}\label{B2}
|n^{\rm sup}\rangle = |n\rangle+\sum_{p\neq n}\frac{\langle p|\hat{V}|n\rangle}{E_n-E_p}|p\rangle+\mathcal{O}(V^2)\quad\quad\quad\quad 
E_n^{\rm sup} = E_n+\langle n|\hat{V}|n\rangle +\mathcal{O}(V^2)
\end{eqnarray}
We expand the super block partition function and probability function up to the first order in $\hat{V}$: 
\begin{eqnarray}\label{B3}
e^{-\beta E_n^{\rm sup}} = e^{-\beta E_n}\left(1-\beta\langle n|\hat{V}|n\rangle +\mathcal{O}(V^2)\right)\quad\quad\quad
\mathcal{Z}^{\rm sup} = \sum_le^{-\beta E_l}\left(1-\beta\langle l|\hat{V}|l\rangle +\mathcal{O}(V^2)\right)
\end{eqnarray}
Let us denote 
\begin{eqnarray}\label{B3.1}
\delta|n\rangle = \sum_{p\neq n}\frac{\langle p|\hat{V}|n\rangle}{E_n-E_p}|p\rangle\qquad\quad 
\delta E_n = \langle n|\hat{V}|n\rangle\qquad\quad
\delta\mathcal{Z} = -\sum_le^{-\beta E_l}\beta\langle l|\hat{V}|l\rangle 
\end{eqnarray}
to represent the first order expansions of the eigenstates, eigenvalues and partition functions. The following definitions will be very useful in details of calculations:
\begin{eqnarray}\label{B4}
\chi_{ijkl}^{\rm rel(1)}
 & = & 
\frac{\beta}{V}\int d^3xd^3x'\sum_{nm}P_nP_m\langle n|\hat{T}_{ij}(\vec x)|n\rangle \langle m|\hat{T}_{kl}(\vec x')|m\rangle\nonumber \\
\chi_{ijkl}^{\rm rel(2)}
 & = & 
\frac{\beta}{V}\int d^3xd^3x'\sum_nP_n\langle n|\hat{T}_{ij}(\vec x)|n\rangle \langle n|\hat{T}_{kl}(\vec x')|n\rangle  
\nonumber \\
\chi_{ijkl}^{\rm res}(\omega)
 & = & 
\frac{2}{V\hbar}\int d^3xd^3x'\sum_{nl}\frac{e^{-\beta E_n}}{\mathcal{Z}}\langle l|\hat{T}_{ij}(\vec x)|n\rangle\langle n|\hat{T}_{kl}(\vec x')|l\rangle\frac{\omega_{ln}}{(\omega+i\eta)^2-\omega_{ln}^{2}}
\end{eqnarray}
where $\omega_{ln}=(E_l-E_n)/\hbar$. Therefore the non-elastic susceptibility is written as follows,
\begin{eqnarray}\label{B5}
\chi_{ijkl}(\omega)
 & = & 
\frac{1}{1-i\omega\tau}\left(\chi_{ijkl}^{\rm rel(1)}-\chi_{ijkl}^{\rm rel(2)} \right)
+\chi_{ijkl}^{\rm res}(\omega)
\end{eqnarray}
In the rest of this appendix we want to expand the three parts of super block non-elastic susceptibility, $\chi_{ijkl}^{\rm sup \, rel(1)}$, $\chi_{ijkl}^{\rm sup \, rel(2)}$ and $\chi_{ijkl}^{\rm sup \,res}(\omega)$ up to the first order of interaction $\hat{V}$ (i.e., up to the second order of single block susceptibility).

From Eq.(\ref{19}) we know there is an extra term in super block stress tensor which is generated by the strain field dependence of coefficient $\Lambda_{ijkl}(\vec x-\vec x')$. We have super block stress tensor $\hat{T}^{\rm sup}_{ij}(\vec x)=\hat{T}_{ij}(\vec x)+\int dx_sdx_s'\sum_{abcd}\frac{\delta\Lambda_{abcd}(\vec x_s-\vec x_s')}{\delta e_{ij}(x)}\hat{T}_{ab}(\vec x_s)\hat{T}_{cd}(\vec x_s')$. We will discuss higher order expansions from this extra term $\int dx_sdx_s'\sum_{abcd}\frac{\delta\Lambda_{abcd}(\vec x_s-\vec x_s')}{\delta e_{ij}(x)}\hat{T}_{ab}(\vec x_s)\hat{T}_{cd}(\vec x_s')$ in the last subsection of this Appendix (B). Currently we expand super block susceptibility without considering this extra term.

\subsection{Expansion details for $\chi_{ijkl}^{\rm sup \,rel(1)}$}
\begin{eqnarray}\label{B6}
\chi_{ijkl}^{\rm sup \,rel(1)} & = & 
\frac{\beta}{(N_0L)^3}\int d^3x_sd^3x'_s\sum_{n^*m^*}\frac{e^{-\beta (E_n^*+E_m^*)}}{\mathcal{Z}^{*2}}\langle n^*|\hat{T}_{ij}(\vec x_s)|n^*\rangle \langle m^*|\hat{T}_{kl}(\vec x'_s)|m^*\rangle  \nonumber \\
 & = & 
\frac{\beta}{(N_0L)^3}\int d^3x_sd^3x'_s\sum_{nm}\frac{e^{-\beta (E_n+E_m)}}{\mathcal{Z}^2}\langle n|\hat{T}_{ij}(\vec x_s)|n\rangle \langle m|\hat{T}_{kl}(\vec x_s')|m\rangle  \nonumber \\
 & + & \frac{\beta}{(N_0L)^3}\int d^3x_sd^3x'_s\sum_{nm}\frac{e^{-\beta(E_n+E_m)}(-\beta \delta E_n-\beta\delta E_m)}{\mathcal{Z}^2}\langle n|\hat{T}_{ij}(\vec x_s)|n\rangle\langle m|\hat{T}_{kl}(\vec x_s')|m\rangle
\qquad\qquad\qquad\qquad J_1\nonumber \\
 & + & \frac{\beta}{(N_0L)^3}\int d^3x_sd^3x'_s\sum_{nm}\frac{e^{-\beta(E_n+E_m)}(-2\delta\mathcal{Z})}{\mathcal{Z}^3}\langle n|\hat{T}_{ij}(\vec x_s)|n\rangle\langle m|\hat{T}_{kl}(\vec x_s')|m\rangle\qquad\qquad\qquad\qquad\qquad\qquad\,\,\, J_2\nonumber \\
 & + & \frac{\beta}{(N_0L)^3}\int d^3x_sd^3x'_s\sum_{nm}\frac{e^{-\beta(E_n+E_m)}}{\mathcal{Z}^2} \bigg[\left(\delta\langle n|\right)\hat{T}_{ij}(\vec x_s)|n\rangle\langle m|\hat{T}_{kl}(\vec x_s')|m\rangle+\langle n|\hat{T}_{ij}(\vec x_s)\left(\delta|n\rangle\right)\langle m|\hat{T}_{kl}(\vec x_s')|m\rangle\nonumber \\
 & {} & \qquad\qquad\qquad\qquad\qquad\qquad\quad+\langle n|\hat{T}_{ij}(\vec x_s)|n\rangle\left(\delta\langle m|\right)\hat{T}_{kl}(\vec x_s')|m\rangle+\langle n|\hat{T}_{ij}(\vec x_s)|n\rangle\langle m|\hat{T}_{kl}(\vec x_s')\left(\delta|m\rangle\right)\bigg]\quad\,\,\, J_3\nonumber \\
\end{eqnarray}
where $\delta|n\rangle$, $\delta\mathcal{Z}$ and $\delta E_n$ stand for the first order expansions defined in Eq.(\ref{B3.1}). Now we begin to calculate every expansion terms $J_1, J_2$ and $J_3$ in the above result. \\
Expansion for term $J_1$:
\begin{eqnarray}\label{B6.1}
J_1 & = & -\frac{\beta^2}{(N_0L)^3}\int d^3x_sd^3x'_s\sum_{nm}\frac{e^{-\beta(E_n+E_m)}( \delta E_n+\delta E_m)}{\mathcal{Z}^2}\langle n|\hat{T}_{ij}(\vec x_s)|n\rangle\langle m|\hat{T}_{kl}(\vec x_s')|m\rangle\nonumber \\
 & = & -\frac{\beta^2}{(N_0L)^3}\int d^3x_sd^3x'_s\sum_{nm}\frac{e^{-\beta(E_n+E_m)}}{\mathcal{Z}^2}\left(\langle n|\hat{V}|n\rangle+\langle m|\hat{V}|m\rangle\right)\langle n|\hat{T}_{ij}(\vec x_s)|n\rangle\langle m|\hat{T}_{kl}(\vec x_s')|m\rangle\nonumber \\
 & = & -\frac{\beta^2}{(N_0L)^3}\int d^3x_sd^3x'_s\sum_{nm}\frac{e^{-\beta(E_n+E_m)}}{\mathcal{Z}^2}
\sum_{abcd}\int d^3x_ud^3x_u'\Lambda_{abcd}(\vec x_u-\vec x_u')\nonumber \\
 & {} & 
\left(\langle n|\hat{T}_{ab}(\vec x_u)\hat{T}_{cd}(\vec x_u')|n\rangle+\langle m|\hat{T}_{ab}(\vec x_u)\hat{T}_{cd}(\vec x_u')|m\rangle\right)\langle n|\hat{T}_{ij}(\vec x_s)|n\rangle\langle m|\hat{T}_{kl}(\vec x_s')|m\rangle\nonumber \\
 & = & -\frac{\beta^2}{(N_0L)^3}\int d^3x_sd^3x'_s\sum_{nm}\frac{e^{-\beta(E_n+E_m)}}{\mathcal{Z}^2}
\sum_{abcd}\int d^3x_ud^3x_u'\Lambda_{abcd}(\vec x_u-\vec x_u')\nonumber \\
 & {} & 
\left(\langle n|\hat{T}_{ab}(\vec x_u)\sum_{l}|l\rangle\langle l|\hat{T}_{cd}(\vec x_u')|n\rangle+\langle m|\hat{T}_{ab}(\vec x_u)\sum_{l}|l\rangle\langle l|\hat{T}_{cd}(\vec x_u')|m\rangle\right)\langle n|\hat{T}_{ij}(\vec x_s)|n\rangle\langle m|\hat{T}_{kl}(\vec x_s')|m\rangle
\end{eqnarray}
Let's stop here for a moment and talk about how could we write the above result in terms of single block susceptibility. Please note that we have only defined the single block relaxation and resonance susceptibilities, Eq.(\ref{B4}). The relaxation susceptibility (part 1 and part 2 of relaxation susceptibilities in Eq.(\ref{B4})) is the product of diagonal matrix elements of stress tensors; the resonance susceptibility is the product of off-diagonal matrix elements of stress tensors. So the question is, why do we never define such a term, that is the product between diagonal and off-diagonal matrix element of stress tensors? 

The reason is after averaging over spacial coordinate $\vec x_S$, such kind of product between diagonal and off-diagonal matrix elements of stress tensors $\hat{T}_{IJ}(\vec x_S)$ will vanish, becasue the diagonal and off-diagonal $\hat{T}_{IJ}(\vec x_S)$ matrix elements are random as the function of spacial coordinate $\vec x_S$. In other words, there is no specific relation between diagonal and off-diagonal matrix elements. In addition, the diagonal matrix element $\langle N|\hat{T}_{IJ}(\vec x_S)|N\rangle\propto \delta\langle\hat{H}^{\rm non}\rangle/\delta e_{IJ}$ is defined as the ``non-elastic" stress tensor in glass. It is highly plausible that the non-elastic stress tensor expectation value tends to vanish for large enough block of glass.

There is another problem for the pairing rule of stress tensor matrix elements: can we pair matrix elements between different blocks $\vec x_S\neq \vec x_S'$? For example, does the term $\langle N|\hat{T}_{IJ}(\vec x_S)|M\rangle \langle M|\hat{T}_{KL}(\vec x_S')|N\rangle$ with $S\neq S'$ vanish or not? Again, because the diagonal and off-diagonal stress tensor matrix elements are random as the function of spacial coordinate $\vec x_S$, after the summation over $S, S'$, $\sum_{SS'}\langle N|\hat{T}_{IJ}(\vec x_S)|M\rangle \langle M|\hat{T}_{KL}(\vec x_S')|N\rangle$ turns out to be zero. Therefore, the matrix element of stress tensor at the $S$-th block must be paired with the matrix element at the same $S$-th block. In other words, there is no obvious relation between stress tensors at different blocks.

Based on the above three reasons, we obtain the following two rules of matrix element pairing: suppose we have, for example, a diagonal matrix element $\langle N|\hat{T}_{IJ}(\vec x_S)|N\rangle$ and an off-diagonal matrix element $\langle N|\hat{T}_{IJ}(\vec x_S)|M\rangle$. The diagonal matrix element at the $S$-th block, $\langle N|\hat{T}_{IJ}(\vec x_S)|N\rangle$, is required to be paired with the diagonal matrix element of stress tensor at the same $S$-th block; the off-diagonal matrix element at the $S$-th block, $\langle N|\hat{T}_{IJ}(\vec x_S)|M\rangle$ is required to be pair with the off-diagonal matrix element at the same $S$-th block.

Now let us go back to the final result of Eq.(\ref{B6.1}). There are two summations: 
\begin{eqnarray}\label{B6.2}
 & {} & \int d^3x_sd^3x_s'd^3x_ud^3x_u'\Lambda_{abcd}(\vec x_u-\vec x_u')
\langle n|\hat{T}_{ab}(\vec x_u)\sum_{l}|l\rangle\langle l|\hat{T}_{cd}(\vec x_u')|n\rangle\langle n|\hat{T}_{ij}(\vec x_s)|n\rangle\langle m|\hat{T}_{kl}(\vec x_s')|m\rangle\nonumber \\
 & {} & \int d^3x_sd^3x_s'd^3x_ud^3x_u'\Lambda_{abcd}(\vec x_u-\vec x_u')
\langle m|\hat{T}_{ab}(\vec x_u)\sum_{l}|l\rangle\langle l|\hat{T}_{cd}(\vec x_u')|m\rangle\langle n|\hat{T}_{ij}(\vec x_s)|n\rangle\langle m|\hat{T}_{kl}(\vec x_s')|m\rangle
\end{eqnarray}
The coefficient of non-elastic stress-stress interaction, $\Lambda_{abcd}(\vec x_u-\vec x_u')$, does not allow $\vec x_u$ and $\vec x_u'$ belong to the same single block subspace (i.e., block numbers $u$ and $u'$ does not equal to each other, $u\neq u'$). Therefore, the matrix elements $\langle n|\hat{T}_{ab}(\vec x_u)|l\rangle$, $\langle l|\hat{T}_{cd}(\vec x_u')|n\rangle$, and $\langle m|\hat{T}_{ab}(\vec x_u)|l\rangle$, $\langle l|\hat{T}_{cd}(\vec x_u')|m\rangle$, are not allowed to pair with each other. In the first summation of Eq.(\ref{B6.2}), the only two possible ways of pairing is:
\begin{eqnarray}\label{B6.3}
 & {} & \langle n|\hat{T}_{ab}(\vec x_u)|l\rangle \quad {\rm paired \,\,\, with}\quad \langle n|\hat{T}_{ij}(\vec x_s)|n\rangle; \quad\quad
\langle l|\hat{T}_{cd}(\vec x_u')|n\rangle \quad {\rm paired \,\,\, with}\quad \langle m|\hat{T}_{kl}(\vec x_s')|m\rangle\nonumber \\
 & {\rm or\quad} & 
\langle n|\hat{T}_{ab}(\vec x_u)|l\rangle \quad {\rm paired \,\,\, with}\quad \langle m|\hat{T}_{kl}(\vec x_s')|m\rangle; \quad\quad
\langle l|\hat{T}_{cd}(\vec x_u')|n\rangle \quad {\rm paired \,\,\, with}\quad \langle n|\hat{T}_{ij}(\vec x_s)|n\rangle
\end{eqnarray}
the first pairing candidate requires $\vec x_u, \vec x_s\in V^{(s)}$ and $\vec x_u', \vec x_s'\in V^{(s')}$ (where $V^{(s)}$ and $V^{(s')}$ denotes the $s, s'$-th single block subspaces), and $|l\rangle=|n\rangle$; the second pairing candidate requires $\vec x_u, \vec x_s'\in V^{(s')}$ and $\vec x_u', \vec x_s\in V^{(s)}$, and $|l\rangle=|n\rangle$. In the second summation of Eq.(\ref{B6.2}) the possible pairings require $\vec x_u, \vec x_s\in V^{(s)}$ and $\vec x_u', \vec x_s'\in V^{(s')}$ or $\vec x_u, \vec x_s'\in V^{(s')}$ and $\vec x_u',\vec x_s\in V^{(s)}$, and $|l\rangle=|m\rangle$. With the above pairing rules, finally we can procede the calculation of term $J_1$ from Eq.(\ref{B6.1}) as follows,

\begin{eqnarray}\label{B7}
J_1 & = & -\frac{\beta^2}{(N_0L)^3}\int d^3x_sd^3x'_sd^3x_ud^3x_u'
\sum_{abcd}\sum_{n^{(s)}n^{(s')}m^{(s')}}\frac{e^{-\beta(E_n^{(s)}+E_n^{(s')}+E_m^{(s')})}}{\mathcal{Z}^{(s)}\mathcal{Z}^{(s')2}}\Lambda_{abcd}(\vec x_u-\vec x_u')\nonumber \\
 & {} & \bigg\{\langle n^{(s)}|\hat{T}_{cd}(\vec x_u')|n^{(s)}\rangle\langle n^{(s)}|\hat{T}_{ij}(\vec x_s)|n^{(s)}\rangle\langle m^{(s')}|\hat{T}_{kl}(\vec x_s')|m^{(s')}\rangle\langle n^{(s')}|\hat{T}_{ab}(\vec x_u)|n^{(s')}\rangle \nonumber \\
 & {} & +
\langle n^{(s)}|\hat{T}_{ij}(\vec x_s)|n^{(s)}\rangle\langle m^{(s)}|\hat{T}_{ab}(\vec x_u)|m^{(s)}\rangle\langle m^{(s')}|\hat{T}_{cd}(\vec x_u')|m^{(s')}\rangle\langle m^{(s')}|\hat{T}_{kl}(\vec x_s')|m^{(s')}\rangle 
\bigg\}\nonumber \\
 & = & -\frac{1}{(N_0L)^3}\int d^3x_sd^3x'_sd^3x_ud^3x_u'\sum_{abcd}\Lambda_{abcd}(\vec x_u-\vec x_u')\left(\chi_{cdij}^{\rm rel(2)}(\vec x_u', \vec x_s)\chi_{abkl}^{\rm rel(1)}(\vec x_s', \vec x_u)+\chi_{cdkl}^{\rm rel(2)}(\vec x_u', \vec x_s')\chi_{abij}^{\rm rel(1)}(\vec x_s, \vec x_u)\right)\nonumber \\
\end{eqnarray}
Next, we consider the expansion for term $J_2$:
\begin{eqnarray}\label{B8}
J_2 & = & -\frac{2\beta}{(N_0L)^3}\int d^3x_sd^3x'_s\sum_{nm}\frac{e^{-\beta(E_n+E_m)}}{\mathcal{Z}^3}\delta\mathcal{Z}\langle n|\hat{T}_{ij}(\vec x_s)|n\rangle \langle m|\hat{T}_{kl}(\vec x_s')|m\rangle\nonumber \\
 & = & \frac{2\beta^2}{(N_0L)^3}\int d^3x_sd^3x'_s\sum_{lmn}\frac{e^{-\beta(E_n+E_m+E_l)}}{\mathcal{Z}^3}\sum_{abcd}\int d^3x_ud^3x_u'\Lambda_{abcd}(\vec x_u-\vec x_u')\nonumber \\
 & {} & \langle l|\hat{T}_{ab}(\vec x_u)\hat{T}_{cd}(\vec x_u')|l\rangle\langle n|\hat{T}_{ij}(\vec x_s)|n\rangle\langle m|\hat{T}_{kl}(\vec x_s')|m\rangle\nonumber \\
 & = & \frac{2\beta^2}{(N_0L)^3}\int d^3x_sd^3x'_s\sum_{lmn}\frac{e^{-\beta(E_n+E_m+E_l)}}{\mathcal{Z}^3}\sum_{abcd}\int d^3x_ud^3x_u'\Lambda_{abcd}(\vec x_u-\vec x_u')\nonumber \\
 & {} & \langle l|\hat{T}_{ab}(\vec x_u)\sum_k|k\rangle\langle k|\hat{T}_{cd}(\vec x_u')|l\rangle\langle n|\hat{T}_{ij}(\vec x_s)|n\rangle\langle m|\hat{T}_{kl}(\vec x_s')|m\rangle\nonumber \\
 & = & \frac{2\beta^2}{(N_0L)^3}\int d^3x_sd^3x'_sd^3x_ud^3x_u'\sum_{l^{(s)}l^{(s')}m^{(s')}n^{(s)}}\frac{e^{-\beta(E_n^{(s)}+E_m^{(s')}+E_l^{(s)}+E_l^{(s')})}}{\mathcal{Z}^{(s)2}\mathcal{Z}^{(s')2}}\sum_{abcd}\Lambda_{abcd}(\vec x_u-\vec x_u')\nonumber \\
 & {} & \langle l^{(s')}|\hat{T}_{cd}(\vec x_u')|l^{(s')}\rangle\langle m^{(s')}|\hat{T}_{kl}(\vec x_s')|m^{(s')}\rangle\langle n^{(s)}|\hat{T}_{ij}(\vec x_s)|n^{(s)}\rangle\langle l^{(s)}|\hat{T}_{ab}(\vec x_u)|l^{(s)}\rangle\nonumber \\
 & = & \frac{2}{(N_0L)^3}\int d^3x_sd^3x'_sd^3x_ud^3x_u'\sum_{abcd}\Lambda_{abcd}(\vec x_u-\vec x_u')\chi_{cdij}^{\rm rel(1)}(\vec x_u', \vec x_s')\chi_{abkl}^{\rm rel(1)}(\vec x_s, \vec x_u)
\end{eqnarray}
In the above calculations, because the coefficient $\Lambda_{abcd}(\vec x_u-\vec x_u')$ does not allow $\vec x_u, \vec x_u'$ belong to the same single block subspace, the matrix elements $\langle l|\hat{T}_{ab}(\vec x_u)|k\rangle$, $\langle k|\hat{T}_{cd}(\vec x_u')|l\rangle$ are not allowed to pair with each other. We need to pair matrix elements $\langle l|\hat{T}_{ab}(\vec x_u)|k\rangle$, $\langle k|\hat{T}_{cd}(\vec x_u')|l\rangle$ with diagonal matrix elements $\langle n|\hat{T}_{ij}(\vec x_s)|n\rangle$, $\langle m|\hat{T}_{kl}(\vec x_s')|m\rangle$. Since diagonal matrix elements are only allowed to be paired with diagonal matrix elements, the choice of quantum number $k$ has to be $k=l$, so that $\langle l|\hat{T}_{ab}(\vec x_u)|k\rangle$ is a diagonal matrix element. The diagonal matrix element $\langle l|\hat{T}_{ab}(\vec x_u)|k=l\rangle$ can be paired with $\langle n|\hat{T}_{ij}(\vec x_s)|n\rangle$ or $\langle m|\hat{T}_{kl}(\vec x_s')|m\rangle$. Finally we obtain the result in Eq.(\ref{B8}).

The expansion for term $J_3$:
\begin{eqnarray}\label{B9.1}
J_3 & = & \frac{\beta}{(N_0L)^3}\int d^3x_sd^3x'_s\sum_{nm}\frac{e^{-\beta (E_n+E_m)}}{\mathcal{Z}^2}\bigg[\left(\delta\langle n|\right)\hat{T}_{ij}|n\rangle\langle m|\hat{T}_{kl}|m\rangle+\langle n|\hat{T}_{ij}\left(\delta|n\rangle \right)\langle m|\hat{T}_{kl}|m\rangle\nonumber \\
 & {} & \qquad\qquad\qquad\qquad\qquad+\langle n|\hat{T}_{ij}|n\rangle\left(\delta\langle m|\right)\hat{T}_{kl}|m\rangle+\langle n|\hat{T}_{ij}|n\rangle\langle m|\hat{T}_{kl}\left(\delta|m\rangle\right)\bigg]\nonumber \\
 & = & \frac{\beta}{(N_0L)^3}\int d^3x_sd^3x'_s\sum_{lmn}\sum_{abcd}\int d^3x_ud^3x_u'\sum_{ss'}\frac{1}{E_n-E_l}\frac{e^{-\beta (E_n+E_m)}}{\mathcal{Z}^2}\Lambda_{abcd}(\vec x_u-\vec x_u')\nonumber \\
 & {} & \left(\langle n|\hat{T}_{ab}(\vec x_u)\hat{T}_{cd}(\vec x_u')|l\rangle\langle l|\hat{T}_{ij}(\vec x_s)|n\rangle\langle m|\hat{T}_{kl}(\vec x_s')|m\rangle+\langle n|\hat{T}_{ij}(\vec x_s)|l\rangle \langle l|\hat{T}_{ab}(\vec x_u)\hat{T}_{cd}(\vec x_u')|n\rangle \langle m|\hat{T}_{kl}(\vec x_s')|m\rangle\right)\nonumber \\
 & + & \frac{\beta}{(N_0L)^3}\int d^3x_sd^3x'_s\sum_{lmn}\sum_{abcd}\int d^3x_ud^3x_u'\sum_{ss'}\frac{1}{E_m-E_l}\frac{e^{-\beta (E_n+E_m)}}{\mathcal{Z}^2}\Lambda_{abcd}(\vec x_u-\vec x_u')\nonumber \\
 & {} & \left(\langle n|\hat{T}_{ij}(\vec x_s)|n\rangle \langle m|\hat{T}_{ab}(\vec x_u)\hat{T}_{cd}(\vec x_u')|l\rangle \langle l|\hat{T}_{kl}(\vec x_s')|m\rangle +\langle n|\hat{T}_{ij}(\vec x_s)|n\rangle \langle m|\hat{T}_{kl}(\vec x_s')|l\rangle\langle l|\hat{T}_{ab}(\vec x_u)\hat{T}_{cd}(\vec x_u')|m\rangle\right)
\end{eqnarray}
At this stage we insert the identity between $\hat{T}_{ab}(\vec x_u)$ and $\hat{T}_{cd}(\vec x_u')$: $\sum_k|k\rangle \langle k|$. 
\begin{eqnarray}\label{B9}
J_3 & = & \frac{\beta}{(N_0L)^3}\int d^3x_sd^3x'_s\sum_{lmn}\sum_{abcd}\int d^3x_ud^3x_u'\sum_{ss'}\frac{1}{E_n-E_l}\frac{e^{-\beta (E_n+E_m)}}{\mathcal{Z}^2}\Lambda_{abcd}(\vec x_u-\vec x_u')\nonumber \\
 & {} & \bigg(\langle n|\hat{T}_{ab}(\vec x_u)\sum_k|k\rangle\langle k|\hat{T}_{cd}(\vec x_u')|l\rangle\langle l|\hat{T}_{ij}(\vec x_s)|n\rangle\langle m|\hat{T}_{kl}(\vec x_s')|m\rangle\nonumber \\
 & {} & +\langle n|\hat{T}_{ij}(\vec x_s)|l\rangle \langle l|\hat{T}_{ab}(\vec x_u)\sum_k|k\rangle\langle k|\hat{T}_{cd}(\vec x_u')|n\rangle \langle m|\hat{T}_{kl}(\vec x_s')|m\rangle\bigg)\nonumber \\
 & + & \frac{\beta}{(N_0L)^3}\int d^3x_sd^3x'_s\sum_{lmn}\sum_{abcd}\int d^3x_ud^3x_u'\sum_{ss'}\frac{1}{E_m-E_l}\frac{e^{-\beta (E_n+E_m)}}{\mathcal{Z}^2}\Lambda_{abcd}(\vec x_u-\vec x_u')\nonumber \\
 & {} & \bigg(\langle n|\hat{T}_{ij}(\vec x_s)|n\rangle \langle m|\hat{T}_{ab}(\vec x_u)\sum_k|k\rangle\langle k|\hat{T}_{cd}(\vec x_u')|l\rangle \langle l|\hat{T}_{kl}(\vec x_s')|m\rangle \nonumber \\
 & {} & +\langle n|\hat{T}_{ij}(\vec x_s)|n\rangle \langle m|\hat{T}_{kl}(\vec x_s')|l\rangle\langle l|\hat{T}_{ab}(\vec x_u)\sum_k|k\rangle\langle k|\hat{T}_{cd}(\vec x_u')|m\rangle\bigg)\nonumber \\
 & = & \frac{2\beta}{(N_0L)^3}\int d^3x_sd^3x'_s\sum_{abcd}\int d^3x_ud^3x_u'\sum_{l^{(s)}m^{(s')}n^{(s)}n^{(s')}}\frac{e^{-\beta (E_n^{(s)}+E_n^{(s')}+E_m^{(s')})}}{\mathcal{Z}^{(s)}\mathcal{Z}^{(s')2}}\Lambda_{abcd}(\vec x_u-\vec x_u')\nonumber \\
 & {} & \langle m^{(s')}|\hat{T}_{kl}(\vec x_s')|m^{(s')}\rangle \langle n^{(s')}|\hat{T}_{cd}(\vec x_u')|n^{(s')}\rangle\frac{\langle l^{(s)}|\hat{T}_{ij}(\vec x_s)|n^{(s)}\rangle\langle n^{(s)}|\hat{T}_{ab}(\vec x_u)|l^{(s)}\rangle}{E_n^{(s)}-E_l^{(s)}+i\eta}\nonumber \\
 & + & \frac{2\beta}{(N_0L)^3}\int d^3x_sd^3x'_s\sum_{abcd}\int d^3x_ud^3x_u'\sum_{l^{(s')}m^{(s')}n^{(s)}m^{(s)}}\frac{e^{-\beta (E_n^{(s)}+E_m^{(s)}+E_m^{(s')})}}{\mathcal{Z}^{(s)2}\mathcal{Z}^{(s')}}\Lambda_{abcd}(\vec x_u-\vec x_u')
\nonumber \\
 & {} &
\langle n^{(s)}|\hat{T}_{ij}(\vec x_s)|n^{(s)}\rangle \langle m^{(s)}|\hat{T}_{ab}(\vec x_u)|m^{(s)}\rangle\frac{\langle m^{(s')}|\hat{T}_{kl}(\vec x_s')|l^{(s')}\rangle\langle l^{(s')}|\hat{T}_{cd}(\vec x_u')|m^{(s')}\rangle}{E_m^{(s')}-E_l^{(s')}+i\eta}\nonumber \\
 & = & \frac{1}{(N_0L)^3}\int d^3x_sd^3x'_sd^3x_ud^3x_u'\sum_{abcd}\Lambda_{abcd}(\vec x_u-\vec x_u')\nonumber \\
 & {} & \left(\chi_{klcd}^{\rm rel(1)}(\vec x_u', \vec x_s')\chi_{abij}^{\rm res}(\vec x_s, \vec x_u; \omega=0)+\chi_{abij}^{\rm rel(1)}(\vec x_u, \vec x_s)\chi_{klcd}^{\rm res}(\vec x_s', \vec x_u';\omega=0)\right)
\end{eqnarray}
In the above calculations, coefficient $\Lambda_{abcd}(\vec x_u-\vec x_u')$ does not allow $\vec x_u, \vec x_u'$ belong to the same single block subspace. In the first step of Eq.(\ref{B9}) calculations, we have 4 summations:
\begin{eqnarray}\label{B9.2}
 & {} & \sum_{lmnk}\int d^3x_ud^3x_u'\sum_{ss'}\Lambda_{abcd}(\vec x_u-\vec x_u')\bigg(\langle n|\hat{T}_{ab}(\vec x_u)|k\rangle\langle k|\hat{T}_{cd}(\vec x_u')|l\rangle\langle l|\hat{T}_{ij}(\vec x_s)|n\rangle\langle m|\hat{T}_{kl}(\vec x_s')|m\rangle\nonumber \\
 & {} & +\langle n|\hat{T}_{ij}(\vec x_s)|l\rangle \langle l|\hat{T}_{ab}(\vec x_u)|k\rangle\langle k|\hat{T}_{cd}(\vec x_u')|n\rangle \langle m|\hat{T}_{kl}(\vec x_s')|m\rangle\bigg)\nonumber \\
 & {} & \sum_{lmnk}\int d^3x_ud^3x_u'\sum_{ss'}\Lambda_{abcd}(\vec x_u-\vec x_u')
\bigg(\langle n|\hat{T}_{ij}(\vec x_s)|n\rangle \langle m|\hat{T}_{ab}(\vec x_u)|k\rangle\langle k|\hat{T}_{cd}(\vec x_u')|l\rangle \langle l|\hat{T}_{kl}(\vec x_s')|m\rangle\nonumber \\
 & {} &  +\langle n|\hat{T}_{ij}(\vec x_s)|n\rangle \langle m|\hat{T}_{kl}(\vec x_s')|l\rangle\langle l|\hat{T}_{ab}(\vec x_u)|k\rangle\langle k|\hat{T}_{cd}(\vec x_u')|m\rangle\bigg)
\end{eqnarray}
For example, we discuss the pairing rule of the first summation only. The pairing rule for the other three summations is the same. In the first summation, the matrix elements $\langle n|\hat{T}_{ab}(\vec x_u)|k\rangle$, $\langle k|\hat{T}_{cd}(\vec x_u')|l\rangle$ cannot be paired with each other. Therefore, we need to pair $\langle n|\hat{T}_{ab}(\vec x_u)|k\rangle$, $\langle k|\hat{T}_{cd}(\vec x_u')|l\rangle$ with matrix elements $\langle l|\hat{T}_{ij}(\vec x_s)|n\rangle$, $\langle m|\hat{T}_{kl}(\vec x_s')|m\rangle$. There are two candidates: first, $\langle n|\hat{T}_{ab}(\vec x_u)|k\rangle$ is paired with $\langle l|\hat{T}_{ij}(\vec x_s)|n\rangle$, and $\langle k|\hat{T}_{cd}(\vec x_u')|l\rangle$ is paired with $\langle m|\hat{T}_{kl}(\vec x_s')|m\rangle$; second, $\langle n|\hat{T}_{ab}(\vec x_u)|k\rangle$ is paired with $\langle m|\hat{T}_{kl}(\vec x_s')|m\rangle$, and $\langle k|\hat{T}_{cd}(\vec x_u')|l\rangle$ is paired with $\langle l|\hat{T}_{ij}(\vec x_s)|n\rangle$. The first candidate forces $\vec x_u, \vec x_s\in V^{(s)}, \vec x_u', \vec x_s'\in V^{(s')}$. Since the matrix element $\langle l|\hat{T}_{ij}(\vec x_s)|n\rangle$ is off-diagonal, $\langle n|\hat{T}_{ab}(\vec x_u)|k\rangle$ has to be off-diagonal as well. Therefore we choose $k=l$. In the second candidate forces $\vec x_u, \vec x_s'\in V^{(s')}, \vec x_u', \vec x_s\in V^{(s)}$. We also need to choose $k=n$ so that $\langle n|\hat{T}_{ab}(\vec x_u)|k\rangle$ is diagonal. Repeat the same process for the other three summations in Eq.(\ref{B9.2}), we finally obtain the result in Eq.(\ref{B9}).

\subsection{Expansion details for $\chi_{ijkl}^{\rm sup \,rel(2)}$}
\begin{eqnarray}\label{B10}
\chi_{ijkl}^{\rm sup \,rel(2)} & = & -\frac{\beta }{(N_0L)^3}\int d^3x_sd^3x'_s\sum_{n^*}\frac{e^{-\beta E_n^*}}{\mathcal{Z}^*}\langle n^*|\hat{T}_{ij}(\vec x_s)|n^*\rangle\langle n^*|\hat{T}_{kl}(\vec x_s')|n^*\rangle\nonumber \\
 & = & -\frac{\beta }{(N_0L)^3}\int d^3x_sd^3x'_s\sum_{n}\frac{e^{-\beta E_n}(1-\beta E_n)}{\mathcal{Z}+\delta \mathcal{Z}}\left(\langle n|+\delta \langle n|\right)\hat{T}_{ij}(\vec x_s)\left(|n\rangle+\delta|n\rangle\right)\left(\langle n|+\delta\langle n|\right)\hat{T}_{kl}(\vec x_s')\left(|n\rangle+\delta|n\rangle\right)\nonumber \\
 & = & -\frac{\beta }{(N_0L)^3}\int d^3x_sd^3x'_s\sum_{n}\frac{e^{-\beta E_n}}{\mathcal{Z}}\langle n|\hat{T}_{ij}(\vec x_s)|n\rangle\langle n|\hat{T}_{kl}(\vec x_s')|n\rangle\nonumber \\
 & {} & -\frac{\beta}{(N_0L)^3}\int d^3x_sd^3x'_s\sum_n\frac{e^{-\beta E_n}(-\beta \delta E_n)}{\mathcal{Z}}\langle n|\hat{T}_{ij}(\vec x_s)|n\rangle \langle n|\hat{T}_{kl}(\vec x_s')|n\rangle \nonumber \qquad\qquad\qquad\qquad\qquad\qquad J_4\\
 & {} & -\frac{\beta }{(N_0L)^3}\int d^3x_sd^3x'_s\sum_n\frac{-e^{-\beta E_n}}{\mathcal{Z}^2}\delta\mathcal{Z}\langle n|\hat{T}_{ij}(\vec x_s)|n\rangle \langle n|\hat{T}_{kl}(\vec x_s')|n\rangle \qquad\qquad\qquad\qquad\qquad\qquad\qquad J_5\nonumber \\
 & {} & -\frac{\beta }{(N_0L)^3}\int d^3x_sd^3x'_s\sum_n\frac{e^{-\beta E_n}}{\mathcal{Z}}\bigg[\left(\delta\langle n|\right)\hat{T}_{ij}(\vec x_s)|n\rangle \langle n|\hat{T}_{kl}(\vec x_s')|n\rangle+\langle n|\hat{T}_{ij}(\vec x_s)\left(\delta |n\rangle\right)\langle n|\hat{T}_{kl}(\vec x_s')|n\rangle\nonumber \\
 & {} & \qquad\qquad\qquad\qquad\qquad\qquad\,\,\,\,+\langle n|\hat{T}_{ij}(\vec x_s) |n\rangle\left(\delta\langle n|\right)\hat{T}_{kl}(\vec x_s')|n\rangle+\langle n|\hat{T}_{ij}(\vec x_s) |n\rangle\langle n|\hat{T}_{kl}(\vec x_s')\left(\delta|n\rangle\right)\bigg]\quad J_6
\end{eqnarray}
Expansion for term $J_4$:
\begin{eqnarray}\label{B11}
J_4 & = & \frac{\beta^2 }{(N_0L)^3}\int d^3x_sd^3x'_s\sum_n\frac{e^{-\beta E_n}}{\mathcal{Z}}\delta E_n\langle n|\hat{T}_{ij}(\vec x_s)|n\rangle \langle n|\hat{T}_{kl}(\vec x_s')|n\rangle\nonumber \\
 & = & \frac{\beta^2}{(N_0L)^3}\int d^3x_sd^3x'_s\sum_n\frac{e^{-\beta E_n}}{\mathcal{Z}}\sum_{abcd}\int d^3x_ud^3x_u'\Lambda_{abcd}(\vec x_u-\vec x_u')\langle n|\hat{T}_{ab}(\vec x_u)\hat{T}_{cd}(\vec x_u')|n\rangle \langle n|\hat{T}_{ij}(\vec x_s)|n\rangle \langle n|\hat{T}_{kl}(\vec x_s')|n\rangle\nonumber \\
 & = & \frac{\beta^2}{(N_0L)^3}\int d^3x_sd^3x'_s\sum_n\frac{e^{-\beta E_n}}{\mathcal{Z}}\sum_{abcd}\int d^3x_ud^3x_u'\Lambda_{abcd}(\vec x_u-\vec x_u')\nonumber \\
 & {} & \langle n|\hat{T}_{ab}(\vec x_u)\sum_k|k\rangle\langle k|\hat{T}_{cd}(\vec x_u')|n\rangle \langle n|\hat{T}_{ij}(\vec x_s)|n\rangle \langle n|\hat{T}_{kl}(\vec x_s')|n\rangle\nonumber \\
 & = & \frac{\beta^2}{(N_0L)^3}\int d^3x_sd^3x'_s\sum_n\frac{e^{-\beta E_n}}{\mathcal{Z}}\sum_{abcd}\Lambda_{abcd}(\vec x_u-\vec x_u')\nonumber \\
 & {} & \langle n^{(s)}|\hat{T}_{ab}(\vec x_u)|n^{(s)}\rangle\langle n^{(s)}|\hat{T}_{ij}(\vec x_s)|n^{(s)}\rangle \langle n^{(s')}|\hat{T}_{kl}(\vec x_s')|n^{(s')}\rangle \langle n^{(s')}|\hat{T}_{cd}(\vec x_u')|n^{(s')}\rangle\nonumber \\
 & = & \frac{1}{(N_0L)^3}\int d^3x_sd^3x'_sd^3x_ud^3x_u'\sum_{abcd}\Lambda_{abcd}(\vec x_u-\vec x_u')\chi_{abij}^{\rm rel(2)}(\vec x_u, \vec x_s)\chi_{cdkl}^{\rm rel(2)}(\vec x_s', \vec x_u')
\end{eqnarray}
where in the above calculation we insert the identity $\sum_k|k\rangle\langle k|$. Again, the coefficient $\Lambda_{abcd}(\vec x_u-\vec x_u')$ does not allow $\vec x_u, \vec x_u'$ belong to the same single block subspace. We need to pair matrix elements $\langle n|\hat{T}_{ab}(\vec x_u)|k\rangle$, $\langle k|\hat{T}_{cd}(\vec x_u')|n\rangle $ with $\langle n|\hat{T}_{ij}(\vec x_s)|n\rangle$, $\langle n|\hat{T}_{kl}(\vec x_s')|n\rangle$. Therefore,  we choose $\vec x_u, \vec x_s\in V^{(s)}, \vec x_u', \vec x_s'\in V^{(s')}$, or $\vec x_u, \vec x_s'\in V^{(s')}, \vec x_u', \vec x_s\in V^{(s)}$. Since the matrix elements $\langle n|\hat{T}_{ij}(\vec x_s)|n\rangle$, $\langle n|\hat{T}_{kl}(\vec x_s')|n\rangle$ are diagonal, the only choice for quantum number $k$ is $k=n$.

Expansion for term $J_5$:
\begin{eqnarray}\label{B12}
J_5 & = & \frac{\beta}{(N_0L)^3}\int d^3x_sd^3x'_s\sum_n\frac{e^{-\beta E_n}}{\mathcal{Z}^2}\delta \mathcal{Z}\langle n|\hat{T}_{ij}(\vec x_s)|n\rangle \langle n|\hat{T}_{kl}(\vec x_s')|n\rangle\nonumber \\
 & = & -\frac{\beta^2}{(N_0L)^3}\int d^3x_sd^3x'_s\sum_{nl}\frac{e^{-\beta( E_n+E_l)}}{\mathcal{Z}^2}\langle l|\hat{V}|l\rangle \langle n|\hat{T}_{ij}(\vec x_s)|n\rangle \langle n|\hat{T}_{kl}(\vec x_s')|n\rangle\nonumber \\
 & = & -\frac{\beta^2}{(N_0L)^3}\int d^3x_sd^3x'_s\sum_{nl}\frac{e^{-\beta( E_n+E_l)}}{\mathcal{Z}^2}\sum_{abcd}\int d^3x_ud^3x_u'\Lambda_{abcd}(\vec x_u-\vec x_u')\langle l| \hat{T}_{ab}(\vec x_u)\hat{T}_{cd}(\vec x_u')|l\rangle \langle n|\hat{T}_{ij}(\vec x_s)|n\rangle \langle n|\hat{T}_{kl}(\vec x_s')|n\rangle\nonumber \\
 & = & -\frac{\beta^2}{(N_0L)^3}\int d^3x_sd^3x'_s\sum_{nl}\frac{e^{-\beta( E_n+E_l)}}{\mathcal{Z}^2}\sum_{abcd}\int d^3x_ud^3x_u'\Lambda_{abcd}(\vec x_u-\vec x_u')\nonumber \\
 & {} & \langle l| \hat{T}_{ab}(\vec x_u)\sum_k|k\rangle\langle k|\hat{T}_{cd}(\vec x_u')|l\rangle \langle n|\hat{T}_{ij}(\vec x_s)|n\rangle \langle n|\hat{T}_{kl}(\vec x_s')|n\rangle\nonumber \\
 & = & -\frac{\beta^2}{(N_0L)^3}\int d^3x_sd^3x'_s\sum_{lmn}\frac{e^{-\beta( E_n+E_l)}}{\mathcal{Z}^2}\sum_{abcd}\int d^3x_ud^3x_u'\Lambda_{abcd}(\vec x_u-\vec x_u')\nonumber \\
 & {} & \langle l| \hat{T}_{ab}(\vec x_u)|m\rangle \langle m|\hat{T}_{cd}(\vec x_u')|l\rangle \langle n|\hat{T}_{ij}(\vec x_s)|n\rangle \langle n|\hat{T}_{kl}(\vec x_s')|n\rangle\nonumber \\
 & = & -\frac{\beta^2}{(N_0L)^3}\int d^3x_sd^3x'_s\sum_{l^{(s)}l^{(s')}n^{(s)}n^{(s')}}\frac{e^{-\beta( E_n^{(s)}+E_n^{(s')}+E_l^{(s)}+E_l^{(s')})}}{\mathcal{Z}^{(s)2}\mathcal{Z}^{(s')2}}\sum_{abcd}\int d^3x_ud^3x_u'\Lambda_{abcd}(\vec x_u-\vec x_u')\nonumber \\
 & {} & \langle l^{(s)}| \hat{T}_{cd}(\vec x_u')|l^{(s)}\rangle \langle n^{(s)}|\hat{T}_{ij}(\vec x_s)|n^{(s)}\rangle \langle l^{(s')}|\hat{T}_{ab}(\vec x_u)|l^{(s')}\rangle \langle n^{(s')}|\hat{T}_{kl}(\vec x_s')|n^{(s')}\rangle\nonumber \\
 & = & -\frac{1}{(N_0L)^3}\int d^3x_sd^3x'_sd^3x_ud^3x_u'\sum_{abcd}\Lambda_{abcd}(\vec x_u-\vec x_u')\chi_{abij}^{\rm rel(1)}(\vec x_u, \vec x_s)\chi_{cdkl}^{\rm rel(1)}(\vec x_s', \vec x_u')
\end{eqnarray}
where in the above calculations we have inserted the identity $\sum_k|k\rangle\langle k|$. The only choice for quantum number $k$ is $k=l$.

Expansion for term $J_6$:
\begin{eqnarray}\label{B13.1}
J_6 & = & -\frac{\beta}{(N_0L)^3}\int d^3x_sd^3x'_s\sum_{nm}\frac{1}{E_n-E_m}\frac{e^{-\beta E_n}}{\mathcal{Z}}\bigg[
   \langle n|\hat{V}|m\rangle \langle m|\hat{T}_{ij}|n\rangle\langle n|\hat{T}_{kl}|n\rangle
+ \langle n|\hat{T}_{ij}|m\rangle\langle m|\hat{V}|n\rangle\langle n|\hat{T}_{kl}|n\rangle\nonumber \\
 & {} & \qquad\qquad\qquad\qquad\qquad\qquad\quad+ \langle n|\hat{T}_{ij}|n\rangle\langle n|\hat{V}|m\rangle\langle m|\hat{T}_{kl}|n\rangle
+ \langle n|\hat{T}_{ij}|n\rangle\langle n|\hat{T}_{kl}|m\rangle\langle m|\hat{V}|n\rangle
\bigg]\nonumber \\
 & = & 
-\frac{\beta}{(N_0L)^3}\int d^3x_sd^3x'_s\sum_{nml}\sum_{abcd}\int d^3x_ud^3x_u'\frac{1}{E_n-E_m}\frac{e^{-\beta E_n}}{\mathcal{Z}}\Lambda_{abcd}(\vec x_u-\vec x_u')\nonumber \\
 & {} & \bigg\{\langle n|\hat{T}_{ab}(\vec x_u)\sum_l|l\rangle \langle l|\hat{T}_{cd}(\vec x_u')|m\rangle\langle m|\hat{T}_{ij}(\vec x_s)|n\rangle\langle n|\hat{T}_{kl}(\vec x_s')|n\rangle+\langle n|\hat{T}_{ij}(\vec x_s)|m\rangle\langle m|\hat{T}_{ab}(\vec x_u)\sum_l|l\rangle \langle l|\hat{T}_{cd}(\vec x_u')|n\rangle\langle n|\hat{T}_{kl}(\vec x_s')|n\rangle\nonumber \\
 & {} & +\langle n|\hat{T}_{ij}(\vec x_s)|n\rangle\langle n|\hat{T}_{ab}(\vec x_u)\sum_l|l\rangle \langle l|\hat{T}_{cd}(\vec x_u')|m\rangle\langle m|\hat{T}_{kl}(\vec x_s')|n\rangle+\langle n|\hat{T}_{ij}(\vec x_s)|n\rangle\langle n|\hat{T}_{kl}(\vec x_s')|m\rangle\langle m|\hat{T}_{ab}(\vec x_u)\sum_l|l\rangle \langle l|\hat{T}_{cd}(\vec x_u')|n\rangle\bigg\}\nonumber \\
 & = & 
-\frac{\beta}{(N_0L)^3}\int d^3x_sd^3x'_s\sum_{nml}\sum_{abcd}\int d^3x_ud^3x_u'\frac{1}{E_n-E_m}\frac{e^{-\beta E_n}}{\mathcal{Z}}\Lambda_{abcd}(\vec x_u-\vec x_u')\nonumber \\
 & {} & \bigg\{\langle n|\hat{T}_{ab}(\vec x_u)|l\rangle \langle l|\hat{T}_{cd}(\vec x_u')|m\rangle\langle m|\hat{T}_{ij}(\vec x_s)|n\rangle\langle n|\hat{T}_{kl}(\vec x_s')|n\rangle+\langle n|\hat{T}_{ij}(\vec x_s)|m\rangle\langle m|\hat{T}_{ab}(\vec x_u)|l\rangle\langle l|\hat{T}_{cd}(\vec x_u')|n\rangle\langle n|\hat{T}_{kl}(\vec x_s')|n\rangle\nonumber \\
 & {} & +\langle n|\hat{T}_{ij}(\vec x_s)|n\rangle\langle n|\hat{T}_{ab}(\vec x_u)|l\rangle \langle l|\hat{T}_{cd}(\vec x_u')|m\rangle\langle m|\hat{T}_{kl}(\vec x_s')|n\rangle+\langle n|\hat{T}_{ij}(\vec x_s)|n\rangle\langle n|\hat{T}_{kl}(\vec x_s')|m\rangle\langle m|\hat{T}_{ab}(\vec x_u)|l\rangle\langle l|T_{cd}^{(u')}|n\rangle\bigg\}\nonumber \\
 & = & 
-\frac{\beta}{(N_0L)^3}\int d^3x_sd^3x'_s\sum_{nml}\sum_{abcd}\int d^3x_ud^3x_u'\frac{1}{E_n-E_m}\frac{e^{-\beta E_n}}{\mathcal{Z}}\Lambda_{abcd}(\vec x_u-\vec x_u')\nonumber \\
 & {} & \bigg\{\langle n|\hat{T}_{ab}(\vec x_u)|l\rangle \langle l|\hat{T}_{cd}(\vec x_u')|m\rangle\langle m|\hat{T}_{ij}(\vec x_s)|n\rangle\langle n|\hat{T}_{kl}(\vec x_s')|n\rangle+\langle n|\hat{T}_{ij}(\vec x_s)|m\rangle\langle m|\hat{T}_{ab}(\vec x_u)|l\rangle\langle l|\hat{T}_{cd}(\vec x_u')|n\rangle\langle n|\hat{T}_{kl}(\vec x_s')|n\rangle\nonumber \\
 & {} & +\langle n|\hat{T}_{ij}(\vec x_s)|n\rangle\langle n|\hat{T}_{ab}(\vec x_u)|l\rangle \langle l|\hat{T}_{cd}(\vec x_u')|m\rangle\langle m|\hat{T}_{kl}(\vec x_s')|n\rangle+\langle n|\hat{T}_{ij}(\vec x_s)|n\rangle\langle n|\hat{T}_{kl}(\vec x_s')|m\rangle\langle m|\hat{T}_{ab}(\vec x_u)|l\rangle\langle l|\hat{T}_{cd}(\vec x_u')|n\rangle\bigg\}
\end{eqnarray}
In the above calculations, we insert the identity $\sum_l|l\rangle\langle l|$. 

There are four summations in the above result. As an example, we discuss the pairing rule for the first summation, $\sum_{nml}\int d^3x_ud^3x_u'\Lambda_{abcd}(\vec x_u-\vec x_u')\langle n|\hat{T}_{ab}(\vec x_u)|l\rangle \langle l|\hat{T}_{cd}(\vec x_u')|m\rangle\langle m|\hat{T}_{ij}(\vec x_s)|n\rangle\langle n|\hat{T}_{kl}(\vec x_s')|n\rangle$. The coefficient $\Lambda_{abcd}(\vec x_u-\vec x_u')$ does not allow $\vec x_u, \vec x_u'$ belong to the same single block subspace. Therefore the matrix elements $\langle n|\hat{T}_{ab}(\vec x_u)|l\rangle$, $ \langle l|\hat{T}_{cd}(\vec x_u')|m\rangle$ cannot be paired with each other. We need to pair $\langle n|\hat{T}_{ab}(\vec x_u)|l\rangle$, $ \langle l|\hat{T}_{cd}(\vec x_u')|m\rangle$ with $\langle m|\hat{T}_{ij}(\vec x_s)|n\rangle$, $\langle n|\hat{T}_{kl}(\vec x_s')|n\rangle$. There are two pairing candidates: first, $\langle n|\hat{T}_{ab}(\vec x_u)|l\rangle$ is paired with $\langle m|\hat{T}_{ij}(\vec x_s)|n\rangle$, and $ \langle l|\hat{T}_{cd}(\vec x_u')|m\rangle$ is paired with $\langle n|\hat{T}_{kl}(\vec x_s')|n\rangle$; second, $\langle n|\hat{T}_{ab}(\vec x_u)|l\rangle$ is paired with $\langle n|\hat{T}_{kl}(\vec x_s')|n\rangle$, and $ \langle l|\hat{T}_{cd}(\vec x_u')|m\rangle$ is paired with $\langle m|\hat{T}_{ij}(\vec x_s)|n\rangle$. In the first candidate of pairing matrix elements, we have $\vec x_u, \vec x_s\in V^{(s)}, \vec x_u', \vec x_s'\in V^{(s')}$. Since $\langle n|\hat{T}_{kl}(\vec x_s')|n\rangle$ is diagonal, the matrix element $ \langle l|\hat{T}_{cd}(\vec x_u')|m\rangle$ which is pair to it must be diagonal as well. Therefore the only choice for quantum number $l$ is $l=m$. In the second candidate, we have $\vec x_u, \vec x_s'\in V^{(s')}, \vec x_u', \vec x_s\in V^{(s)}$. Again, since $\langle n|\hat{T}_{kl}(\vec x_s')|n\rangle$ is diagonal, the matrix element $\langle n|\hat{T}_{ab}(\vec x_u)|l\rangle$ which is paired to it must be diagonal as well. The only choice for quantum number $l$ is $l=n$. With the previous pairing rule, we continue our calculation as follows,  

\begin{eqnarray}\label{B13}
J_6
 & = & 
-\frac{2\beta}{(N_0L)^3}\int d^3x_sd^3x'_s\sum_{abcd}\sum_{n^{(s)}n^{(s')}m^{(s)}}\int d^3x_ud^3x_u'\frac{1}{E_n^{(s)}-E_m^{(s)}}\frac{e^{-\beta (E_n^{(s)}+E_n^{(s')})}}{\mathcal{Z}^{(s)}\mathcal{Z}^{(s')}}\Lambda_{abcd}(\vec x_u-\vec x_u')\nonumber \\
 & {} & 
\langle n^{(s')}|\hat{T}_{kl}(\vec x_s')|n^{(s')}\rangle \langle n^{(s')}|\hat{T}_{ab}(\vec x_u)|n^{(s')}\rangle\langle m^{(s)}|\hat{T}_{cd}(\vec x_u')|n^{(s)}\rangle\langle n^{(s)}|\hat{T}_{ij}(\vec x_s)|m^{(s)}\rangle\nonumber \\
 & + & -\frac{2\beta}{(N_0L)^3}\int d^3x_sd^3x'_s\sum_{abcd}\sum_{n^{(s)}n^{(s')}m^{(s')}}\int d^3x_ud^3x_u'\frac{1}{E_n^{(s')}-E_m^{(s')}}\frac{e^{-\beta (E_n^{(s)}+E_n^{(s')})}}{\mathcal{Z}^{(s)}\mathcal{Z}^{(s')}}\Lambda_{abcd}(\vec x_u-\vec x_u')\nonumber \\
 & {} & \langle n^{(s)}|\hat{T}_{ij}(\vec x_s)|n^{(s)}\rangle\langle n^{(s)}|\hat{T}_{ab}(\vec x_u)|n^{(s)}\rangle\langle n^{(s')}|\hat{T}_{cd}(\vec x_u')|m^{(s')}\rangle\langle m^{(s')}|\hat{T}_{kl}(\vec x_s')|n^{(s')}\rangle\nonumber \\
 & = & -\frac{1}{(N_0L)^3}\int d^3x_sd^3x'_s\sum_{abcd}\int d^3x_ud^3x_u'\Lambda_{abcd}(\vec x_u-\vec x_u')\nonumber \\
 & {} & \left(\chi_{abkl}^{\rm rel(2)}(\vec x_u, \vec x_s')\chi_{cdij}^{\rm res}(\vec x_s, \vec x_u'; \omega=0)+\chi_{abij}^{\rm rel(2)}(\vec x_u, \vec x_s)\chi_{cdkl}^{\rm res}(\vec x_s', \vec x_u'; 0)\right)
\end{eqnarray}

\subsection{Expansion details for $\chi_{ijkl}^{\rm sup \, res}(\omega)$}
\begin{eqnarray}\label{B14}
\chi_{ijkl}^{\rm sup \, res}(\omega) & = & \frac{2}{\hbar (N_0L)^3}\int d^3x_sd^3x'_s\sum_{n^*l^*}\frac{e^{-\beta E_n^*}}{\mathcal{Z}^*}\langle l^*|\hat{T}_{ij}(\vec x_s)|n^*\rangle\langle n^*|\hat{T}_{kl}(\vec x_s')|l^*\rangle\frac{\omega_{ln}^*}{(\omega+i\eta)^2-\omega_{ln}^{*2}}\nonumber \\
 & = & \frac{2}{\hbar (N_0L)^3}\int d^3x_sd^3x'_s\sum_{nl}\frac{e^{-\beta E_n}}{\mathcal{Z}}\langle l|\hat{T}_{ij}(\vec x_s)|n\rangle\langle n|\hat{T}_{kl}(\vec x_s')|l\rangle\frac{\omega_{ln}}{(\omega+i\eta)^2-\omega_{ln}^{2}}\nonumber \\
 & + & \frac{2}{\hbar (N_0L)^3}\int d^3x_sd^3x'_s\sum_{nl}\frac{e^{-\beta E_n}(-\beta \delta E_n)}{\mathcal{Z}}\langle l|\hat{T}_{ij}(\vec x_s)|n\rangle\langle n|\hat{T}_{kl}(\vec x_s')|l\rangle\frac{\omega_{ln}}{(\omega+i\eta)^2-\omega_{ln}^2}\qquad\qquad\qquad J_7\nonumber \\
 & + & \frac{2}{\hbar (N_0L)^3}\int d^3x_sd^3x'_s\sum_{nl}\frac{e^{-\beta E_n}(-\delta \mathcal{Z})}{\mathcal{Z}}\langle l|\hat{T}_{ij}(\vec x_s)|n\rangle\langle n|\hat{T}_{kl}(\vec x_s')|l\rangle\frac{\omega_{ln}}{(\omega+i\eta)^2-\omega_{ln}^2}
\qquad\qquad\qquad\quad\, J_8
\nonumber \\
 & + & \frac{2}{\hbar (N_0L)^3}\int d^3x_sd^3x'_s\sum_{nl}\frac{e^{-\beta E_n}}{\mathcal{Z}}\langle l|\hat{T}_{ij}(\vec x_s)|n\rangle\langle n|\hat{T}_{kl}(\vec x_s')|l\rangle\frac{(\omega+i\eta)^2+\omega_{ln}^2}{[(\omega+i\eta)^2-\omega_{ln}^2]^2}\delta\omega_{ln}
\qquad\qquad\qquad\quad\, J_9
\nonumber \\
 & + & \frac{2}{\hbar (N_0L)^3}\int d^3x_sd^3x'_s\sum_{nl}\frac{e^{-\beta E_n}}{\mathcal{Z}}\bigg[\left(\delta\langle l|\right)\hat{T}_{ij}(\vec x_s)|n\rangle\langle n|\hat{T}_{kl}(\vec x_s')|l\rangle+\langle l|\hat{T}_{ij}(\vec x_s)\left(\delta|n\rangle\right)\langle n|\hat{T}_{kl}(\vec x_s')|l\rangle
\nonumber \\
 & {} & \qquad+\langle l|\hat{T}_{ij}(\vec x_s)|n\rangle\left(\delta\langle n|\right)\hat{T}_{kl}(\vec x_s')|l\rangle+\langle l|\hat{T}_{ij}(\vec x_s)|n\rangle\langle n|\hat{T}_{kl}(\vec x_s')\left(\delta|l\rangle\right)\bigg]\frac{\omega_{ln}}{(\omega+i\eta)^2-\omega_{ln}^2} \qquad J_{10}
\end{eqnarray}
where please note we use the simplified notation $(E_l-E_n)/\hbar=\omega_l-\omega_n=\omega_{ln}$. We denote the change of $\omega_{ln}$ to be $\delta\omega_{ln}=(
\delta E_l-\delta E_n)/\hbar$.\\
Expansion for term $J_7$:
\begin{eqnarray}\label{B15}
J_7 & = & -\frac{2\beta}{\hbar (N_0L)^3}\int d^3x_sd^3x'_s\sum_{nl}\frac{e^{-\beta E_n}}{\mathcal{Z}}\langle n|\hat{V}|n\rangle\langle l|\hat{T}_{ij}(\vec x_s)|n\rangle\langle n|\hat{T}_{kl}(\vec x_s')|l\rangle\frac{\omega_{ln}}{(\omega+i\eta)^2-\omega_{ln}^2}\nonumber \\
 & = & -\frac{2\beta}{\hbar (N_0L)^3}\int d^3x_sd^3x'_s\sum_{nl}\frac{e^{-\beta E_n}}{\mathcal{Z}}\sum_{abcd}\int d^3x_ud^3x_u'\Lambda_{abcd}(\vec x_u-\vec x_u')\nonumber \\
 & {} & \langle n|\hat{T}_{ab}(\vec x_u)\hat{T}_{cd}(\vec x_u')|n\rangle\langle l|\hat{T}_{ij}(\vec x_s)|n\rangle\langle n|\hat{T}_{kl}(\vec x_s')|l\rangle\frac{\omega_{ln}}{(\omega+i\eta)^2-\omega_{ln}^2}\nonumber \\
 & = & -\frac{2\beta}{\hbar (N_0L)^3}\int d^3x_sd^3x'_s\sum_{nl}\frac{e^{-\beta E_n}}{\mathcal{Z}}\sum_{abcd}\int d^3x_ud^3x_u'\Lambda_{abcd}(\vec x_u-\vec x_u')\nonumber \\
 & {} & \langle n|\hat{T}_{ab}(\vec x_u)\sum_k|k\rangle\langle k|\hat{T}_{cd}(\vec x_u')|n\rangle\langle l|\hat{T}_{ij}(\vec x_s)|n\rangle\langle n|\hat{T}_{kl}(\vec x_s')|l\rangle\frac{\omega_{ln}}{(\omega+i\eta)^2-\omega_{ln}^2}\nonumber \\
 & = & -\frac{4\beta}{\hbar (N_0L)^3}\int d^3x_sd^3x'_s\sum_{nl}\frac{e^{-\beta E_n}}{\mathcal{Z}}\sum_{abcd}\int d^3x_ud^3x_u'\Lambda_{abcd}(\vec x_u-\vec x_u')\frac{\omega_{ln}}{(\omega+i\eta)^2-\omega_{ln}^2}\nonumber \\
 & {} & {\rm Tr}\,\left[\hat{T}_{cd}(\vec x_u')|n\rangle\langle l|\hat{T}_{ij}(\vec x_s)|n\rangle\langle n|\hat{T}_{kl}(\vec x_s')|l\rangle\langle n|\hat{T}_{ab}(\vec x_u)\right]\nonumber \\
 & = & 
-\frac{4\beta}{\hbar (N_0L)^3}\int d^3x_sd^3x'_s\sum_{nl}\frac{e^{-\beta E_n}}{\mathcal{Z}}\sum_{abcd}\int d^3x_ud^3x_u'\Lambda_{abcd}(\vec x_u-\vec x_u')\frac{\omega_{ln}}{(\omega+i\eta)^2-\omega_{ln}^2}\nonumber \\
 & {} & 
\langle n^{(s)}|\hat{T}_{ij}(\vec x_s)|n^{(s)}\rangle \langle n^{(s)}|\hat{T}_{ab}(\vec x_u)|n^{(s)}\rangle\langle n^{(s')}|\hat{T}_{cd}(\vec x_u')|n^{(s')}\rangle\langle n^{(s')}|\hat{T}_{kl}(\vec x_s')|n^{(s')}\rangle
\langle n^{(s)}|l^{(s)}\rangle\langle n^{(s')}|l^{(s')}\rangle\langle n^{(r)}|l^{(r)}\rangle\nonumber \\
 & = & 0
\end{eqnarray}
In the above calculations we insert the identity $\sum_k|k\rangle\langle k|$. The coefficient $\Lambda_{abcd}(\vec x_u-\vec x_u')$ does not allow $\vec x_u, \vec x_u'$ belong to the same single block subspace. Therefore we need to pair matrix elements $\langle n|\hat{T}_{ab}(\vec x_u)|k\rangle$, $\langle k|\hat{T}_{cd}(\vec x_u')|n\rangle$ with $\langle l|\hat{T}_{ij}(\vec x_s)|n\rangle$, $\langle n|\hat{T}_{kl}(\vec x_s')|l\rangle$. There are two candidates: first, $\langle n|\hat{T}_{ab}(\vec x_u)|k\rangle$ is paired with $\langle l|\hat{T}_{ij}(\vec x_s)|n\rangle$, and $\langle k|\hat{T}_{cd}(\vec x_u')|n\rangle$ is paired with $\langle n|\hat{T}_{kl}(\vec x_s')|l\rangle$; second, $\langle n|\hat{T}_{ab}(\vec x_u)|k\rangle$ is paired with $\langle n|\hat{T}_{kl}(\vec x_s')|l\rangle$, and $\langle k|\hat{T}_{cd}(\vec x_u')|n\rangle$ is paired with $\langle l|\hat{T}_{ij}(\vec x_s)|n\rangle$. The first candidate allows $\vec x_u, \vec x_s\in V^{(s)}, \vec x_u', \vec x_s'\in V^{(s')}$. Since the matrix elements $\langle l|\hat{T}_{ij}(\vec x_s)|n\rangle$ and $\langle n|\hat{T}_{kl}(\vec x_s')|l\rangle$ are off-diagonal, the pairing rule requires the quantum number $k$ to be $k=l$. The matrix element product is therefore given by 
\begin{eqnarray}\label{B15.1}
 & {} & \sum_{nl}\frac{e^{-\beta E_n}}{\mathcal{Z}}\langle n|\hat{T}_{ab}(\vec x_u)|l\rangle\langle l|\hat{T}_{ij}(\vec x_s)|n\rangle\langle n|\hat{T}_{kl}(\vec x_s')|l\rangle\langle l|\hat{T}_{cd}(\vec x_u')|n\rangle\frac{\omega_{ln}}{(\omega+i\eta)^2-\omega_{ln}^2}\nonumber \\
 & = & \frac{e^{-\beta \left(E_n^{(s)}+E_n^{(s')}\right)}}{\mathcal{Z}^{(s)}\mathcal{Z}^{(s')}}\sum_{n^{(s)}n^{(s')}l^{(s)}l^{(s')}}\langle n^{(s)}|\hat{T}_{ab}(\vec x_u)|l^{(s)}\rangle\langle l^{(s)}|\hat{T}_{ij}(\vec x_s)|n^{(s)}\rangle\langle n^{(s')}|l^{(s')}\rangle \langle l^{(s')}|n^{(s')}\rangle\nonumber \\
 & {} & \langle n^{(s')}|\hat{T}_{kl}(\vec x_s')|l^{(s')}\rangle\langle l^{(s')}|\hat{T}_{cd}(\vec x_u')|n^{(s')}\rangle\langle n^{(s)}|l^{(s)}\rangle\langle l^{(s)}|n^{(s)}\rangle\frac{\omega_{ln}^{(s)}+\omega_{ln}^{(s')}}{(\omega+i\eta)^2-\left(\omega_{ln}^{(s)}+\omega_{ln}^{(s')}\right)^2}\nonumber \\
 & = & 0 \quad\quad {\rm since}\quad\quad \omega_{ln}^{(s)}=\omega_{ln}^{(s')}=0
\end{eqnarray}
The second candidate allows $\vec x_u, \vec x_s'\in V^{(s')}, \vec x_u', \vec x_s\in V^{(s)}$ and $k=l=n$, so that all of the matrix elements are diagonal. We also have $\omega_{ln}=0$. Finally, the above result Eq.(\ref{B15}) is zero.

There is an additional qualitative argument which leads us to the same result for term(7) very quickly: suppose $s\neq s'$ and $l\neq n$ in the second step of Eq.(\ref{B15}). The operator for the $s$-th block stress tensor $\hat{T}_{ij}(\vec x_s)$ changes state from wavefunction $|l\rangle$ to $|n\rangle$, and the $s'$-th block stress tensor $\hat{T}_{kl}(\vec x_s')$ changes state from $|n\rangle$ to $|l\rangle$. However, since the $s$-th block stress tensor only acts on the $s$-th block wavefunction, and the $s'$-th block stress tensor only acts on the $s'$-th block wavefunction, it is impossible to change state $|l\rangle$ to state $|n\rangle$ and state $|n\rangle$ to state $|l\rangle$ simultaneously by stress tensor operators from two different blocks. The only possibility is that states $|l\rangle$ and $|n\rangle$ are the same state: $|n\rangle=|l\rangle$, which means the wave functions $|l\rangle$ and $|n\rangle$ are not changed by stress tensors $\hat{T}_{ij}(\vec x_s)$ and $\hat{T}_{kl}(\vec x_s')$. This argument also leads to the same result of term(7), because the factor $\omega_{ln}=\omega_l-\omega_n=0$ makes term $J_7$ to vanish. 

Expansion for term $J_8$:
\begin{eqnarray}\label{B16}
J_8 & = & -\frac{2}{\hbar (N_0L)^3}\int d^3x_sd^3x'_s\sum_{nl}\frac{e^{-\beta E_n}}{\mathcal{Z}^2}\delta\mathcal{Z}\langle l|\hat{T}_{ij}(\vec x_s)|n\rangle\langle n|\hat{T}_{kl}(\vec x_s')|l\rangle\frac{\omega_{ln}}{(\omega+i\eta)^2-\omega_{ln}^2}\nonumber \\
 & = & \frac{2\beta}{\hbar (N_0L)^3}\int d^3x_sd^3x'_s\sum_{lmn}\frac{e^{-\beta (E_n+E_m)}}{\mathcal{Z}^2}\langle m|V|m\rangle\langle l|\hat{T}_{ij}(\vec x_s)|n\rangle\langle n|\hat{T}_{kl}(\vec x_s')|l\rangle\frac{\omega_{ln}}{(\omega+i\eta)^2-\omega_{ln}^2}\nonumber \\
 & = & \frac{2\beta}{\hbar (N_0L)^3}\int d^3x_sd^3x'_s\sum_{lmn}\frac{e^{-\beta (E_n+E_m)}}{\mathcal{Z}^2}\sum_{abcd}\int d^3x_ud^3x_u'\Lambda_{abcd}(\vec x_u-\vec x_u')\frac{\omega_{ln}}{(\omega+i\eta)^2-\omega_{ln}^2}\nonumber \\
 & {} & \langle m|\hat{T}_{ab}(\vec x_u)\hat{T}_{cd}(\vec x_u')|m\rangle\langle l|\hat{T}_{ij}(\vec x_s)|n\rangle\langle n|\hat{T}_{kl}(\vec x_s')|l\rangle\nonumber \\
 & = & \frac{2\beta}{\hbar (N_0L)^3}\int d^3x_sd^3x'_s\sum_{lmn}\frac{e^{-\beta (E_n+E_m)}}{\mathcal{Z}^2}\sum_{abcd}\int d^3x_ud^3x_u'\Lambda_{abcd}(\vec x_u-\vec x_u')\frac{\omega_{ln}}{(\omega+i\eta)^2-\omega_{ln}^2}\nonumber \\
 & {} & \langle m|\hat{T}_{ab}(\vec x_u)\sum_k|k\rangle \langle k|\hat{T}_{cd}(\vec x_u')|m\rangle\langle l|\hat{T}_{ij}(\vec x_s)|n\rangle\langle n|\hat{T}_{kl}(\vec x_s')|l\rangle\nonumber \\
 & = & \frac{2\beta}{\hbar (N_0L)^3}\int d^3x_sd^3x'_s\sum_{lmn}\frac{e^{-\beta (E_n+E_m)}}{\mathcal{Z}^2}\sum_{abcd}\int d^3x_ud^3x_u'\Lambda_{abcd}(\vec x_u-\vec x_u')\frac{\omega_{ln}}{(\omega+i\eta)^2-\omega_{ln}^2}\nonumber \\
 & {} & {\rm Tr}\,\left[\hat{T}_{cd}(\vec x_u')|m\rangle\langle l|\hat{T}_{ij}(\vec x_s)|n\rangle\langle n|\hat{T}_{kl}(\vec x_s')|l\rangle\langle m|\hat{T}_{ab}(\vec x_u)\right]\nonumber \\
 & = & \frac{4\beta}{\hbar (N_0L)^3}\int d^3x_sd^3x'_s\sum_{lmn}\frac{e^{-\beta (E_n+E_m)}}{\mathcal{Z}^2}\sum_{abcd}\int d^3x_ud^3x_u'\Lambda_{abcd}(\vec x_u-\vec x_u')\frac{\omega_{ln}}{(\omega+i\eta)^2-\omega_{ln}^2}\nonumber \\
 & {} & \langle m^{(s)}|\hat{T}_{cd}(\vec x_u')|m^{(s)}\rangle\langle l^{(s)}|\hat{T}_{ij}(\vec x_s)|n^{(s)}\rangle \langle n^{(s')}|\hat{T}_{kl}(\vec x_s')|n^{(s')}\rangle \langle m^{(s')}|\hat{T}_{ab}(\vec x_u)|m^{(s')}\rangle\nonumber \\
 & {} & 
\langle n^{(s)}|l^{(s)}\rangle\langle n^{(s')}|l^{(s')}\rangle\langle n^{(r)}|l^{(r)}\rangle\nonumber \\
 & = & 0
\end{eqnarray}
Again we insert the identity $\sum_k|k\rangle\langle k|$ in the fourth step of the above calculations. We use the same argument in term $J_7$ calculatins: suppose $s\neq s'$ and $l\neq n$ in the third step of Eq.(\ref{B16}). The operator for the $s$-th block stress tensor $\hat{T}_{ij}(\vec x_s)$ changes state from wavefunction $|l\rangle$ to $|n\rangle$, and the $s'$-th block stress tensor $\hat{T}_{kl}(\vec x_s')$ changes state from $|n\rangle$ to $|l\rangle$. However, since the $s$-th block stress tensor only acts on the $s$-th block wavefunction, and the $s'$-th block stress tensor only acts on the $s'$-th block wavefunction, it is impossible to change state $|l\rangle$ to state $|n\rangle$ and state $|n\rangle$ to state $|l\rangle$ simultaneously, except for the only possibility that $|l\rangle$ and $|n\rangle$ are the same. Finally, the factor $\omega_{ln}=\omega_l-\omega_n=0$ makes term $J_8$ to vanish. \\

Expansion for term $J_9$:
\begin{eqnarray}\label{B17}
J_9 & = & \frac{2}{\hbar (N_0L)^3}\int d^3x_sd^3x'_s\sum_{nl}\frac{e^{-\beta E_n}}{\mathcal{Z}}\langle l|\hat{T}_{ij}(\vec x_s)|n\rangle\langle n|\hat{T}_{kl}(\vec x_s')|l\rangle\frac{(\omega+i\eta)^2+\omega_{ln}^2}{[(\omega+i\eta)^2-\omega_{ln}^2]^2}\delta\omega_{ln}\nonumber \\
 & = & \frac{2}{\hbar^2 (N_0L)^3}\int d^3x_sd^3x'_s\sum_{nl}\frac{e^{-\beta E_n}}{\mathcal{Z}}\langle l|\hat{T}_{ij}(\vec x_s)|n\rangle\langle n|\hat{T}_{kl}(\vec x_s')|l\rangle\frac{(\omega+i\eta)^2+\omega_{ln}^2}{[(\omega+i\eta)^2-\omega_{ln}^2]^2}\left(\langle l|\hat{V}|l\rangle -\langle n|\hat{V}|n\rangle\right)\nonumber \\
 & = & \frac{2}{\hbar^2 (N_0L)^3}\int d^3x_sd^3x'_s\sum_{nl}\frac{e^{-\beta E_n}}{\mathcal{Z}}\sum_{abcd}\int d^3x_ud^3x_u'\Lambda_{abcd}(\vec x_u-\vec x_u')\frac{(\omega+i\eta)^2+\omega_{ln}^2}{[(\omega+i\eta)^2-\omega_{ln}^2]^2}\nonumber \\
 & {} & \langle l|\hat{T}_{ij}(\vec x_s)|n\rangle\langle n|\hat{T}_{kl}^{(\vec x_s')}|l\rangle\left(\langle l|\hat{T}_{ab}(\vec x_u)\hat{T}_{cd}(\vec x_u')|l\rangle -\langle n|\hat{T}_{ab}(\vec x_u)\hat{T}_{cd}(\vec x_u')|n\rangle\right)\nonumber \\
 & = & \frac{2}{\hbar^2 (N_0L)^3}\int d^3x_sd^3x'_s\sum_{nl}\frac{e^{-\beta E_n}}{\mathcal{Z}}\sum_{abcd}\int d^3x_ud^3x_u'\Lambda_{abcd}(\vec x_u-\vec x_u')\frac{(\omega+i\eta)^2+\omega_{ln}^2}{[(\omega+i\eta)^2-\omega_{ln}^2]^2}\nonumber \\
 & {} & \left[{\rm Tr}\,\left(\hat{T}_{cd}(\vec x_u')|l\rangle\langle l|\hat{T}_{ij}(\vec x_s)|n\rangle\langle n|\hat{T}_{kl}(\vec x_s')|l\rangle \langle l|\hat{T}_{ab}^{(u')}\right)-{\rm Tr}\,\left(\hat{T}_{cd}(\vec x_u')|n\rangle\langle l|\hat{T}_{ij}(\vec x_s)|n\rangle\langle n|\hat{T}_{kl}(\vec x_s')|l\rangle\langle n|\hat{T}_{ab}(\vec x_u)\right)\right]\nonumber \\
 & = & \frac{1}{\hbar^2 (N_0L)^3}\int d^3x_sd^3x'_s\sum_{nl}\frac{e^{-\beta E_n}}{\mathcal{Z}}\sum_{abcd}\int d^3x_ud^3x_u'\Lambda_{abcd}(\vec x_u-\vec x_u')\frac{(\omega+i\eta)^2+\omega_{ln}^2}{[(\omega+i\eta)^2-\omega_{ln}^2]^2}
\langle n^{(s)}|l^{(s)}\rangle\langle n^{(s')}|l^{(s')}\rangle\langle n^{(r)}|l^{(r)}\rangle\nonumber \\
 & {} & \bigg[\langle n^{(s)}|\hat{T}_{ab}(\vec x_u)|n^{(s)}\rangle\langle n^{(s)}|\hat{T}_{ij}(\vec x_s)|n^{(s)}\rangle\langle n^{(s')}|\hat{T}_{cd}(\vec x_u')|n^{(s')}\rangle\langle n^{(s')}|\hat{T}_{kl}(\vec x_s')|n^{(s')}\rangle\nonumber \\
 & {} & -\langle n^{(s)}|\hat{T}_{ab}(\vec x_u)|n^{(s)}\rangle\langle n^{(s)}|\hat{T}_{ij}(\vec x_s)|n^{(s)}\rangle\langle n^{(s')}|\hat{T}_{cd}(\vec x_u')|n^{(s')}\rangle\langle n^{(s')}|\hat{T}_{kl}(\vec x_s')|n^{(s')}\rangle\bigg]\nonumber \\
 & = & 0
\end{eqnarray}
To obtain the above vanishing result of $J_9$ we use the same argument in calculating $J_7$ and $J_8$.

Expansion for term $J_{10}$:
\begin{eqnarray}\label{B18.1}
J_{10} & = & \frac{2}{\hbar (N_0L)^3}\int d^3x_sd^3x'_s\sum_{nl}\frac{e^{-\beta E_n}}{\mathcal{Z}}\bigg[\left(\delta\langle l|\right)\hat{T}_{ij}(\vec x_s)|n\rangle\langle n|\hat{T}_{kl}(\vec x_s')|l\rangle+\langle l|\hat{T}_{ij}(\vec x_s)\left(\delta|n\rangle\right)\langle n|\hat{T}_{kl}(\vec x_s')|l\rangle\nonumber \\
 & {} & \qquad\qquad\qquad
+\langle l|\hat{T}_{ij}(\vec x_s)|n\rangle\left(\delta\langle n|\right)\hat{T}_{kl}(\vec x_s')|l\rangle+\langle l|\hat{T}_{ij}(\vec x_s)|n\rangle\langle n|\hat{T}_{kl}(\vec x_s')\left(\delta|l\rangle\right)\bigg]\frac{\omega_{ln}}{(\omega+i\eta)^2-\omega_{ln}^2}\nonumber \\
 & = & \frac{2}{\hbar (N_0L)^3}\int d^3x_sd^3x'_s\sum_{nl}\sum_{abcd}\int d^3x_ud^3x_u'\frac{e^{-\beta E_n}}{\mathcal{Z}}\Lambda_{abcd}(\vec x_u-\vec x_u')\frac{\omega_{ln}}{(\omega+i\eta)^2-\omega_{ln}^2}\sum_m\nonumber \\
 & {} & \bigg\{\frac{1}{E_l-E_m}\left(\langle l|\hat{T}_{ab}(\vec x_u)\hat{T}_{cd}(\vec x_u')|m\rangle\langle m|\hat{T}_{ij}(\vec x_s)|n\rangle\langle n|\hat{T}_{kl}(\vec x_s')|l\rangle+\langle l|\hat{T}_{ij}(\vec x_s)|n\rangle\langle n|\hat{T}_{kl}(\vec x_s')|m\rangle\langle m|\hat{T}_{ab}(\vec x_u)\hat{T}_{cd}(\vec x_u')|l\rangle\right)\nonumber \\
 & {} & +\frac{1}{E_n-E_m}\left(\langle l|\hat{T}_{ij}(\vec x_s)|m\rangle\langle m|\hat{T}_{ab}(\vec x_u)\hat{T}_{cd}(\vec x_u')|n\rangle\langle n|\hat{T}_{kl}(\vec x_s')|l\rangle+\langle l|\hat{T}_{ij}(\vec x_s)|n\rangle \langle n|\hat{T}_{ab}(\vec x_u)\hat{T}_{cd}(\vec x_u')|m\rangle\langle m|\hat{T}_{kl}(\vec x_s')|l\rangle\right)\bigg\}\nonumber \\
 & = & \frac{2}{\hbar (N_0L)^3}\int d^3x_sd^3x'_s\sum_{nl}\sum_{abcd}\int d^3x_ud^3x_u'\frac{e^{-\beta E_n}}{\mathcal{Z}}\Lambda_{abcd}(\vec x_u-\vec x_u')\frac{\omega_{ln}}{(\omega+i\eta)^2-\omega_{ln}^2}\sum_m\sum_k\nonumber \\
 & \bigg\{ & \frac{1}{E_l-E_m}\bigg(\langle l|\hat{T}_{ab}(\vec x_u)|k\rangle\langle k|\hat{T}_{cd}(\vec x_u')|m\rangle\langle m|\hat{T}_{ij}(\vec x_s)|n\rangle\langle n|\hat{T}_{kl}(\vec x_s')|l\rangle\nonumber \\
 & {} & \qquad\qquad+\langle l|\hat{T}_{ij}(\vec x_s)|n\rangle\langle n|\hat{T}_{kl}(\vec x_s')|m\rangle\langle m|\hat{T}_{ab}(\vec x_u)|k\rangle\langle k|\hat{T}_{cd}(\vec x_u')|l\rangle\bigg)\nonumber \\
 & + & \frac{1}{E_n-E_m}\bigg(\langle l|\hat{T}_{ij}(\vec x_s)|m\rangle\langle m|\hat{T}_{ab}(\vec x_u)|k\rangle\langle k|\hat{T}_{cd}(\vec x_u')|n\rangle\langle n|\hat{T}_{kl}(\vec x_s')|l\rangle\nonumber \\
 & {} & \qquad\qquad+\langle l|\hat{T}_{ij}(\vec x_s)|n\rangle \langle n|\hat{T}_{ab}(\vec x_u)|k\rangle\langle k|\hat{T}_{cd}(\vec x_u')|m\rangle\langle m|\hat{T}_{kl}(\vec x_s')|l\rangle\bigg)\bigg\}
\end{eqnarray}
where in the third step of the above calculation we insert the identity $\sum_k|k\rangle\langle k|$. Because of the coefficient $\Lambda_{abcd}(\vec x_u-\vec x_u')$, we have $u\neq u'$. In the final result of the above Eq.(\ref{B18.1}) we get 4 summations. Let us discuss the first  summation for example. 
\begin{eqnarray}\label{B18.2}
 & {} & \frac{2}{\hbar (N_0L)^3}\int d^3x_sd^3x'_s\sum_{mnkl}\sum_{abcd}\int d^3x_ud^3x_u'\frac{e^{-\beta E_n}}{\mathcal{Z}}\Lambda_{abcd}(\vec x_u-\vec x_u')\frac{\omega_{ln}}{(\omega+i\eta)^2-\omega_{ln}^2}\frac{1}{E_l-E_m}\nonumber \\
 & {} & \langle l|\hat{T}_{ab}(\vec x_u)|k\rangle\langle k|\hat{T}_{cd}(\vec x_u')|m\rangle\langle m|\hat{T}_{ij}(\vec x_s)|n\rangle\langle n|\hat{T}_{kl}(\vec x_s')|l\rangle
\end{eqnarray}
The pairing rule for the other three summations are the same. In the summation, Eq.(\ref{B18.2}), the matrix elements $\langle l|\hat{T}_{ab}(\vec x_u)|k\rangle$, $\langle k|\hat{T}_{cd}(\vec x_u')|m\rangle$ must be paired with $\langle m|\hat{T}_{ij}(\vec x_s)|n\rangle$, $\langle n|\hat{T}_{kl}(\vec x_s')|l\rangle$. We get two candidates of pairing: first, $\langle l|\hat{T}_{ab}(\vec x_u)|k\rangle$ is paired with $\langle m|\hat{T}_{ij}(\vec x_s)|n\rangle$, and $\langle k|\hat{T}_{cd}(\vec x_u')|m\rangle$ is paired with $\langle n|\hat{T}_{kl}(\vec x_s')|l\rangle$; second, $\langle l|\hat{T}_{ab}(\vec x_u)|k\rangle$ is paired with $\langle n|\hat{T}_{kl}(\vec x_s')|l\rangle$, and $\langle k|\hat{T}_{cd}(\vec x_u')|m\rangle$ is paired with $\langle m|\hat{T}_{ij}(\vec x_s)|n\rangle$.

In the first candidate, we have $\vec x_u, \vec x_s\in V^{(s)}, \vec x_u', \vec x_s'\in V^{(s')}$. According to the factor ${\omega_{ln}}/({(\omega+i\eta)^2-\omega_{ln}^2})$ which requires $l\neq n$, the matrix element $\langle n|\hat{T}_{kl}(\vec x_s')|l\rangle$ must be off-diagonal. Therefore, the matrix element $\langle k|\hat{T}_{cd}(\vec x_u')|m\rangle$ which is paired to it must be off-diagonal as well. The wavefunctions of $|k\rangle=\prod_{r=1}^{N_0^3}|k^{(r)}\rangle$ and $|m\rangle=\prod_{r=1}^{N_0^3}|m^{(r)}\rangle$ are required to be $|k^{(s')}\rangle=|l^{(s')}\rangle$, $|m^{(s')}\rangle=|n^{(s')}\rangle$, and $\prod_{r\neq s'}|k^{(r)}\rangle=\prod_{r\neq s'}|m^{(r)}\rangle$, $\prod_{r\neq s'}|n^{(r)}\rangle=\prod_{r\neq s'}|l^{(r)}\rangle$. Therefore in the first candidate case, the first term in Eq.(\ref{B18.2}) is simplified as 
\begin{eqnarray}\label{B18.3}
 & {} & \frac{2}{\hbar (N_0L)^3}\int d^3x_sd^3x'_s\sum_{mnkl}\sum_{abcd}\frac{e^{-\beta E_n}}{\mathcal{Z}}\int d^3x_ud^3x_u'\Lambda_{abcd}(\vec x_u-\vec x_u')\frac{\omega_{ln}}{(\omega+i\eta)^2-\omega_{ln}^2}\frac{1}{E_l-E_m}\nonumber \\
 & {} & \langle l|\hat{T}_{ab}(\vec x_u)|k\rangle\langle m|\hat{T}_{ij}(\vec x_s)|n\rangle\langle k|\hat{T}_{cd}(\vec x_u')|m\rangle\langle n|\hat{T}_{kl}(\vec x_s')|l\rangle\nonumber \\
 & = & \frac{2}{\hbar (N_0L)^3}\int d^3x_sd^3x'_s\sum_{abcd}\sum_{l^{(s')}m^{(s)}n^{(s)}n^{(s')}}\frac{e^{-\beta (E_n^{(s)}+E_n^{(s')})}}{\mathcal{Z}^{(s)}\mathcal{Z}^{(s')}}\int d^3x_ud^3x_u'\Lambda_{abcd}(\vec x_u-\vec x_u')\frac{\omega_{ln}^{(s')}}{(\omega+i\eta)^2-\omega_{ln}^{(s')2}}\nonumber \\
 & {} & 
\frac{\langle n^{(s)}|\hat{T}_{ab}(\vec x_u)|m^{(s)}\rangle\langle m^{(s)}|\hat{T}_{ij}(\vec x_s)|n^{(s)}\rangle\langle n^{(s')}|\hat{T}_{kl}(\vec x_s')|l^{(s')}\rangle\langle l^{(s')}|\hat{T}_{cd}(\vec x_u')|n^{(s')}\rangle}{(E_n^{(s)}-E_m^{(s)})+(E_l^{(s')}-E_n^{(s')})}\nonumber \\
 & = & \frac{2}{\hbar (N_0L)^3}\int d^3x_sd^3x'_s\sum_{abcd}\sum_{l^{(s')}m^{(s)}n^{(s)}n^{(s')}}\frac{e^{-\beta (E_n^{(s)}+E_n^{(s')})}}{\mathcal{Z}^{(s)}\mathcal{Z}^{(s')}}\int d^3x_ud^3x_u'\Lambda_{abcd}(\vec x_u-\vec x_u')\frac{\omega_{ln}^{(s')}}{(\omega+i\eta)^2-\omega_{ln}^{(s')2}}\nonumber \\
 & {} & 
\frac{\langle n^{(s)}|\hat{T}_{cd}(\vec x_u')|m^{(s)}\rangle\langle m^{(s)}|\hat{T}_{ij}(\vec x_s)|n^{(s)}\rangle\langle n^{(s')}|\hat{T}_{kl}(\vec x_s')|l^{(s')}\rangle\langle l^{(s')}|\hat{T}_{ab}(\vec x_u)|n^{(s')}\rangle}{(E_n^{(s)}-E_m^{(s)})+(E_l^{(s')}-E_n^{(s')})}
\end{eqnarray}
where in the last step, we exchange the indices $(ab)$ and $(cd)$ in the stress tensors $\hat{T}_{ab}(\vec x_u)$ and $\hat{T}_{cd}(\vec x_u')$. The exchange of indices is correct, because the coefficient $\Lambda_{abcd}(\vec x_u-\vec x_u')$ have the symmetry property: $\Lambda_{abcd}(\vec x_u-\vec x_u')=\Lambda_{cdab}(\vec x_u-\vec x_u')$.

Next we consider the second candidate, with $\vec x_u, \vec x_s'\in V^{(s')}, \vec x_u', \vec x_s\in V^{(s)}$. Actually the second candidate equals to first candidate, because with the exchange of indices $(ab)$, $(cd)$, $(s)$, $(s')$ and $(u)$, $(u')$, the coefficient $\Lambda_{abcd}(\vec x_u-\vec x_u')$ keeps invariant: $\Lambda_{abcd}(\vec x_u-\vec x_u')=\Lambda_{cdab}(\vec x_u'-\vec x_u)$, and the stress tensor operators commute: $\left[\hat{T}_{ab}(\vec x_u),\hat{T}_{cd}(\vec x_u')\right]_{u\neq u'}=0$.

Repeat the same process for the other three summations in Eq.(\ref{B18.3}), we procede our calculation as follows,
\begin{eqnarray}\label{B18}
J_{10} & = & \frac{4}{\hbar (N_0L)^3}\int d^3x_sd^3x'_s\sum_{abcd}\sum_{l^{(s')}m^{(s)}n^{(s)}n^{(s')}}\frac{e^{-\beta (E_n^{(s)}+E_n^{(s')})}}{\mathcal{Z}^{(s)}\mathcal{Z}^{(s')}}\int d^3x_ud^3x_u'\Lambda_{abcd}(\vec x_u-\vec x_u')\frac{\omega_{ln}^{(s')}}{(\omega+i\eta)^2-\omega_{ln}^{(s')2}}\nonumber \\
 & {} & 
\frac{\langle n^{(s)}|\hat{T}_{cd}(\vec x_u')|m^{(s)}\rangle\langle m^{(s)}|\hat{T}_{ij}(\vec x_s)|n^{(s)}\rangle\langle n^{(s')}|\hat{T}_{kl}(\vec x_s')|l^{(s')}\rangle\langle l^{(s')}|\hat{T}_{ab}(\vec x_u)|n^{(s')}\rangle}{(E_n^{(s)}-E_m^{(s)})+(E_l^{(s')}-E_n^{(s')})}\nonumber \\
 & + & \frac{4}{\hbar (N_0L)^3}\int d^3x_sd^3x'_s\sum_{abcd}\sum_{l^{(s)}m^{(s')}n^{(s)}n^{(s')}}\frac{e^{-\beta (E_n^{(s)}+E_n^{(s')})}}{\mathcal{Z}^{(s)}\mathcal{Z}^{(s')}}\int d^3x_ud^3x_u'\Lambda_{abcd}(\vec x_u-\vec x_u')\frac{\omega_{ln}^{(s)}}{(\omega+i\eta)^2-\omega_{ln}^{(s)2}}\nonumber \\
 & {} & 
\frac{\langle n^{(s)}|\hat{T}_{cd}(\vec x_u')|l^{(s)}\rangle\langle l^{(s)}|\hat{T}_{ij}(\vec x_s)|n^{(s)}\rangle\langle n^{(s')}|\hat{T}_{kl}(\vec x_s')|m^{(s')}\rangle\langle m^{(s')}|\hat{T}_{ab}(\vec x_u)|n^{(s')}\rangle}{(E_n^{(s')}-E_m^{(s')})+(E_l^{(s)}-E_n^{(s)})}\nonumber \\
 & + & \frac{4}{\hbar (N_0L)^3}\int d^3x_sd^3x'_s\sum_{abcd}\sum_{l^{(s')}m^{(s)}n^{(s)}n^{(s')}}\frac{e^{-\beta (E_n^{(s)}+E_n^{(s')})}}{\mathcal{Z}^{(s)}\mathcal{Z}^{(s')}}\int d^3x_ud^3x_u'\Lambda_{abcd}(\vec x_u-\vec x_u')\frac{\omega_{ln}^{(s')}}{(\omega+i\eta)^2-\omega_{ln}^{(s')2}}\nonumber \\
 & {} & 
\frac{\langle n^{(s)}|\hat{T}_{cd}(\vec x_u')|m^{(s)}\rangle\langle m^{(s)}|\hat{T}_{ij}(\vec x_s)|n^{(s)}\rangle\langle n^{(s')}|\hat{T}_{kl}(\vec x_s')|l^{(s')}\rangle\langle l^{(s')}|\hat{T}_{ab}(\vec x_u)|n^{(s')}\rangle}{(E_n^{(s)}-E_m^{(s)})-(E_l^{(s')}-E_n^{(s')})}\nonumber \\
 & + & \frac{4}{\hbar (N_0L)^3}\int d^3x_sd^3x'_s\sum_{abcd}\sum_{l^{(s)}m^{(s')}n^{(s)}n^{(s')}}\frac{e^{-\beta (E_n^{(s)}+E_n^{(s')})}}{\mathcal{Z}^{(s)}\mathcal{Z}^{(s')}}\int d^3x_ud^3x_u'\Lambda_{abcd}(\vec x_u-\vec x_u')\frac{\omega_{ln}^{(s)}}{(\omega+i\eta)^2-\omega_{ln}^{(s)2}}\nonumber \\
 & {} & 
\frac{\langle n^{(s)}|\hat{T}_{cd}(\vec x_u')|l^{(s)}\rangle\langle l^{(s)}|\hat{T}_{ij}(\vec x_s)|n^{(s)}\rangle\langle n^{(s')}|\hat{T}_{kl}(\vec x_s')|m^{(s')}\rangle\langle m^{(s')}|\hat{T}_{ab}(\vec x_u)|n^{(s')}\rangle}{(E_n^{(s')}-E_m^{(s')})-(E_l^{(s)}-E_n^{(s)})}
\end{eqnarray}

There are 4 terms above. The third and fourth terms are similar with the first and second terms. Therefore let us focus on the calculations of the first and second terms. To calculate the first term, we exchange the indices $l,m$, $s,s'$ and $u,u'$ in it. Because $\Lambda_{abcd}(\vec x_u-\vec x_u')=\Lambda_{cdab}(\vec x_u'-\vec x_u)$, the first term keeps invariant with the exchange of indices $l$, $m$, $s$, $s'$, $u$, $u'$:
\begin{eqnarray}\label{B19}
 & {} & \frac{4}{\hbar (N_0L)^3}\int d^3x_sd^3x'_s\sum_{abcd}\sum_{l^{(s')}m^{(s)}n^{(s)}n^{(s')}}\frac{e^{-\beta (E_n^{(s)}+E_n^{(s')})}}{\mathcal{Z}^{(s)}\mathcal{Z}^{(s')}}\int d^3x_ud^3x'_u\Lambda_{abcd}(\vec x_u-\vec x_u')\frac{\omega_{ln}^{(s')}}{(\omega+i\eta)^2-\omega_{ln}^{(s')2}}\nonumber \\
 & {} & 
\frac{\langle n^{(s)}|\hat{T}_{cd}(\vec x_u')|m^{(s)}\rangle\langle m^{(s)}|\hat{T}_{ij}(\vec x_s)|n^{(s)}\rangle\langle n^{(s')}|\hat{T}_{kl}(\vec x_s')|l^{(s')}\rangle\langle l^{(s')}|\hat{T}_{ab}(\vec x_u)|n^{(s')}\rangle}{(E_n^{(s)}-E_m^{(s)})+(E_l^{(s')}-E_n^{(s')})}\nonumber \\
 & = &
 \frac{4}{\hbar (N_0L)^3}\int d^3x_sd^3x'_s\sum_{abcd}\sum_{l^{(s')}m^{(s)}n^{(s)}n^{(s')}}\frac{e^{-\beta (E_n^{(s)}+E_n^{(s')})}}{\mathcal{Z}^{(s)}\mathcal{Z}^{(s')}}\int d^3x_ud^3x'_u\Lambda_{abcd}(\vec x_u-\vec x_u')\frac{\omega_{mn}^{(s)}}{(\omega+i\eta)^2-\omega_{mn}^{(s)2}}\nonumber \\
 & {} & 
\frac{\langle n^{(s')}|\hat{T}_{cd}(\vec x_u')|l^{(s')}\rangle\langle l^{(s')}|\hat{T}_{ij}^{(s')}|n^{(s')}\rangle\langle n^{(s)}|\hat{T}_{kl}^{(s)}|m^{(s)}\rangle\langle m^{(s)}|\hat{T}_{ab}(\vec x_u)|n^{(s)}\rangle}{(E_n^{(s')}-E_l^{(s')})+(E_m^{(s)}-E_n^{(s)})}\nonumber \\
 & = &
 \frac{4}{\hbar (N_0L)^3}\int d^3x_sd^3x'_s\sum_{abcd}\sum_{l^{(s')}m^{(s)}n^{(s)}n^{(s')}}\frac{e^{-\beta (E_n^{(s)}+E_n^{(s')})}}{\mathcal{Z}^{(s)}\mathcal{Z}^{(s')}}\int d^3x_ud^3x'_u\Lambda_{abcd}(\vec x_u-\vec x_u')\frac{\omega_{nm}^{(s)}}{(\omega+i\eta)^2-\omega_{nm}^{(s)2}}\nonumber \\
 & {} & 
\frac{\langle n^{(s')}|\hat{T}_{cd}(\vec x_u')|l^{(s')}\rangle\langle l^{(s')}|\hat{T}_{ij}^{(s')}|n^{(s')}\rangle\langle n^{(s)}|\hat{T}_{kl}^{(s)}|m^{(s)}\rangle\langle m^{(s)}|\hat{T}_{ab}(\vec x_u)|n^{(s)}\rangle}{(E_l^{(s')}-E_n^{(s')})+(E_n^{(s)}-E_m^{(s)})}
\end{eqnarray}
Use the identity 
\begin{eqnarray}\label{B23}
\left(\frac{1}{i\eta-\omega_{ln}^{(s')}}-\frac{1}{i\eta+\omega_{nm}^{(s)}}\right)\frac{1}{\omega_{ln}^{(s')}+\omega_{nm}^{(s)}}=\frac{1}{i\eta-\omega_{ln}^{(s')}}\frac{1}{i\eta+\omega_{nm}^{(s)}}
\end{eqnarray}
Finally the first term equals to 
\begin{eqnarray}\label{B24}
 & {} & \frac{1}{\hbar^2 (N_0L)^3}\int d^3x_sd^3x'_s\sum_{abcd}\sum_{l^{(s')}m^{(s)}n^{(s)}n^{(s')}}\frac{e^{-\beta (E_n^{(s)}+E_n^{(s')})}}{\mathcal{Z}^{(s)}\mathcal{Z}^{(s')}}\int d^3x_ud^3x'_u\Lambda_{abcd}(\vec x_u-\vec x_u')\nonumber \\
 & {} & 
{\langle n^{(s)}|\hat{T}_{cd}(\vec x_u')|m^{(s)}\rangle\langle m^{(s)}|\hat{T}_{ij}(\vec x_s)|n^{(s)}\rangle\langle n^{(s')}|\hat{T}_{kl}(\vec x_s')|l^{(s')}\rangle\langle l^{(s')}|\hat{T}_{ab}(\vec x_u)|n^{(s')}\rangle}\nonumber \\
 & {} & \left[\frac{1}{(i\eta-\omega_{ln}^{(s')})}\frac{1}{(i\eta+\omega_{nm}^{(s)})}+
\frac{1}{(i\eta+\omega_{ln}^{(s')})}\frac{1}{(i\eta-\omega_{nm}^{(s)})}
\right]
\end{eqnarray}
Similarly the second term is 
\begin{eqnarray}\label{B25}
 & {} & \frac{1}{\hbar^2 (N_0L)^3}\int d^3x_sd^3x'_s\sum_{abcd}\sum_{l^{(s')}m^{(s)}n^{(s)}n^{(s')}}\frac{e^{-\beta (E_n^{(s)}+E_n^{(s')})}}{\mathcal{Z}^{(s)}\mathcal{Z}^{(s')}}\int d^3x_ud^3x'_u\Lambda_{abcd}(\vec x_u-\vec x_u')\nonumber \\
 & {} & 
{\langle n^{(s)}|\hat{T}_{cd}(\vec x_u')|l^{(s)}\rangle\langle l^{(s)}|\hat{T}_{ij}(\vec x_s)|n^{(s)}\rangle\langle n^{(s')}|\hat{T}_{kl}(\vec x_s')|m^{(s')}\rangle\langle m^{(s')}|\hat{T}_{ab}(\vec x_u)|n^{(s')}\rangle}\nonumber \\
 & {} & \left[\frac{1}{(i\eta-\omega_{ln}^{(s)})}\frac{1}{(i\eta+\omega_{nm}^{(s')})}+
\frac{1}{(i\eta+\omega_{ln}^{(s)})}\frac{1}{(i\eta-\omega_{nm}^{(s')})}
\right]
\end{eqnarray}
The third term, 
\begin{eqnarray}\label{B26}
 & {} & -\frac{1}{\hbar^2 (N_0L)^3}\int d^3x_sd^3x'_s\sum_{abcd}\sum_{l^{(s')}m^{(s)}n^{(s)}n^{(s')}}\frac{e^{-\beta (E_n^{(s)}+E_n^{(s')})}}{\mathcal{Z}^{(s)}\mathcal{Z}^{(s')}}\int d^3x_ud^3x'_u\Lambda_{abcd}(\vec x_u-\vec x_u')\nonumber \\
 & {} & 
{\langle n^{(s)}|\hat{T}_{cd}(\vec x_u')|m^{(s)}\rangle\langle m^{(s)}|\hat{T}_{ij}(\vec x_s)|n^{(s)}\rangle\langle n^{(s')}|\hat{T}_{kl}(\vec x_s')|l^{(s')}\rangle\langle l^{(s')}|\hat{T}_{ab}(\vec x_u)|n^{(s')}\rangle}\nonumber \\
 & {} & \left[\frac{1}{(i\eta+\omega_{ln}^{(s')})}\frac{1}{(i\eta+\omega_{nm}^{(s)})}+
\frac{1}{(i\eta-\omega_{ln}^{(s')})}\frac{1}{(i\eta-\omega_{nm}^{(s)})}
\right]
\end{eqnarray}
The fourth term
\begin{eqnarray}\label{B27}
 & {} & -\frac{1}{\hbar^2 (N_0L)^3}\int d^3x_sd^3x'_s\sum_{abcd}\sum_{l^{(s')}m^{(s)}n^{(s)}n^{(s')}}\frac{e^{-\beta (E_n^{(s)}+E_n^{(s')})}}{\mathcal{Z}^{(s)}\mathcal{Z}^{(s')}}\int d^3x_ud^3x'_u\Lambda_{abcd}(\vec x_u-\vec x_u')\nonumber \\
 & {} & 
{\langle n^{(s)}|\hat{T}_{cd}(\vec x_u')|l^{(s)}\rangle\langle l^{(s)}|\hat{T}_{ij}(\vec x_s)|n^{(s)}\rangle\langle n^{(s')}|\hat{T}_{kl}(\vec x_s')|m^{(s')}\rangle\langle m^{(s')}|\hat{T}_{ab}(\vec x_u)|n^{(s')}\rangle}\nonumber \\
 & {} & \left[\frac{1}{(i\eta+\omega_{ln}^{(s)})}\frac{1}{(i\eta+\omega_{nm}^{(s')})}+
\frac{1}{(i\eta-\omega_{ln}^{(s)})}\frac{1}{(i\eta-\omega_{nm}^{(s')})}
\right]
\end{eqnarray}
Sum the above 4 terms up we finally obtain the result
\begin{eqnarray}\label{B28}
 & {} & J_{10}=\frac{1}{(N_0L)^3}\sum_{abcd}\int d^3x_sd^3x_s'd^3x_ud^3x'_u\Lambda_{abcd}(\vec x_u-\vec x_u')\chi^{\rm res}_{cdij}(\vec x_u', \vec x_s, \omega)\chi^{\rm res}_{klab}(\vec x_s', \vec x_u,\omega)
\end{eqnarray}

\newpage

\subsection{Super block susceptibility corrections due to the extra term in stress tensor $\int d^3x_sd^3x_s'\sum_{abcd}\frac{\delta\Lambda_{abcd}(\vec x_s-\vec x_s')}{\delta e_{ij}(\vec x)}\hat{T}_{ab}(\vec x_s)\hat{T}_{cd}(\vec x_s')$}
Finally let us consider the higher order corrections to super block non-elastic susceptibility due to super block stress tensor correction in Eq.(2.17). There are two kinds of extra expansions in super block susceptibility, (1) the product between $\hat{T}_{ij}(\vec x)$ and $\int d^3x_sd^3x_s'\sum_{abcd}\frac{\delta\Lambda_{abcd}(\vec x_s-\vec x_s')}{\delta e_{ij}(\vec x)}\hat{T}_{ab}(\vec x_s)\hat{T}_{cd}(\vec x_s')$, and (2) the super block susceptiblity expansion quadratic in the operator $\int d^3x_sd^3x_s'\sum_{abcd}\frac{\delta\Lambda_{abcd}(\vec x_s-\vec x_s')}{\delta e_{ij}(\vec x)}\hat{T}_{ab}(\vec x_s)\hat{T}_{cd}(\vec x_s')$. The susceptibility correction of the first kind is in odd orders of stress tensor matrix elements. Bare in mind that the stress tensors are a highly frustrated system, the expectation values of stress tensors are random quantities functional of spacial coordinates and it's quantum numbers $n,m$ in $\langle n|\hat{T}_{ij}(\vec x_s)|m\rangle$. Those terms in odd orders of stress tensor matrix elements vanish after integrating over spacial coordinates, because it does not come out in pairs of stress tensors matrix element products. For the super block susceptibility expansion of the second kind, we calculate it's contribution to the first, second part of relaxation susceptibility, and the resonance susceptibility separately.

According to the definition of stress tensor operator, the super block stress tensor is given as follows, 
\begin{eqnarray}\label{B40}
\hat{T}^{\rm sup}_{ij}(\vec x)=\frac{\delta \hat{H}^{\rm sup}}{\delta e_{ij}(\vec x)}=\hat{T}_{ij}(\vec x)+\int d^3x_sd^3x_s'\sum_{abcd}\frac{\delta\Lambda_{abcd}(\vec x_s-\vec x_s')}{\delta e_{ij}(\vec x)}\hat{T}_{ab}(\vec x_s)\hat{T}_{cd}(\vec x_s')
\end{eqnarray}
Which means it has a second contribution proportional to the quadratic in $\hat{T}_{ij}^{(s)}$ operators. Since the super block susceptibility by definition is given by 
\begin{eqnarray}\label{B41}
 \chi_{ijkl}^{\rm sup}( \omega) 
 & = & 
\frac{1}{(N_0L)^3}\frac{\beta}{1-i\omega\tau}\int d^3xd^3x'\bigg(\sum_{n^{\rm sup}m^{\rm sup}}\frac{e^{-\beta \left(E_n^{\rm sup}+E_m^{\rm sup}\right)}}{\mathcal{Z}^{\rm sup 2}}\langle n^{\rm sup}|\hat{T}_{ij}^{\rm sup}(\vec x)|n^{\rm sup}\rangle \langle m^{\rm sup}|\hat{T}_{kl}^{\rm sup}(\vec x')|m^{\rm sup}\rangle\nonumber \\
 & {} & \qquad\qquad\qquad\qquad\qquad\qquad\,\,\, -\sum_{n^{\rm sup}}\frac{e^{-\beta E_n^{\rm sup}}}{\mathcal{Z}^{\rm sup}}\langle n^{\rm sup}|\hat{T}_{ij}^{\rm sup}(\vec x)|n^{\rm sup}\rangle \langle n^{\rm sup}|\hat{T}_{kl}^{\rm sup}(\vec x')|n^{\rm sup}\rangle  \bigg)\nonumber \\
 & + & \frac{1}{(N_0L)^3}\frac{2}{\hbar}\sum_{n^{\rm sup}l^{\rm sup}}\frac{e^{-\beta E_n^{\rm sup}}}{\mathcal{Z}^{\rm sup}}\frac{\omega_l^{\rm sup}-\omega_n^{\rm sup}}{(\omega+i\eta)^2-(\omega_l^{\rm sup}-\omega_n^{\rm sup})^2}\int d^3xd^3x'\langle l^{\rm sup}|\hat{T}_{ij}^{\rm sup}(\vec x)|n^{\rm sup}\rangle \langle n^{\rm sup}|\hat{T}_{kl}^{\rm sup}(\vec x')|l^{\rm sup}\rangle \nonumber \\
\end{eqnarray}

We need to take the term which is quadratic in $T_{ij}^{(s)}$ into account as well. Note that we are only interested in the 1st and 2nd order in susceptibility $\chi$, we only take quadratic and quatic order in $T_{ij}$ into account. The contribution from this quadratic operator term results in the change of susceptibility as follows, 
\begin{eqnarray}\label{B42}
 & {} & 
\frac{1}{{(N_0L)^3}}\frac{\beta}{1-i\omega\tau}\int d^3xd^3x'd^3x_sd^3x_s'd^3x_ud^3x_u'\frac{\delta \Lambda_{abcd}(\vec x_s-\vec x_s')}{\delta e_{ij}(\vec x)}\frac{\delta \Lambda_{efgh}(\vec x_u-\vec x_u')}{\delta e_{kl}(\vec x')}\sum_{abcdefgh}\nonumber \\
 & {} & \bigg(\sum_{nm}P_nP_m\langle n|
\hat{T}_{ab}(\vec x_s)\hat{T}_{cd}(\vec x_s')
|n\rangle \langle m|
\hat{T}_{ef}(\vec x_u)\hat{T}_{gh}(\vec x_u')
|m\rangle-\sum_{n}P_n\langle n|\hat{T}_{ab}(\vec x_s)\hat{T}_{cd}(\vec x_s')|n\rangle \langle n|\hat{T}_{ef}(\vec x_u)\hat{T}_{gh}(\vec x_u')
|n\rangle  \bigg)\nonumber \\
 & + & \frac{1}{{(N_0L)^3}}\frac{2}{\hbar}\int d^3xd^3x'd^3x_sd^3x_s'd^3x_ud^3x_u'\sum_{abcdefgh}\frac{\delta \Lambda_{abcd}(\vec x_s-\vec x_s')}{\delta e_{ij}(\vec x)}\frac{\delta \Lambda_{efgh}(\vec x_u-\vec x_u')}{\delta e_{kl}(\vec x')}\nonumber \\
 & {} & \sum_{nl}P_n\langle l|
\hat{T}_{ab}(\vec x_s)\hat{T}_{cd}(\vec x_s')|n\rangle \langle n|
\hat{T}_{ef}(\vec x_u)\hat{T}_{gh}(\vec x_u')
|l\rangle\frac{\omega_l-\omega_n}{(\omega+i\eta)^2-(\omega_l-\omega_n)^2}
\end{eqnarray}

For the corrections to first part of non-elastic relaxation susceptibility $\chi_{ijkl}^{\rm sup\, rel(1)}$, we have
\begin{eqnarray}\label{B43}
 & {} & 
\frac{1}{(N_0L)^3}\frac{\beta}{1-i\omega\tau}\int d^3xd^3x'd^3x_sd^3x_s'd^3x_ud^3x_u'\sum_{abcdefgh}\sum_{nm}P_nP_m\frac{\delta \Lambda_{abcd}(\vec x_s-\vec x_s')}{\delta e_{ij}(\vec x)}\frac{\delta \Lambda_{efgh}(\vec x_u-\vec x_u')}{\delta e_{kl}(\vec x')}\nonumber \\
 & {} & \langle n|\hat{T}_{ab}(\vec x_s)\hat{T}_{cd}(\vec x_s')|n\rangle 
\langle m|\hat{T}_{ef}(\vec x_u)\hat{T}_{gh}(\vec x_u')|m\rangle \nonumber \\
 & = & 
\frac{2}{(N_0L)^3}\frac{\beta}{1-i\omega\tau}\int d^3xd^3x'd^3x_sd^3x_s'd^3x_ud^3x_u'\sum_{abcdefgh}\sum_{n^{(s)}n^{(s')}m^{(s)}m^{(s')}}\nonumber \\
 & {} & \frac{e^{-\beta \left(E_n^{(s)}+E_n^{(s')}+E_m^{(s)}+E_m^{(s')}\right)}}{\mathcal{Z}^{(s)2}\mathcal{Z}^{(s')2}}\frac{\delta \Lambda_{abcd}(\vec x_s-\vec x_s')}{\delta e_{ij}(\vec x)}\frac{\delta \Lambda_{efgh}(\vec x_u-\vec x_u')}{\delta e_{kl}(\vec x')}\nonumber \\
 & {} & \langle n^{(s)}|\hat{T}_{ab}(\vec x_s)|n^{(s)}\rangle \langle m^{(s)}|\hat{T}_{ef}(\vec x_u)|m^{(s)}\rangle \langle n^{(s')}|\hat{T}_{cd}(\vec x_s')|n^{(s')}\rangle 
\langle m^{(s')}|\hat{T}_{gh}(\vec x_u')|m^{(s')}\rangle \nonumber \\
 & = & 
\frac{2}{(N_0L)^3}\frac{\beta^{-1}}{1-i\omega\tau}\int d^3xd^3x'd^3x_sd^3x_s'd^3x_ud^3x_u'\sum_{abcdefgh}\nonumber \\
 & {} & \frac{\delta \Lambda_{abcd}(\vec x_s-\vec x_s')}{\delta e_{ij}(\vec x)}\frac{\delta \Lambda_{efgh}(\vec x_u-\vec x_u')}{\delta e_{kl}(\vec x')}\chi_{abef}^{\rm rel(1)}(\vec x_s,\vec x_u)\chi_{ghcd}^{\rm rel(1)}(\vec x_u',\vec x_s')
\end{eqnarray}
where in the above calculation, since $\int d^3x_ud^3x_u'\Lambda_{efgh}(\vec x_u-\vec x_u')$ is invariant under the indice exchange, $(ef), (gh)$ and $u, u'$, we can either put $\hat{T}_{ef}(\vec x_u)$ in $s$-th single block subspace, and put $\hat{T}_{gh}(\vec x_u')$ in $s'$-th single block subspace, or we can put $\hat{T}_{ef}(\vec x_u)$ in $s'$-th single block subspace, and put $\hat{T}_{gh}(\vec x_u')$ in $s$-th single block subspace. Then we can exchange indices $(ef), (gh)$ and $u, u'$ and prove that these two choices are equavalanet. This is where the factor of 2 comes from.

For the corrections to the second part of non-elastic relaxation susceptibility $\chi_{ijkl}^{\rm sup\, rel(2)}$, we have
\begin{eqnarray}\label{B44}
 & {} & -
\frac{1}{(N_0L)^3}\frac{\beta}{1-i\omega\tau}\sum_{n}\frac{e^{-\beta E_n}}{\mathcal{Z}}
\int d^3xd^3x'd^3x_sd^3x_s'd^3x_ud^3x_u'\sum_{abcdefgh}\nonumber \\
 & {} & \frac{\delta \Lambda_{abcd}(\vec x_s-\vec x_s')}{\delta e_{ij}(\vec x)}\frac{\delta \Lambda_{efgh}(\vec x_u-\vec x_u')}{\delta e_{kl}(\vec x')}\langle n|\hat{T}_{ab}(\vec x_s)\hat{T}_{cd}(\vec x_s')
|n\rangle \langle n|
\hat{T}_{ef}(\vec x_u)\hat{T}_{gh}(\vec x_u')
|n\rangle  \nonumber \\
 & = & -
\frac{2}{(N_0L)^3}\frac{\beta}{1-i\omega\tau}\int d^3xd^3x'd^3x_sd^3x_s'd^3x_ud^3x_u'\sum_{n^{(s)}n^{(s')}}\frac{e^{-\beta \left(E_n^{(s)}+E_n^{(s')}\right)}}{\mathcal{Z}^{(s)}\mathcal{Z}^{(s')}}
\sum_{abcdefgh}\nonumber \\
 & {} & \frac{\delta \Lambda_{abcd}(\vec x_s-\vec x_s')}{\delta e_{ij}(\vec x)}\frac{\delta \Lambda_{efgh}(\vec x_u-\vec x_u')}{\delta e_{kl}(\vec x')}
\langle n^{(s)}|\hat{T}_{ab}(\vec x_s)|n^{(s)}\rangle \langle n^{(s)}|\hat{T}_{ef}(\vec x_u)|n^{(s)}\rangle \langle n^{(s')}|\hat{T}_{cd}(\vec x_s')|n^{(s')}\rangle\langle n^{(s')}|\hat{T}_{gh}(\vec x_u')
|n^{(s')}\rangle  \nonumber \\
 & = & -
\frac{2}{(N_0L)^3}\frac{\beta^{-1}}{1-i\omega\tau}\int d^3xd^3x'd^3x_sd^3x_s'd^3x_ud^3x_u'
\sum_{abcdefgh}\frac{\delta \Lambda_{abcd}(\vec x_s-\vec x_s')}{\delta e_{ij}(\vec x)}\frac{\delta \Lambda_{efgh}(\vec x_u-\vec x_u')}{\delta e_{kl}(\vec x')}
\chi_{abef}^{\rm rel(2)}(\vec x_s,\vec x_u)\chi_{ghcd}^{\rm rel(2)} (\vec x_u',\vec x_s') \nonumber \\
\end{eqnarray}

For the corrections to the non-elastic resonance susceptibility $\chi_{ijkl}^{\rm sup\, res}(\omega)$, we have
\begin{eqnarray}\label{B45}
 & {} & 
\frac{1}{(N_0L)^3}\frac{1}{\hbar}\sum_{nl}(P_n-P_l)\int d^3xd^3x'd^3x_sd^3x_s'd^3x_ud^3x_u'\sum_{abcdefgh}\frac{\delta \Lambda_{abcd}(\vec x_s-\vec x_s')}{\delta e_{ij}(\vec x)}\frac{\delta \Lambda_{efgh}(\vec x_u-\vec x_u')}{\delta e_{kl}(\vec x')}
\nonumber \\
 & {} & \frac{\langle l|\hat{T}_{ab}(\vec x_s)\hat{T}_{cd}(\vec x_s')
|n\rangle \langle n|
\hat{T}_{ef}(\vec x_u)\hat{T}_{gh}(\vec x_u')|l\rangle }{\omega+i\eta+\omega_{nl}}\nonumber \\
 & = & 
\frac{1}{(N_0L)^3}\frac{1}{\hbar}\sum_{nl}(P_n-P_l)\int d^3xd^3x'd^3x_sd^3x_s'd^3x_ud^3x_u'\sum_{abcdefgh}\frac{\delta \Lambda_{abcd}(\vec x_s-\vec x_s')}{\delta e_{ij}(\vec x)}\frac{\delta \Lambda_{efgh}(\vec x_u-\vec x_u')}{\delta e_{kl}(\vec x')}\nonumber \\
 & {} & \langle l|\hat{T}_{ab}(\vec x_s)\hat{T}_{cd}(\vec x_s')
|n\rangle \langle n|
\hat{T}_{ef}(\vec x_u)\hat{T}_{gh}(\vec x_u')|l\rangle
\left(\frac{1 }{\omega+\omega_{nl}}-i\pi\delta(\omega+\omega_{nl})\right)
\nonumber \\
 & = & 
\frac{2}{(N_0L)^3}\frac{1}{\hbar}\sum_{nl}(P_n-P_l)\int d^3x_sd^3x_s'd^3x_ud^3x_u'\sum_{abcdefgh}\frac{\delta \Lambda_{abcd}(\vec x_s-\vec x_s')}{\delta e_{ij}(\vec x)}\frac{\delta \Lambda_{efgh}(\vec x_u-\vec x_u')}{\delta e_{kl}(\vec x')}\nonumber \\
 & {} & 
\langle l^{(s)}|\hat{T}_{ab}(\vec x_s)|n^{(s)}\rangle \langle n^{(s)}|\hat{T}_{ef}(\vec x_u)|l^{(s)}\rangle
\langle l^{(s')}|\hat{T}_{cd}(\vec x_s')|n^{(s')}\rangle \langle n^{(s')}|\hat{T}_{gh}(\vec x_u')|l^{(s')}\rangle
\left(\frac{1 }{\omega+\omega_{nl}}-i\pi\delta(\omega+\omega_{nl})\right)
\nonumber \\
 & = & 
\frac{2}{(N_0L)^3}\int d^3xd^3x'd^3x_sd^3x_s'd^3x_ud^3x_u'\sum_{abcdefgh}\frac{\delta \Lambda_{abcd}(\vec x_s-\vec x_s')}{\delta e_{ij}(\vec x)}\frac{\delta \Lambda_{efgh}(\vec x_u-\vec x_u')}{\delta e_{kl}(\vec x')}\nonumber \\
 & {} & 
\sum_{n^{(s)}n^{(s')}l^{(s)}l^{(s')}}(P_n-P_l)\left(\frac{1 }{\hbar\omega+\hbar\omega_{nl}^{(s)}+\hbar\omega_{nl}^{(s')}}-i\pi\delta(\hbar\omega+\hbar\omega_{nl}^{(s)}+\hbar\omega_{nl}^{(s')})\right)\nonumber \\
 & {} & 
\int \langle n^{(s)}|\hat{T}_{ef}(\vec x_u)|l^{(s)}\rangle \langle l^{(s)}|\hat{T}_{ab}(\vec x_s)|n^{(s)}\rangle\delta(E_l^{(s)}-E_n^{(s)}-\hbar\omega_s)d(\hbar\omega_s)\nonumber \\
 & {} & 
\int \langle n^{(s')}|\hat{T}_{gh}(\vec x_u')|l^{(s')}\rangle \langle l^{(s')}|\hat{T}_{cd}(\vec x_s')|n^{(s')}\rangle \delta(E_l^{(s')}-E_n^{(s')}-\hbar\omega_s')d(\hbar\omega_s')
\nonumber \\
 & = & 
\frac{2L^3}{N_0^3\pi^2}\int d^3xd^3x'd^3x_sd^3x_s'd^3x_ud^3x_u'\sum_{abcdefgh}\frac{\delta \Lambda_{abcd}(\vec x_s-\vec x_s')}{\delta e_{ij}(\vec x)}\frac{\delta \Lambda_{efgh}(\vec x_u-\vec x_u')}{\delta e_{kl}(\vec x')}\nonumber \\
 & {} & 
(1-e^{-\beta\hbar(\omega_s+\omega_s')})\left(\frac{1 }{\hbar\omega-\hbar\omega_{s}-\hbar\omega_{s'}}-i\pi\delta(\hbar\omega-\hbar\omega_s-\hbar\omega_{s'})\right)\nonumber \\
 & {} & 
\int\,{\rm Im}\,\chi_{abef}^{\rm res-absorb}(\vec x_s,\vec x_u;\omega_s)d(\hbar\omega_s)
\int\,{\rm Im}\,\chi_{ghcd}^{\rm res-absorb}(\vec x_u',\vec x_s';\omega_s')d(\hbar\omega_s')
\nonumber \\
 & = & 
-\frac{2}{(N_0L)^3\pi^2}\int d^3xd^3x'd^3x_sd^3x_s'd^3x_ud^3x_u'\sum_{abcdefgh}\frac{\delta \Lambda_{abcd}(\vec x_s-\vec x_s')}{\delta e_{ij}(\vec x)}\frac{\delta \Lambda_{efgh}(\vec x_u-\vec x_u')}{\delta e_{kl}(\vec x')}\nonumber \\
 & {} & 
\int(1-e^{-\beta\hbar(\omega_s+\omega_s')})\frac{{\rm Im}\,\chi_{abef}^{\rm res}(\vec x_s,\vec x_u;\omega_s)
{\rm Im}\,\chi_{cdgh}^{\rm res}(\vec x_u',\vec x_s';\omega_s') }{(1-e^{-\beta\hbar\omega_s})(1-e^{-\beta\hbar\omega_s'})(\hbar\omega_{s}+\hbar\omega_{s'}-\hbar\omega)}d(\hbar\omega_s)d(\hbar\omega_s')
\nonumber \\
 & {} & 
-i\frac{2}{(N_0L)^3\pi^2}\int d^3x_sd^3x_s'd^3x_ud^3x_u'\sum_{abcdefgh}\frac{\delta \Lambda_{abcd}(\vec x_s-\vec x_s')}{\delta e_{ij}(\vec x)}\frac{\delta \Lambda_{efgh}(\vec x_u-\vec x_u')}{\delta e_{kl}(\vec x')}\nonumber \\
 & {} & 
(1-e^{-\beta\hbar\omega})\left(\pi\int\,\frac{{\rm Im}\,\chi_{abef}^{\rm res}(\vec x_s,\vec x_u;\omega_s){\rm Im}\,\chi_{ghcd}^{\rm res}(\vec x_u',\vec x_s';\omega-\omega_s)}{(1-e^{-\beta\hbar\omega_s})(1-e^{-\beta\hbar\omega_s'})}d(\hbar\omega_s)
\right)
\end{eqnarray}
The super block susceptibility extra expansion terms Eq.(\ref{B43}, \ref{B44}, \ref{B45}) are the terms in the $\mathcal{K}_3$ and $\mathcal{K}_4$ of susceptibility renormalization equation, Eq.(\ref{23}).

\section{The Positivity of the Real Part of Non-Elastic Resonance Susceptibility in Zero-Frequency Limit: $\lim_{\omega\to 0}{\rm Re\,}\chi^{\rm res}_{ijkl}(\omega)$}
In this section we want to focus on the positivity of the real part of non-elastic resonance susceptibility, $\chi_{ijkl}^{\rm res}(\omega)=V^{-1}\int d^3xd^3x'\chi_{ijkl}^{\rm res}(\vec x, \vec x'; \omega)$. At the first glance, with the indices $k=i, l=j$, $\lim_{\omega\to 0^+}{\rm Re\,}\chi_{ijkl}^{\rm res}(\omega)$ (see the third line of Eq.(\ref{7})) is negative; on the other hand, if the frequency $\omega$ is large enough, then ${\rm Re\,}\chi_{ijij}^{\rm res}(\omega)$ could be positive. Let us consider a simple example of a two-level-system glass Hamiltonian, with the energy eigenvalues $E_1=0$ and $E_2>0$. The conclusion is similar for an arbitrary multiple-level-system glass Hamiltonian. We use $\int d^3x \,\hat{T}_{ij}(\vec x)=\hat{T}_{ij}$ to denote the uniform stress tensor operator for such a block of glass. The non-elastic resonance susceptibility $\chi_{ijij}^{\rm res}(\omega)$ is therefore given by


\begin{eqnarray}\label{C1}
\chi_{ijij}^{\rm res}( \omega) & = & \frac{2}{\hbar V}(P_1-P_2)|\langle 1|\hat{T}_{ij}|2\rangle|^2\frac{\omega_2}{(\omega+i\eta)^2-\omega_2^2}
\end{eqnarray}
where $\omega_{2}=E_2/\hbar$, and $|1\rangle$, $|2\rangle$ denote the ground state and the excitation state of the glass two-level-system Hamiltonian. $P_1=e^{-\beta E_1}/\mathcal{Z}$ and $P_2=e^{-\beta E_2}/\mathcal{Z}$ are the probability functions of the ground state and the excitation energy level. In the above result Eq.(\ref{C1}), in the limit of $\omega\to0^+$, the real part of two-level-system resonance susceptibility is negative. On the other hand, if $\omega>(\omega_2^2+\eta^2)^{1/2}$, then the real part of two-level-system resonance susceptibility ${\rm Re}\,\chi_{ijij}^{\rm res}(\omega)$ is positive. In fact, the two-level-system resonance susceptibility suggests that there is a critical frequency $\omega_c=(\omega_2^2+\eta^2)^{1/2}$. For $\omega>\omega_c$, ${\rm Re}\,\chi_{ijij}^{\rm res}(\omega)>0$, while for $\omega<\omega_c$, ${\rm Re}\,\chi_{ijij}^{\rm res}(\omega)<0$.

According to the qualitative argument in section 3(A), such positive-negative transition in ${\rm Re}\,\chi_{ijij}^{\rm res}(\omega)$ will lead to unreasonable result of non-elastic susceptibility at macroscopic length scales. For example, for $\omega>\omega_c$, ${\rm Re\,}\chi_{l,t}^{\rm res}(\omega)$ and ${\rm Im\,}\chi_{l,t}^{\rm res}(\omega)$ are renormalization irrelevant, which means at experimental length scales ${\rm Re\,}\chi_{l,t}^{\rm res}(\omega)\propto 1/\ln(R/L_1)$ is a logarithmically small, positive quantity, and ${\rm Im\,}\chi_{l,t}^{\rm res}(\omega)$ is a logarithmically small, negative quantity; for $\omega<\omega_c$, however, ${\rm Re\,}\chi_{l,t}^{\rm res}(\omega)$ and ${\rm Im\,}\chi_{l,t}^{\rm res}(\omega)$ are renormalization relevant, which means ${\rm Re\,}\chi_{l,t}^{\rm res}(\omega), {\rm Im\,}\chi_{l,t}^{\rm res}(\omega) \propto \ln(R/L_1)$ are logarithmically huge, negative quantities. At least to the author's knowledge, such unusual steep transition of glass mechanical response function was never reported\cite{Lasjaunias1975, Hunklinger1981, Hunklinger1982, Schickfus1990, Pohl2002, Cahill1996}. This is not the only problem. The more dangerous problem inferred from the above positive-negative transition of ${\rm Re}\,\chi_{ijij}^{\rm res}(\omega)$, is that any amorphous material must be mechanically unstable against arbitrary infinitesimal static perturbation, because $\chi_{ijkl}^{\rm res}(\omega=0)$ is a negative, logarithmically large quantity (see the non-elastic part of free energy $F^{\rm non}(e)=F^{\rm non}_0+\frac{1}{2}\int dxdx'\chi(x-x') e(x, t)e(x', t)$).

To solve the above disaster, it is necessary to consider higher order corrections of $\chi_{ijkl}^{\rm res}(\omega)$ due to the coupling between intrinsic phonon strain field and non-elastic stress tensor $\int d^3x\,e_{ij}(\vec x)\hat{T}_{ij}(\vec x)$. Let us expand $\chi_{ijkl}^{\rm res}(\omega)$ in orders of stress-strain coupling $\int d^3x\,e_{ij}(\vec x)\hat{T}_{ij}(\vec x)$. The higher order expansions can be written in terms of the product between $\chi_{ijkl}^{\rm res}(\omega)$ and phonon-phonon correlation function $\chi^{\rm ph}_{ijkl}(\omega)$. According to Dyson equation, the full non-elastic susceptibility which contains higher order corrections of stress-strain coupling is given as follows, 
\begin{eqnarray}\label{C5}
\left(\chi_{ijkl}^{\rm res}\right)^{-1}(\omega)=\left(\chi_{ijkl}^{\rm res}\right)^{-1}_0( \omega)-\chi_{ijkl}^{\rm ph}(\omega)
\end{eqnarray}
where $\chi_{ijkl}^{\rm res}(\omega)=V^{-1}\int d^3xd^3x'\,\chi_{ijkl}^{\rm res}(\vec x, \vec x'; \omega)$ and $\chi_{ijkl}^{\rm ph}(\omega)=V^{-1}\int d^3xd^3x'\, \chi_{ijkl}^{\rm ph}(\vec x, \vec x'; \omega)$ are spacce-averaged non-elastic susceptibility and phonon-phonon correlation function, respectively. The phonon-phonon correlation function $\chi_{ijkl}^{\rm ph}(\vec x, \vec x'; \omega)$ is defined as follows:
\begin{eqnarray}\label{C6}
\chi_{ijkl}^{\rm ph}(\vec x, \vec x'; \omega, \omega') & = & (2\pi)\delta(\omega+\omega')\chi_{ijkl}^{\rm ph}(\vec x, \vec x'; \omega)\nonumber \\
\chi_{ijkl}^{\rm ph}(\vec x, \vec x'; \omega, \omega') & = & -\frac{i}{\hbar}\int dtdt' \, e^{i\omega t+i\omega't'}\Theta(t-t')\langle n|\left[e_{ij}(\vec x, t),e_{kl}(\vec x', t')\right]|n\rangle
\end{eqnarray}
In the above definition, we use the Heisenberg picture with the Hamiltonian $\hat{H}_{\rm ph}=\sum_{k\alpha}\hbar\omega_{k\alpha}(\hat{a}_{k\alpha}^{\dag}\hat{a}_{k\alpha}+\frac{1}{2})$. $e_{ij}(\vec x, t)=e^{i\hat{H}_{\rm ph}t/\hbar}e_{ij}(\vec x)e^{-i\hat{H}_{\rm ph}t/\hbar}$ is the Heisenberg picture phonon strain field operator. A direct calculation of phonon-phonon correlation function gives the result as follows, 
\begin{eqnarray}\label{C7}
\chi_{ijkl}^{\rm ph}(\vec x, \vec x'; \omega)
 & = &  \int \frac{d^3k}{(2\pi)^3}\, e^{i\vec k\cdot (\vec x-\vec x')}\chi_{ijkl}^{\rm ph}(\vec k, \omega)\nonumber \\
\chi_{ijkl}^{\rm ph}(\vec k, \omega) & = & \frac{1}{\rho c_{l}^2}\kappa_i\kappa_j\kappa_k\kappa_l\frac{c_l^2k^2}{(\omega+i\eta)^2-c_l^2k^2}\nonumber \\
 & + & \frac{1}{4\rho c_{t}^2}\left(\delta_{ik}\kappa_j\kappa_l+\delta_{il}\kappa_j\kappa_k+\delta_{jk}\kappa_i\kappa_l+\delta_{jl}\kappa_i\kappa_k-4\kappa_i\kappa_j\kappa_k\kappa_l\right)\frac{c_t^2k^2}{(\omega+i\eta)^2-c_t^2k^2}
\end{eqnarray}
where $\vec \kappa$ is the unit vector of momentum $\vec k$: $\vec k=k\vec\kappa$. In the limit of $\omega\to0$, the real part of intrinsic phonon-phonon correlation function $\lim_{\omega\to 0^+}{\rm Re\,}\chi^{\rm ph}_{ijkl}(\omega)$ is negative. On the other hand, the imaginary part of intrinsic phonon-phonon correlation function, ${\rm Im}\,\chi^{\rm ph}_{ijkl}(\omega)$ is always negative for arbitrary positive $\omega$. This will give rise to a self-energy correction to the non-elastic resonance susceptibility in Eq.(\ref{C5}). Both of the real and the imaginary parts of self-energy correction are positive. Let us illustrate this point of view by considering the two-level-system resonance susceptibility. The Dyson equation of two-level-system resonance susceptibility is given by  
\begin{eqnarray}\label{C8}
\left(\chi_{ijij}^{\rm res}\right)^{-1}(\omega)
 & = & \frac{\hbar V}{2|\langle 1|\hat{T}_{ij}|2\rangle|^{2}}\left(\frac{1}{P_1-P_2}\right)\frac{\omega^2-\omega_2^2}{2\omega_2}+\left|{\rm Re\,}\chi^{\rm ph}_{ijkl}(\omega)\right|+i\left|{\rm Im\,}\chi^{\rm ph}_{ijkl}(\omega)\right|
\end{eqnarray}
The two-level-system resonance susceptibility therefore receives a self-energy correction with both of the real and imaginary parts positive. The above result has two important implications: (1) back to the non-elastic resonance susceptibility in the third line of Eq.(\ref{7}), the term $(\omega+i\eta)^2=\omega^2-\eta^2+2i\omega\eta$ gives an imaginary, positive correction to the denominator of non-elastic resonance susceptibility. This positive imaginary term is self-consistent with the positive, imaginary correction of ``$-{\rm Im}\,\chi_{ijkl}^{\rm ph}(\omega)$" in the Dyson equation of Eq.(\ref{C5}). In ultrasonic experiments with the input phonon frequency $\omega\sim 10^7$rad/s, usually we assume that the phonon frequency is much greater than the real part of self-energy correction. Therefore we neglect the real part of self-energy correction, while keep the imaginary part only. We use the symbol $\eta$ to represent the imaginary part of self-energy correction. In other words, we can use Dyson equation (\ref{C5}) to derive $\eta$, and it is proportional to $-{\rm Im}\,\chi_{ijkl}^{\rm ph}(\omega)$; (2) in the static limit when $\omega\to 0^+$, the real part of self-energy correction is no longer negligible compared to $\omega$. The denominator of non-elastic resonance susceptibility receives a positive self-energy correction, ``$-{\rm Re}\,\chi_{ijkl}^{\rm ph}(\omega)$".

As we have mentioned above, usually we neglect the real part of self-energy correction, $-{\rm Re}\,\chi_{ijkl}^{\rm ph}(\omega)$, because we are interested in the ultrasonic experiments with the phonon frequency $\omega\sim 10^7$rad/s, so that $-{\rm Re}\,\chi_{ijkl}^{\rm ph}(\omega)$ is likely to be negligible. However, in the $\omega\to 0^+$ static limit, the approximation to drop the real part of self-energy correction $-{\rm Re}\,\chi_{ijkl}^{\rm ph}(\omega)$ in the denominator of non-elastic resonance susceptibility is no longer correct. One should also be very careful with the notation ``$(\omega+i\eta)$" in the denominator of non-elastic resonance susceptibility, because it is based on the assumption that $-{\rm Re}\,\chi_{ijkl}^{\rm ph}(\omega)$ is negligible. In conclusion, we have to use Dyson equation to re-consider the denominator of non-elastic resonance susceptibility.


Although we are not able to calculate the self-energy correction for the resonance susceptibility of an arbitrary multiple-level-systen, we would like to argue that in the static limit, the full resonance susceptibility is positive due to the positive self-energy correction $-{\rm Re\,}\chi_{ijkl}^{\rm ph}(\omega)$. To further strengthen our point of view, we give a further discussion on the positivity of non-elastic resonance susceptibility in zero-frequency limit. Let us apply Fourier Transformation to convert the non-elastic susceptibility from real space to momentum space as follows, (where we assume that for a large enough block of glass, the non-elastic susceptibility has translational invariance $\chi_{ijkl}(\vec x, \vec x'; \omega)=\chi_{ijkl}(\vec x-\vec x'; \omega)$)
\begin{eqnarray}\label{D1}
\int d^3xd^3x' \, e^{i\vec k\cdot \vec x+i\vec k'\cdot \vec x'}\chi_{ijkl}(\vec x-\vec x'; \omega)=(2\pi)^3\delta(\vec k+\vec k')\chi_{ijkl}(\vec k; \omega)
\end{eqnarray}
At macroscopic experimental length scales, the space-averaged non-elastic susceptibility $\chi_{ijkl}(\omega)=V^{-1}\int d^3xd^3x'\,\chi_{ijkl}(\vec x-\vec x'; \omega)$ can be written as the function of momentum $\vec k$, with $\vec k\to0$:
 \begin{eqnarray}\label{D2}
\chi_{ijkl}(\omega)=\frac{1}{V}\int d^3xd^3x'\chi_{ijkl}(\vec x-\vec x'; \omega)=\chi_{ijkl}(\vec k\to0, \omega)
\end{eqnarray}
In the denominator of non-elastic resonance susceptibility in Eq.(\ref{7}), the energy spacing $E_{n+1}(L)-E_{n}(L)$ decreases as the length scale increases. Because more and more single block Hamiltonians join in the super block Hamiltonian, the level spacing decreases exponentially as the function of length scale $L$. Although we have no idea whether such a set of energy eigenvalues $E_n(L)$ is gapless or not, we think at the late stages of real space renormalization procedures, the energy spacing $E_{n+1}(\vec k\to 0)-E_n(\vec k\to 0)$ is so small, that the real part of self-energy correction $-{\rm Re\,}\chi_{ijkl}^{\rm res}(\omega)$ is large enough to prevent ${\rm Re\,}\chi_{ijkl}^{\rm res}(\vec k\to0; \omega)$ from being negative. It is at this point that we believe for arbitrary frequency $\omega$, ${\rm Re\,}\chi_{ijkl}^{\rm res}(\omega)$ is positive and logarithmically decreases as the function of length scale $L$.


Dyson equation is also useful in discussing phonon-phonon correlation function. The full phonon-phonon correlation function is given by Dyson equation as follows, 
\begin{eqnarray}\label{C9}
\left(\chi_{ijkl}^{\rm ph}\right)^{-1}(\omega)=\left(\chi_{ijkl}^{\rm ph}\right)^{-1}_0( \omega)-\chi_{ijkl}(\omega)
\end{eqnarray}
where $\chi_{ijkl}(\omega)$ is the non-elastic susceptibility. The imaginary part of non-elastic susceptibility is always negative for arbitrary $\omega>0$: 
\begin{eqnarray}\label{C10}
{\rm Im}\, \chi_{ijkl}(\omega) & = & {\rm Im}\, \chi_{ijkl}^{\rm res}(\omega)+{\rm Im}\, \chi^{\rm rel}_{ijkl}(\omega)\nonumber \\
{\rm Im}\,\chi_{ijkl}^{\rm res}(\omega) & = & 
-\frac{\pi}{V\hbar}\sum_{nm}(P_n-P_m)\langle n|\hat{T}_{ij}|m\rangle\langle m|\hat{T}_{kl}|n\rangle\delta(\omega+\omega_{n}-\omega_m)\nonumber \\
{\rm Im}\,\chi_{ijkl}^{\rm rel}(\omega) & = & \frac{\beta}{V}\frac{\omega\tau}{{1+\omega^2\tau^2}}\bigg(\sum_{nm}P_nP_m\langle n|\hat{T}_{ij}|n\rangle\langle m|\hat{T}_{kl}|m\rangle-\sum_n P_n\langle n|\hat{T}_{ij}|n\rangle \langle n|\hat{T}_{kl}|n\rangle\bigg)
\end{eqnarray}
With the assumption that non-elastic susceptibility is invariant under SO(3) rotational group, $\chi_{ijkl}(\omega)=\left(\chi_{l}(\omega)-2\chi_t(\omega)\right)\delta_{ij}\delta_{kl}+\chi_t(\omega)\left(\delta_{ik}\delta_{jl}+\delta_{il}\delta_{jk}\right)$. The longitudinal and transverse phonon-phonon correlation functions receive the self-energy corrections from the non-elastic susceptibility, 
\begin{eqnarray}\label{C11}
\frac{\omega'^2-\omega_k^2}{\omega_k^2} & = & 
\frac{\omega^2-\omega_k^2}{\omega_k^2}-\frac{{\rm Re}\,\chi_{l,t}(\omega)}{\rho c_{l,t}^2}-i\frac{\,{\rm Im}\,\chi_{l,t}(\omega)}{\rho c_{l,t}^2}
\end{eqnarray}
The poles of the phonon-phonon correlation functions are therefore shifted away from their original positions. According to our renormalization argument, both of the real and imaginary parts of non-elastic susceptibitilities are much smaller than $\rho c^2$ for at least one order of magnitude: $\frac{{\rm Re}\,\chi_{l,t}(\omega)}{\rho c_{l,t}^2}\sim 1/\ln(R/L_1)$, $\frac{\,{\rm Im}\,\chi_{l,t}(\omega)}{\rho c_{l,t}^2}\sim 1/\ln^2(R/L_1)$ (see Eq.(\ref{24.3})). We use the approximation $1+\frac{\chi_{l,t}(\omega)}{\rho c_{l,t}^2}\approx (1+\frac{\chi_{l,t}(\omega)}{2\rho c_{l,t}^2})^2$ to calculate the new pole of full phonon-phonon correlation function:
\begin{eqnarray}\label{C12}
\omega=\left(1+\frac{\chi_{l,t}(\omega)}{2\rho c_{l,t}^2}\right)\omega_k\quad \Rightarrow \quad \frac{\omega-\omega_k}{\omega_k}=\frac{{\rm Re}\,\chi_{l,t}(\omega)}{2\rho c_{l,t}^2}+i\frac{{\rm Im}\,\chi_{l,t}(\omega)}{2\rho c_{l,t}^2}
\end{eqnarray}
This is the result which appears in Eq.(\ref{11}). The real part of non-elastic susceptibility corresponds to the sound velocity shift $\Delta c_{l,t}$, while the imaginary part corresponds to the mean free path of phonon propagation. The propagating phonon wave is given by 
\begin{eqnarray}\label{C13}
\vec u(\vec x, t)=\vec A e^{ik\cdot x-i\omega t}=\vec A e^{ik\cdot x-i\omega_k(1+{\rm Re}\chi_{l,t}/2\rho c_{l,t}^2) t}e^{(\omega_k\,{\rm Im}\,\chi_{l,t}(\omega)/2\rho c_{l,t}^2)t}
\end{eqnarray}
Since the imaginary part of non-elastic susceptibility is always negative, the phonon intensity exponentially decays, with the mean free path $l=(-k \,{\rm Im}\, \chi_{l,t}(\omega)/\rho c_{l,t}^2)^{-1}$. Finally, the sound velocity shift $\Delta c_{l,t}$ is given by the mean free path via Kramers-Kronig relation as follows, 
\begin{eqnarray}\label{C14}
\Delta c_{l,t}=\frac{1}{\pi}\mathcal{P}\int_{0}^{\infty}\frac{c_{l,t}^2l^{-1}(\Omega)}{\omega^2-\Omega^2}d\Omega
\end{eqnarray}
This result was first obtained by Landau and Lifshitz\cite{Landau1984} in 1984.


\section{Details of Calculations of Sound Velocity Shift as the Logarithmic Function of Temperature}
In this chapter we want to give a detailed calculation of Eq.(\ref{28}) in section 4, with the assumption that the reduced imaginary part of resonance susceptibility ${\rm Im}\,\tilde{\chi}_{l,t}^{\rm res}(\omega, T)=\left(1-e^{-\beta\hbar\omega}\right)^{-1}\,{\rm Im}\,\chi_{l,t}^{\rm res}(\omega, T)$ is approximately a constant of frequency and temperature up to $\omega_c\sim 10^{15}$Hz and around the temperature of order 10K\cite{Pohl2002, Leggett2011}. We write Eq.(\ref{28}) in the following,
\begin{eqnarray}\label{E1}
\frac{\Delta c_{l,t}(T)-\Delta c_{l,t}(T_0)}{c_{l,t}( T_0)}\bigg|_{\rm res}= \frac{2}{2\pi\rho c_{l,t}^2}\mathcal{P}\int_0^{\infty}\frac{\Omega\left({\rm Im}\,\chi_{l,t}^{\rm res}(\Omega, T)-{\rm Im}\,\chi_{l,t}^{\rm res}(\Omega, T_0)\right)}{\Omega^2-\omega^2}d\Omega=\mathcal{C}_{l,t}\ln\left(\frac{T}{T_0}\right)
\end{eqnarray}
Let us use the reduced imaginary resonance susceptibility $\,{\rm Im}\,\tilde{\chi}_{l,t}^{\rm res}(\omega, T)\approx \,{\rm Im}\,\tilde{\chi}_{l,t}^{\rm res}$ in the above integral. Eq.(\ref{E1}) can be simplified as 
\begin{eqnarray}\label{E2}
\frac{1}{\pi\rho c_{l,t}^2}\mathcal{P}\int_0^{\infty}\frac{\Omega\left({\rm Im}\,\chi_{l,t}^{\rm res}(\Omega, T)-{\rm Im}\,\chi_{l,t}^{\rm res}(\Omega, T_0)\right)}{\Omega^2-\omega^2}d\Omega
 & = & \frac{\,{\rm Im\,}\tilde{\chi}_{l,t}^{\rm res}}{\pi\rho c_{l,t}^2}\mathcal{P}\int_0^{\infty}\frac{\Omega}{\Omega^2-\omega^2}\left[\left(1-e^{-\beta\hbar\Omega}\right)-\left(1-e^{-\beta_0\hbar\Omega}\right)\right]d\Omega\nonumber \\
 & = & \frac{\,{\rm Im\,}\tilde{\chi}_{l,t}^{\rm res}}{\pi\rho c_{l,t}^2}\mathcal{P}\int_0^{\infty}\frac{\Omega}{\Omega^2-\omega^2}\left(e^{-\beta_0\hbar\Omega}-e^{-\beta\hbar\Omega}\right)d\Omega
\end{eqnarray}
where we define $\beta_0=(k_BT_0)^{-1}$ and take the ``frequency and temperature independent quantity" $\,{\rm Im\,}\tilde{\chi}_{l,t}^{\rm res}$ out of the integral. Such frequency and temperature independence of $\,{\rm Im\,}\tilde{\chi}_{l,t}^{\rm res}(\omega, T)$ is observed in experiments\cite{Pohl2002, Hunklinger1974}. Also, according to the argument by D. C. Vural and A. J. Leggett\cite{Leggett2011}, the frequency dependence of imagnary part of non-elastic susceptibility $\,{\rm Im\,}\chi_{l,t}^{\rm res}(\omega)$ does not differ much between $\tanh(\omega/2k_BT)$ function (derived by TTLS model) and $\left(1-e^{-\beta\hbar\omega}\right)$ function (derived by multiple-level-system model). Therefore one would intuitive expect that the ``logarithmic temperature dependence of sound velocity shift" can be proved in arbitrary multiple-level-system as well. Please note that the above integral Eq.(\ref{E2}) has two nice properties: (1) it converges exponentially fast with the increase of frequency variable $\Omega$; (2) the ``principle value" removes the divengence when $\Omega$ approaches $\omega$. We will evaluate this principle integral in details as follows.

The ultrasonic sound velocity shift experiments are measured around the temperatures of order 10K, which means the input ultrasonic phonon energy $\hbar\omega\sim 10^{-28}$J is much smaller than $k_BT\sim 10^{-22}$J. Thus when the integral variable $\Omega$ approaches the singularity $\omega$, we have the approximation $\lim_{\Omega\to \omega}e^{-\beta \hbar \Omega}\approx 1$. The principle integral in Eq.(\ref{E2}) is therefore given as 
\begin{eqnarray}\label{E3}
\mathcal{P}\int_0^{\omega_{>}}\frac{\Omega}{\Omega^2-\omega^2}e^{-\beta\hbar\Omega}d\Omega & = & 
\lim_{\epsilon\to 0}\left(\int_0^{\omega-\epsilon}\frac{\Omega}{\Omega^2-\omega^2}e^{-\beta\hbar\Omega}d\Omega
+
\int_{\omega+\epsilon}^{\omega_>}\frac{\Omega}{\Omega^2-\omega^2}e^{-\beta\hbar\Omega}d\Omega\right)\nonumber \\
 & = & \frac{1}{2}\lim_{\epsilon\to 0}\left(\ln\frac{\epsilon}{\omega}+\ln\frac{\omega_>}{\epsilon}\right)
=\frac{1}{2}\ln\left(\frac{\omega_>}{\omega}\right)
\end{eqnarray}
where $\omega_>$ is ``some" upper cut-off of $\Omega$ integration. Since the function $e^{-\beta\hbar\Omega}$ exponentially decays with the increase of $\beta\Omega$, we know the upper cut-off of $\Omega$ must be some constant times the temperature $T$: $\omega_>(T)\propto T$. Thus Eq.(\ref{E2}) turns out to be 
\begin{eqnarray}\label{E4}
\frac{\Delta c_{l,t}(T)-\Delta c_{l,t}(T_0)}{c_{l,t}( T_0)}\bigg|_{\rm res} & = & \frac{\,{\rm Im\,}\tilde{\chi}_{l,t}^{\rm res}}{\pi\rho c_{l,t}^2}\mathcal{P}\int_0^{\infty}\frac{\Omega}{\Omega^2-\omega^2}\left(e^{-\beta_0\hbar\Omega}-e^{-\beta\hbar\Omega}\right)d\Omega\nonumber \\
 & = & \frac{\,{\rm Im\,}\tilde{\chi}_{l,t}^{\rm res}}{2\pi\rho c_{l,t}^2}\left[\ln\left(\frac{\omega_>(T_0)}{\omega}\right)-\ln\left(\frac{\omega_>(T)}{\omega}\right)\right]\nonumber \\
 & = & -\frac{\,{\rm Im\,}\tilde{\chi}_{l,t}^{\rm res}}{2\pi\rho c_{l,t}^2}\ln\left(\frac{\omega_>(T)}{\omega_>(T_0)}\right)
=-\frac{\,{\rm Im\,}\tilde{\chi}_{l,t}^{\rm res}}{2\pi\rho c_{l,t}^2}\ln\left(\frac{T}{T_0}\right)
\end{eqnarray}
in the above result, the coefficient $-\frac{\,{\rm Im\,}\tilde{\chi}_{l,t}^{\rm res}}{2\pi\rho c_{l,t}^2}$ is the constant $\mathcal{C}_{l,t}$ which appears in the final result of Eq.(\ref{28}).

\endwidetext

\end{document}